# Transforming Acidic Corrosion and Embrittlement into a Hydrogen-Trapping Cage

Ankang Chen[1]†, Jiewen Liu[1]†, Zihao Huo[1,5]†, Chuang Liu[1]† ,Yongming Sui[1]*, Xuan Liu[1], Qingkun Yuan[1], Yan Li[1], Guangtong Wang[1], Bao Yuan[4], Defang Duan[1,2]*, Gang Liu[3]*, Bo Zou[1,6]*

[1]State Key Laboratory of High Pressure and Superhard Materials, Synergetic Extreme Condition High-Pressure Science Center, College of Physics, Jilin University, Changchun, China.

[2]Key Laboratory of Material Simulation Methods and Software of Ministry of Education, College of Physics, Jilin University, Changchun, China.

[3]State Key Laboratory for Mechanical Behavior of Materials, Xi'an Jiaotong University, Xi'an, China.

[4]Spallation Neutron Source Science Center, Institute of High Energy Physics, Chinese Academy of Sciences (CAS), Dongguan, China.

[5]Key Laboratory of Functional Materials and Devices for Informatics of Anhui Educational Institutions, Fuyang Normal University, Fuyang, China.

[6]Department of Physics, College of Science, Yanbian University, Yanji, China.

*Corresponding authors: suiym@jlu.edu.cn ; duandf@jlu.edu.cn ; lgsammer@mail.xjtu.edu.cn ; zoubo@jlu.edu.cn.

## Abstract

The vision of a hydrogen economy demands efficient platforms to close the gap between sustainable proton sources and solid-state hydrogen carriers. Metal hydrides serve as key carriers, yet their synthesis remains constrained by the energy-intensive use of high-pressure $H_2$, which fragments the hydrogen chain. Here, we overturn this paradigm by transforming two classic degradation mechanisms, acidic corrosion and hydrogen embrittlement, into a constructive materials-design strategy. We demonstrate that synergistic control of these processes in acid enables the in-situ engineering of a "hydrogen-trapping cage" (HTC) microstructure within metals. Composed of a dense defect network, this cage directly captures and stabilizes protons as hydrides under mild conditions, guided by the universal criterion $|\Delta P_{eq}| > \Delta P_{ph}$. Using this platform, we synthesize over 20 hydrides, including challenging targets such as LiH and NaH, and showcase its functional power with a cage-rich titanium hydride electrocatalyst. This catalyst achieves an exceptional current density of 1.07 A $cm^{-2}$ for nitrate-to-ammonia conversion, attributed to rapid $H^-$ transport within the engineered cage. This work establishes a transformative "failure-to-function" paradigm, delivering an integrated platform that unifies hydrogen capture, stabilization, and conversion.

**Main Text:**

The central vision of a hydrogen economy lies in the efficient conversion of protons ($H^+$) derived from sustainable energy into readily storable and utilizable chemical energy[1,2]. Metal hydrides, serving as ideal solid-state hydrogen carriers, constitute a key materials platform for realizing this vision[3,4]. However, their synthesis has long been constrained by a fundamental chemical paradigm: metals are believed to react only with molecular hydrogen ($H_2$) to form hydrides, but cannot directly combine with protons ($H^+$) in solution[5,6]. This paradigm is rooted in classical electrochemical interfacial behavior: on a clean, intact metal surface, the hydrogen atom intermediate (H•(ads)) generated by proton reduction has an extremely short residence time, and the energy barrier for its recombination into $H_2$ molecules and desorption is very low[7,8]. This leads to a rapid and irreversible reaction pathway culminating in gas evolution. This basic chemical principal locks hydride synthesis into an indirect and energy-intensive route: protons must first be converted into high-purity, high-pressure $H_2$ gas via energy-intensive processes before they can react with metals[9-12]. This unavoidable "gaseous intermediate step" not only results in significant systemic energy losses and safety risks but also fundamentally fragments the hydrogen production, storage, and utilization chain, constituting a paradigmatic obstacle to overall efficiency[13].

To overcome this paradigm, a mechanism capable of intercepting interfacial hydrogen atoms (H•) and guiding their stable incorporation into the metal lattice, ultimately leading to hydride formation, is essential. Intriguingly, the core physical process of a classic failure mechanism in metallic materials, hydrogen embrittlement (HE), bears direct relevance to this challenge[14]. Traditionally, research on HE has focused on the mechanical degradation caused by stress-driven ingress of hydrogen into the lattice[15]. However, re-examining this "destructive" process from a materials-synthesis perspective reveals two critical synergistic effects: first, the aggregation of hydrogen atoms at defects induces substantial local lattice distortion, generating an equivalent pressure ($\Delta P_{eq}$) on the gigapascal scale, which thermodynamically creates the high-pressure environment required to stabilize the hydride phase[16,17]. Simultaneously, the micro-cracks and dislocation networks generated during this process provide channels for rapid penetration of hydrogen into the bulk metal, thereby kinetically lowering the energy barriers for hydrogen transport and hydride formation[18,19]. This insight highlights that the "destructive" and "constructive" aspects of HE is two sides of the same coin. A key challenge, however, remains: transforming this stochastic, uncontrollable failure mode into an actively guided and precisely regulated synthesis strategy. Achieving this would open a revolutionary pathway for converting protons ($H^+$) directly into stable hydrogen species within the lattice under mild conditions.

In this work, we introduce the paradigm of transforming acidic corrosion and hydrogen embrittlement into a designed functional microstructure, the in-situ engineered hydrogen-trapping cage (HTC). By synergistically controlling these coupled processes in acid, a three-dimensional HTC can be architected within various metals. This cage serves as an integrated functional unit: its defect-rich framework enables efficient proton capture; the local stress field ($\Delta P_{eq}$) stabilizes them as hydrides; and its interconnected channels facilitate rapid transport of active hydrogen species. We establish the quantitative criterion $|\Delta P_{eq}| > \Delta P_{ph}$ as the governing principle for this cage stabilization and hydride formation. Utilizing this platform, we demonstrate direct

conversion of acidic protons into stable hydrides for a broad range of metals via HTC construction. As a pivotal functional demonstration, we engineer a cage-rich titanium hydride (HTC-$TiH_2$) electrocatalyst that exhibits outstanding performance for nitrate-to-ammonia conversion. This approach transcends classical synthesis, offering a versatile platform that directly couples hydrogen capture, stabilization, and conversion into a single, designed material.

## Fabrication and General Synthesis of the Hydrogen-Trapping Cage

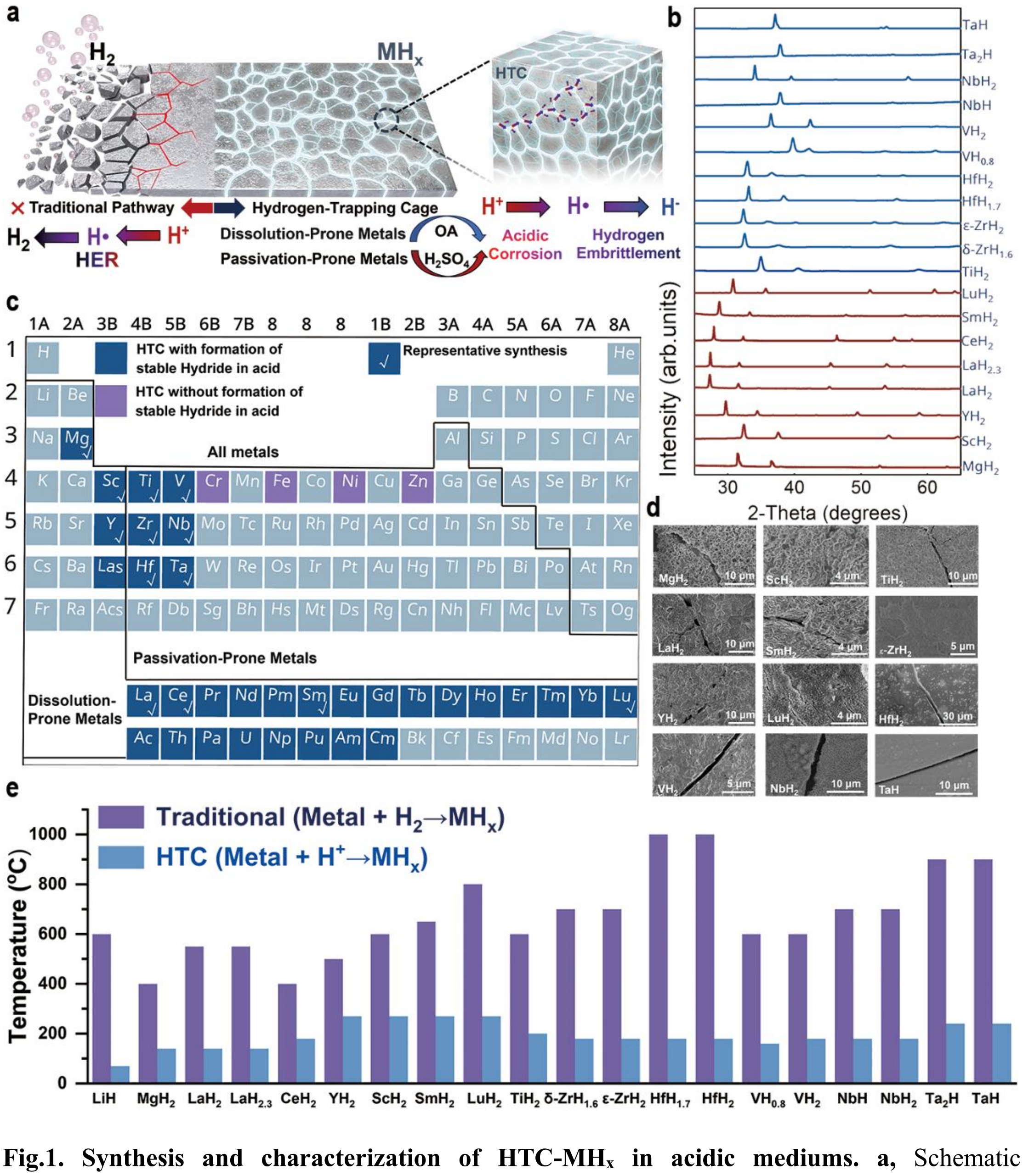

**Fig.1. Synthesis and characterization of HTC-$MH_x$ in acidic mediums. a,** Schematic representation of the synthesis of HTC-$MH_x$. **b**, XRD patterns of HTC-$MH_x$. **c,** Classification of hydrogen embrittlement mechanisms in critical metallic materials. **d**, Representative HTC SEM images of $MgH_2$, $LaH_2$, $YH_2$, $ScH_2$, $SmH_2$ and $LuH_2$ synthesized in the presence of oleic acid and $TiH_2$, ε-$ZrH_2$, $HfH_2$, $VH_2$, $NbH_2$ and TaH synthesized in the presence of $H_2SO_4$. **e**, Synthesizing temperature comparison between this work and traditional approaches.

The key to intercepting interfacial hydrogen atoms (H•) and guiding their stable embedding into the metal lattice lies in the active orchestration of two coupled processes: corrosion-driven hydrogen evolution and subsequent hydrogen embrittlement (HE). To this end, we developed an acid-mediated synthesis strategy that transcends mere phase formation; it is designed to in situ engineer a functional microstructure, a "hydrogen-trapping cage" (HTC), within the metal matrix (Fig.1a). The core of this strategy is to control the initiation of defects and the flux of hydrogen atoms by designing acid media tailored to the intrinsic reactivity of different metals. This transforms the traditionally random and deleterious HE processes into a controllable pathway for directed materials architecture.

Guided by this principle, we designed differentiated acid media strategies. For Group IVB/VB metals (e.g., Ti, Zr) that form dense oxide films, we used inorganic strong acids (e.g., $H_2SO_4$) with $F^-$ as a chemical etchant. This combination removes the passivation layer, exposes a fresh metal surface, and triggers intense localized hydrogen evolution[20]. The synergy of rapid hydrogen release and chemical corrosion intentionally introduces a controllable network of micrometer-scale cracks (Fig. 1d), which act as prefabricated channels for bulk hydrogen penetration (Supplementary Fig. 14 to 24).

| **Met.** | **SG** | **HTC-MH$_x$** | **SG** | **H$^+$** | **Conc.** | **S** | **RT** | **T** |
|---|---|---|---|---|---|---|---|---|
| Mg | $P6_3/mmc$ | $MgH_2$ | $P4_2/mnm$ | OA | ⩾99%(CG) | None | 140 ℃ | 4 h |
| Sc | $P6_3/mmc$ | $ScH_2$ | $Fm\bar{3}m$ | OA | ⩾99%(CG) | None | 270 ℃ | 60 h |
| Y | $P6_3/mmc$ | $YH_2$ | $Fm\bar{3}m$ | OA | ⩾99%(CG) | None | 270 ℃ | 10 h |
| La | $P6_3/mmc$ | $LaH_2$ | $Fm\bar{3}m$ | OA | ⩾99%(CG) | None | 140 ℃ | 10 h |
| La | $P6_3/mmc$ | $LaH_{2.3}$ | $Fm\bar{3}m$ | OA | ⩾99%(CG) | None | 140 ℃ | 12 h |
| Ce | $Fm\bar{3}m$ | $CeH_2$ | $Fm\bar{3}m$ | OA | ⩾99%(CG) | None | 180 ℃ | 10 h |
| Lu | $P6_3/mmc$ | $LuH_2$ | $Fm\bar{3}m$ | OA | ⩾99%(CG) | None | 270 ℃ | 80 h |
| Sm | $P6_3/mmc$ | $SmH_2$ | $Fm\bar{3}m$ | OA | ⩾99%(CG) | None | 270 ℃ | 20 h |
| Ti | $P6_3/mmc$ | $TiH_2$ | $Fm\bar{3}m$ | $H_2SO_4$ | 0.1 mol/L | None | 200 ℃ | 6 h |
| Zr | $P6_3/mmc$ | δ-$ZrH_{1.6}$ | $Fm\bar{3}m$ | $H_2SO_4$ | 0.1 mol/L | NaF | 180 ℃ | 8 h |
| Zr | $P6_3/mmc$ | ε-$ZrH_2$ | $I4/mmm$ | $H_2SO_4$ | 0.1 mol/L | NaF | 180 ℃ | 10 h |
| Hf | $P6_3/mmc$ | $HfH_{1.7}$ | $Fm\bar{3}m$ | $H_2SO_4$ | 1 mol/L | NaF | 180 ℃ | 8 h |
| Hf | $P6_3/mmc$ | $HfH_2$ | I4/mmm | $H_2SO_4$ | 1 mol/L | NaF | 180 ℃ | 10 h |
| V | $Im\bar{3}m$ | $VH_{0.8}$ | $Fm\bar{3}m$ | $H_2SO_4$ | 1 mol/L | NaF | 160 ℃ | 2 h |
| V | $Im\bar{3}m$ | $VH_2$ | $Fm\bar{3}m$ | $H_2SO_4$ | 1 mol/L | None | 180 ℃ | 2 h |
| Nb | $Im\bar{3}m$ | NbH | $Fm\bar{3}m$ | $H_2SO_4$ | 0.1 mol/L | None | 180 ℃ | 7 h |
| Nb | $Im\bar{3}m$ | $NbH_2$ | $I4/mmm$ | $H_2SO_4$ | 0.1 mol/L | NaF | 180 ℃ | 10 h |
| Ta | $Im\bar{3}m$ | $Ta_2H$ | *C*222 | $H_2SO_4$ | 0.2 mol/L | NaF | 240 ℃ | 8 h |
| Ta | $Im\bar{3}m$ | TaH | *C*222 | $H_2SO_4$ | 0.2 mol/L | NaF | 240 ℃ | 10 h |

**Table . 1. Synthesis conditions of HTC-$MH_x$. Met.**: Metals used in the reaction; **SG**: Space group of Metal and hydride; **HTC-$MH_x$**: Hydride formed in acids mediums. **$H^+$:** Hydrogen source. **Conc.:** Concentration of reagents. **S**: Solute added to treat surface oxides. **RT**: Reaction Temperature **T**: Reaction Time. Except for metals with highly electrochemical activity such as Mg and La, most metal systems can be synthesized under air environment conditions, and no significant oxide species is detected in the products.

For highly active metals like Mg and those in Group IIIB (e.g., La, Y), which are prone to overall dissolution, we employed organic weak acids such as oleic acid as mild etchants and proton sources[21]. The carboxylate group (-$COO^-$) modulates the reaction via surface coordination, inhibiting excessive dissolution while enabling a sustained $H^+$ supply[22]. This ensures that hydrogen atom generation is synchronized with controlled, progressive corrosion, preventing material pulverization and favoring guided microstructural evolution (Supplementary Fig. 2 to 9). Deuterated oleic acid experiments combined with neutron powder diffraction confirmed that the product hydrogen originates entirely from the acid's carboxyl group, mechanistically clarifying this pathway[23] (Supplementary Fig. 30).

This targeted regulation does more than initiate hydride formation; it actively orchestrates corrosion and embrittlement to construct a distinct microarchitecture, the Hydrogen-Trapping Cage (HTC). The HTC is defined by its three-dimensional network of micro-cracks, dislocations, and defects (Fig. 1d), which are not byproducts but essential structural features intentionally forged during synthesis (Supplementary Fig. 2 to 24). The cage serves a dual function: its interconnected channels enable rapid hydrogen transport throughout the bulk, while the intense local stress fields generated at defect sites (quantified as $\Delta P_{eq}$) provide the thermodynamic driving force to trap and stabilize hydrogen. This cage-engineering strategy successfully converts protons directly into a library of over 20 high-purity metal hydrides spanning Groups 1A, 2A, and 3-5B transition metals (Fig. 1c, Table 1), as confirmed by XRD (Fig. 1b). The universality of this platform is further underscored by a clear correlation between the acid's pKa and reaction kinetics, and by the practical synthesis of $MgH_2$ using inexpensive soybean oil (Supplementary Fig. 1 to 23).

| Dimension | Parameter | Conventional High-Pressure $H_2$ Method | This Work: HTC | Key Implications & Advantages |
|---|---|---|---|---|
| Reaction Pathway | Representative Equation | $M + (x/2)\ H_2 \rightarrow MH_x$ | $2M + x\ H^+ \rightarrow M^{x+} + MH_x$ | Fundamentally different chemistry: Uses cheap protons instead of gaseous $H_2$; Proton capture via an acid- and embrittlement-formed hydrogen-trapping cage. |
| Reaction Conditions | Typical Temperature | 400 – 800 °C | 70 – 270 °C | >~70% reduction in energy consumption; markedly milder conditions. |
| | Typical Pressure | 2 – 10 MPa $H_2$ | Ambient (~0.1 MPa) | Inherently safer; eliminates costly high-pressure apparatus. |
| | Hydrogen Source | High-purity $H_2$ (≥99.999%) | Low-cost acids (e.g., $H_2SO_4$, Oleic Acid) | Dramatic reduction in hydrogen source cost. |
| Synthesis Efficiency | Yield (e.g., $TiH_2$) | ~90% | >98% (This work) | Near-quantitative conversion to the target hydride. |
| | Metal Atom Economy | ~100% (Theoretical) | ~50% (for $MH_x$ formation) | A strategic trade-off for system-level savings in cost and safety. |
| Economic & Scalability | Primary Energy Cost | $H_2$ Compression + High-Temp Heating | Mild Heating Only | Eliminates the most energy-intensive step. |
| | Capital Expenditure (CAPEX) | High (High-Pressure Reactors, Safety Systems) | Low (Ambient-Pressure Reactors) | Significantly lower initial investment. |
| | Scalability Demonstration | Established | Successful synthesis using crude soybean oil | High potential for cost-effective industrial scaling. |
| Product Characteristics | Product Purity | High | High-purity phase (XRD/NPD confirmed) | Comparable purity without significant oxide impurities. |
| | In-situ Microstructure Engineering | Typically Dense | Rich in defects (micro-cracks, dislocations) | The synthesis path concurrently constructs a unique microstructure, which is highly beneficial for functional applications (e.g., enhanced ion transport, see later sections). |

**Table 2. Comparison of Synthesis Routes for Metal Hydrides: Conventional High-Pressure $H_2$ vs. HTC-$MH_x$ in acid.**

The hydrogen-trapping cage (HTC) engineering strategy fundamentally reshapes the energy landscape of hydride manufacturing. In contrast to traditional routes that rely on external high-pressure $H_2$, our approach leverages the intrinsic local stress ($\Delta P_{eq}$) generated within the cage to drive hydride formation, enabling significantly milder conditions (70-270 °C) at ambient pressure (Table 2). This eliminates the energy-intensive compression and handling of gaseous $H_2$, drastically reducing systemic energy consumption, cost, and safety hazards. More profoundly, by directly utilizing acidic protons, the strategy decouples hydride synthesis from the centralized hydrogen infrastructure, offering a more direct and potentially distributed pathway.

Crucially, the comparison in Table 2 reveals that our method's advantage extends beyond mild conditions. The HTC is not merely a synthetic byproduct but the core functional output. Unlike conventional synthesis, which typically yields dense microstructures, our process is uniquely defined by its in-situ construction of a defect-rich, cage-like architecture. This intentional structural design is what elevates the HTC platform from a synthetic alternative to an integrated materials design and functionalization system. The cage's inherent network of channels and active interfaces, as engineered during synthesis, lays the direct foundation for the exceptional catalytic performance demonstrated in the following sections.

**HTC Stabilization: The $|\Delta P_{eq}| > \Delta P_{ph}$ Criterion**

We have confirmed the general ability to construct hydrogen-trapping cage (HTC) structures in many metals. However, a fundamental selectivity emerges: Group III–V transition metals (e.g., Sc, Ti, V) together with Mg readily form stable HTC-hydride systems, whereas metals such as Fe and Ni do not (Fig. 1c)[24]. This stark divergence points to a deeper question: what determines whether a robust HTC can be established. To move beyond phenomenological correlation and establish a predictive design rule, we turn to the microscopic physical essence of hydrogen embrittlement, the very process that builds the cage. During HE, stress-driven hydrogen enrichment at defects induces severe local lattice distortion ($\Delta V/V_0 \neq 0$)[25], which is thermodynamically equivalent to an internal hydrostatic pressure. This cage-generated local pressure is quantified as an equivalent pressure $\Delta P_{eq} = K \cdot \Delta V/V_0$ (where K is the bulk modulus), reaching gigapascal levels. In this section, we combine this concept with direct high-pressure experiments using a diamond-anvil cell (DAC) to calibrate the critical threshold required for cage stability. Our aim is to elucidate a unified mechanism and a quantitative criterion that explain why an effective HTC can be formed in some metals but not in others, thereby providing a physical principle for the design of cage-stabilized hydride materials.

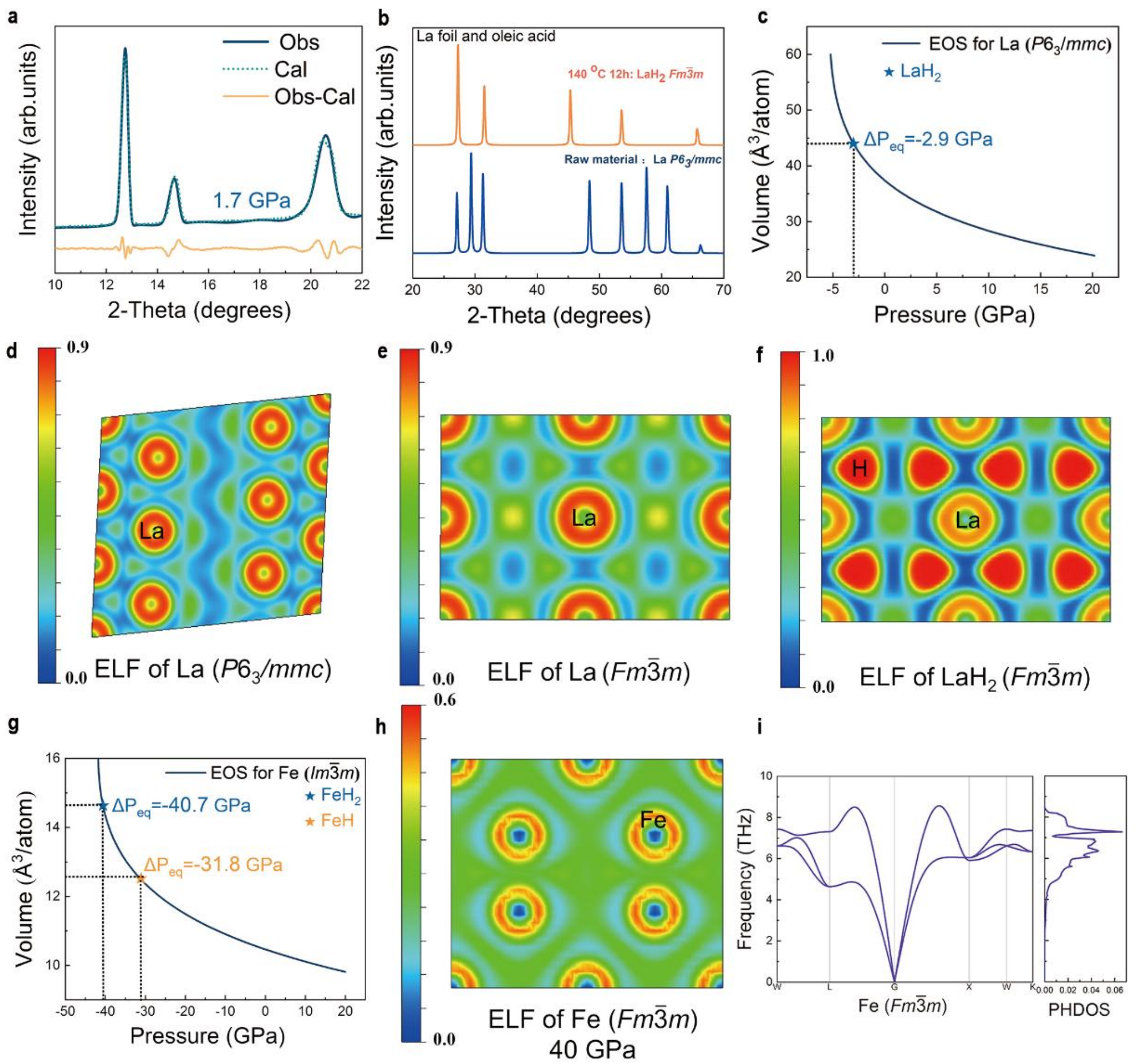

**Fig.2 The mechanism of synthesis. a**, XRD refinement pattern of $LaH_2$ at 1.7 GPa. **b**, XRD pattern of $LaH_2$. **c**, $\Delta P_{eq}$ of $LaH_2$. **d**, ELFs of La ($P6_3/mmc$). **e**, La ($Fm\bar{3}m$). **f**, $LaH_2$ ($Fm\bar{3}m$). **g**, $\Delta P_{eq}$ of FeH and $FeH_2$. **h**, ELFs of La ($P6_3/mmc$). **i**, Computational simulation of the Fe ($Fm\bar{3}m$) phonon spectrum.

To decipher the stabilization mechanism of the hydrogen-trapping cage (HTC), we begin with the La-H system as a representative case. X-ray diffraction (XRD) analysis of the sample formed via acid-mediated hydrogen embrittlement reveals that the process drives a structural transformation from the initial La ($P6_3$/mmc) phase (Fig. 2d) to a distorted $Fm\bar{3}m$ sublattice (Fig. 2e). Calculating the volume change of this sublattice and its equation of state yields an equivalent local pressure of $\Delta P_{eq}$ = – 2.9 GPa (Fig. 2c). This quantifies the cage-generated tensile stress that actively participates in the HTC's self-assembly. We then employed a diamond-anvil cell (DAC) to determine the physical critical pressure required to form $LaH_2$, obtaining $\Delta P_{ph}$ = 1.7 GPa (Fig. 2a). The result clearly shows that the intrinsic pressure generated within the cage ($|\Delta P_{eq}|$) exceeds the external pressure threshold ($\Delta P_{ph}$) needed for phase stability. This establishes $|\Delta P_{eq}| > \Delta P_{ph}$ as the fundamental criterion for the successful stabilization of a HTC, explaining why La forms a stable cage-hydride system while other metals (e.g., Fe) may not (Supplementary Figs 30 to 48, Table 1)[26-28].

Electron-localization-function (ELF) analysis further reveals how this local pressure orchestrates the atomic-scale locking mechanism within the cage. In metallic La ($P6_3/mmc$), the ELF is ~0.5 (Fig. 2d), indicating delocalized electrons. Under the cage-generated $\Delta P_{eq}$, hydrogen incorporation induces a transition to a metastable $Fm\bar{3}m$ phase, around which the ELF rises sharply to ~0.7 (Fig. 2f). This signifies strong electron localization. These localized electrons, through La-5d and H-1s orbital hybridization, enhance the ionic character of the La-H bond, thereby locking hydrogen into specific lattice sites while simultaneously reinforcing the metastable cage framework, self-consistent electron-structure feedback that defines the cage's core stability (Fig. 2g). Crucially, this feedback is activated only when $|\Delta P_{eq}| \geq \Delta P_{ph}$, establishing a dynamic balance between lattice distortion and electron redistribution that drives the system from a reversible hydrogen-adsorption state toward a stable, integrated HTC. This mechanism is general and underpins the stabilization of diverse HTC-hydride systems (Supplementary Figs 50 to 62).

| $MH_x$ | $\Delta P_{eq}$ | $\Delta P_{ph}$ | Stable at 1 atm |
|---|---|---|---|
| $MgH_2$ | -4.7 GPa | ≤0.1 GPa[29] | √ |
| LaH | 0.65 GPa | 11 GPa[30] | ×, 8 GPa[30] |
| $LaH_2$ | -2.9 GPa | ≤1.7 GPa[31] | √ |
| $LaH_{2.3}$ | -3.0 GPa | ≤1.7 GPa[31] | √ |
| $CeH_2$ | -6.2 GPa | ≤1 GPa[32] | √ |
| $ScH_2$ | -2.6 GPa | ≤0.01 GPa[33] | √ |
| $YH_2$ | -1.33 GPa | ≤0.01 GPa[34] | √ |
| $YH_3$ | -3.46 GPa | ≤4 GPa[34] | √ |
| $SmH_2$ | -4 GPa | ≤0.1 GPa[35] | √ |
| $LuH_2$ | -2.5 GPa | ≤2 GPa[36] | √ |
| $TiH_2$ | -16.1 GPa | ≤0.01 GPa[37] | √ |
| $ZrH_{1.6}$ | -12.25 GPa | ≤1 GPa[28] | √ |
| $ZrH_2$ | -13.6 GPa | ≤1 GPa[28] | √ |
| $HfH_{1.7}$ | -4.6 GPa | ≤4 GPa[29] | √ |
| $HfH_2$ | -5.6 GPa | ≤4 GPa[29] | √ |
| $VH_{0.8}$ | -16.3 GPa | ≤0.01 GPa[38] | √ |
| $VH_2$ | -28.8 GPa | ≤0.01 GPa[38] | √ |
| NbH | -1.24 GPa | ≤1 GPa[39] | √ |
| $NbH_2$ | -26.4 GPa | ≤22 GPa[39] | √ |
| $Ta_2H$ | -6 GPa | ≤5 GPa[27] | √ |
| TaH | -16 GPa | ≤5 GPa[27] | √ |
| FeH | -13.6 | 3.5 GPa[40] | ×, 3.2 GPa[40] |
| $FeH_2$ | -40.7 GPa | 67 GPa[41] | ×, 23 GPa[41] |

**Table. 3. $\Delta P_{eq}$ and $\Delta P_{ph}$ of $MH_x$.**

To elucidate the fundamental inability to form a stable hydrogen-trapping cage (HTC) in iron-based materials via the acid-mediated process, we analyzed the Fe-H system. For stoichiometric $FeH_2$, the physical critical pressure required is $\Delta P_{ph}$ = 67 GPa[41]. Assuming the hypothetical formation of such a cage, we calculated the equivalent pressure $\Delta P_{eq}$ = 40.7 GPa from the lattice distortion of the metal sublattice before and after hydrogenation (Fig. 2g). This value, which reflects the pressure generated by the cage's own deformation, falls below the physical threshold, critically violating the essential stability criterion $|\Delta P_{eq}| > \Delta P_{ph}$. First-principles calculations reveal that iron-based hydrides exhibit insufficient electron localization even at 40 GPa (Fig. 2h). Consequently, the electronic "locking" mechanism within the cage fails, unable to stabilize the crystal framework or firmly retain hydrogen atoms, fundamentally preventing the assembly of a stable, high-hydrogen-content HTC in iron. For the low-hydrogen phase FeH, although $|\Delta P_{eq}|$ meets the $\Delta P_{ph}$ = 3.5 GPa threshold[40], phonon calculations show dynamical stability only above ambient pressure (Fig. 2i). Upon stress release, hydrogen desorption followed by lattice cleavage occurs, explaining the characteristic hydride-free brittle fracture of iron-based materials. These findings are fully consistent with the established consensus that hydrogen embrittlement in Fe-based systems occurs without hydride participation[42,43], while providing a clear physical picture from the HTC perspective: iron cannot provide the simultaneous thermodynamic and kinetic conditions required to assemble an electronically self-consistent and stable hydrogen-trapping cage under mild conditions.

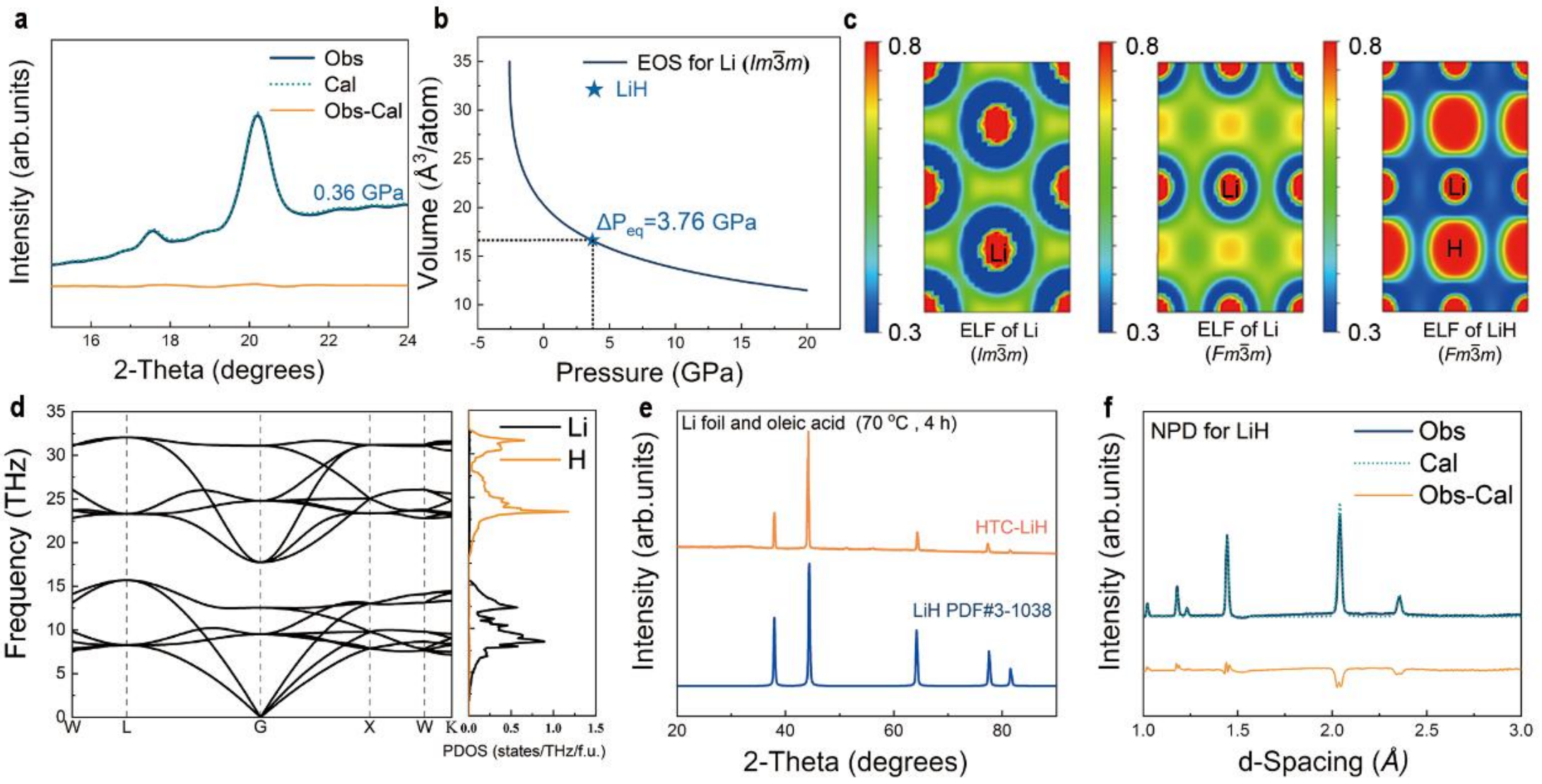

**Fig. 3. Synthesis of LiH. a**, XRD refinement pattern of LiH at 0.36 GPa. **b**, $\Delta P_{eq}$ of LiH. **c**, ELFs of Li ($Im\bar{3}m$), Li ($Fm\bar{3}m$) and LiH ($Fm\bar{3}m$). **d**, Computational simulation of the LiH ($Fm\bar{3}m$) phonon spectrum. **e**, XRD pattern of LiH. **f**, NPD refinement pattern of LiH.

Conventionally, metallic Li undergoes preferential hydrogen evolution over hydride formation in acidic media due to its extreme reactivity. To assess the possibility of constructing a hydrogen-trapping cage (HTC) in lithium, we first determined, via DAC measurements, the physical critical pressure required to form LiH as $\Delta P_{ph} = 0.36$ GPa[44] (Fig. 3a). We then examined the scenario assuming HTC formation: structural calculations of the metal sublattice before and after hydrogenation yield an equivalent local pressure of $\Delta P_{eq} = +3.76$ GPa (Fig. 3b). This value significantly exceeds $\Delta P_{ph}$, satisfying the fundamental criterion $|\Delta P_{eq}| > \Delta Pph$ for cage stabilization. First-principles calculations further confirm that a stable HTC in lithium is feasible:

thermodynamic stability is evidenced by strong electron localization (ELF > 0.75, Fig. 3e), and kinetic stability is verified by phonon spectra without imaginary frequencies at ambient pressure (Fig. 3f). Guided by this theoretical prediction, LiH was successfully synthesized via acid-mediated controlled HE under mild conditions (70 °C, 4 h, oleic acid). XRD (Fig. 3e) and NPD (Fig. 3f) confirmed phase-pure LiH with the standard structure, while SEM revealed the characteristic crack-rich morphology (Supplementary Fig. 26), signifying the in-situ formation of the HTC microstructure as designed. Thus, the quantitative criterion not only deciphers the mechanism but also provides a predictive design rule for cage engineering, as demonstrated by the successful and mild synthesis of this challenging hydride. Importantly, the applicability of this strategy is not limited to LiH; it can also be extended to the synthesis of other reactive hydrides like sodium hydride (NaH), underscoring the generality and power of our HTC platform.

**HTC-Powered Hydrogen Conversion**

The performance of metal hydride-mediated electrocatalytic reactions, such as nitrate reduction to ammonia, is fundamentally governed by the efficiency of active hydrogen species transport[45]. Conventional metal hydrides (e.g., $TiH_2$), while offering the beneficial involvement of lattice hydride ions ($H^-$), suffer from inherently sluggish $H^-$ mobility. This limitation manifests as a severe performance decline under low driving forces, where the hydrogen evolution reaction (HER) outcompetes the desired pathway[4]. We posit that the hydrogen-trapping cage (HTC), constructed in situ during our acid-mediated process, provides an ideal architecture to overcome this bottleneck through its designed multifunctionality: the cage walls and scaffold offer abundant active sites; its internal interconnected channels serve as highways for rapid $H^-$ transport; and the integrated structure enables dynamic hydrogen exchange.

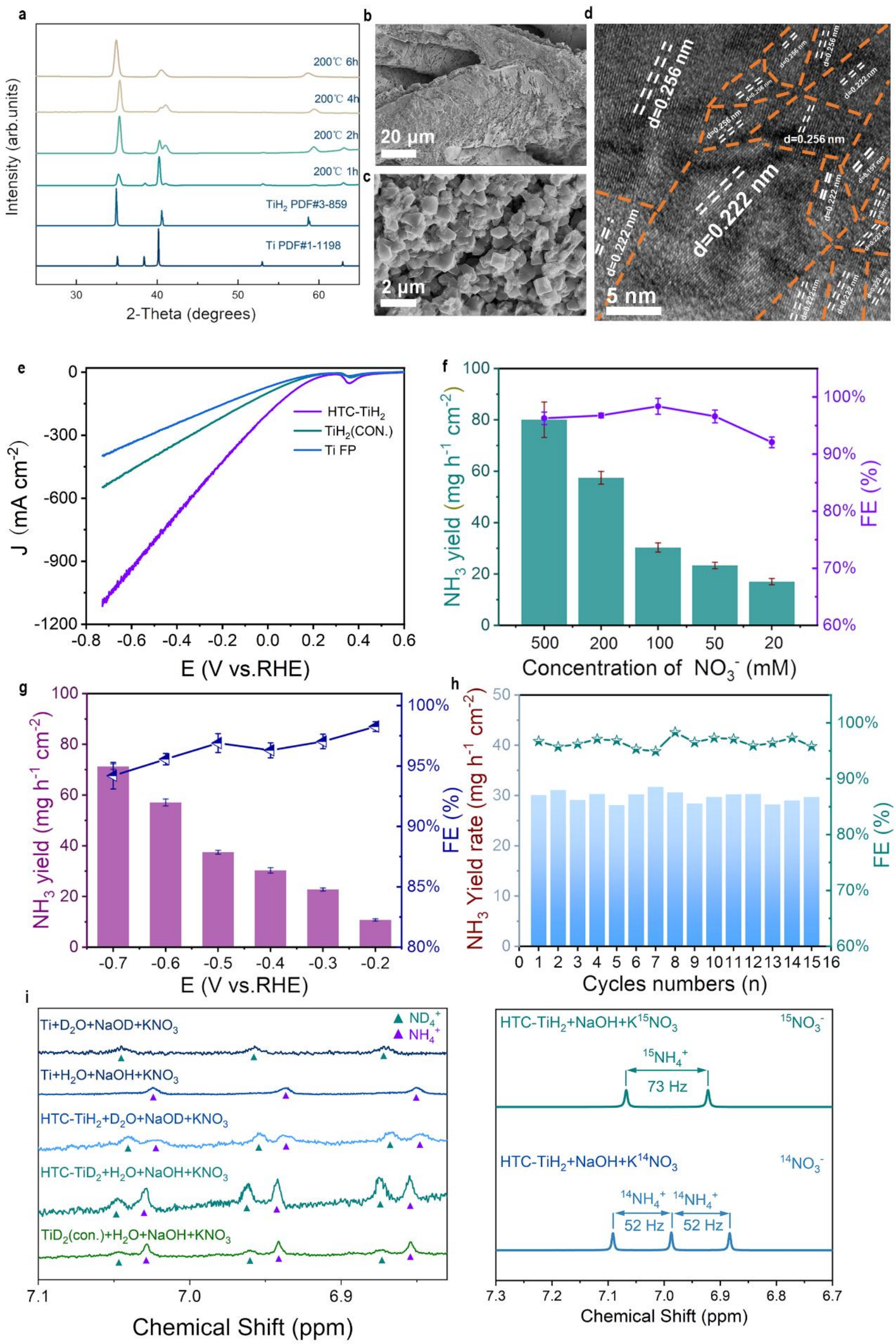

**Fig. 4. Nitrate reduction reaction of HTC-TiH₂. a**, XRD phase analysis under varying synthetic conditions. **b**, Representative SEM image under reaction conditions of 200 ℃ and 4

hours. **c**, Representative SEM image of HTC-$TiH_2$. **d**, Representative TEM image of HTC-$TiH_2$. **e**, LSV curves of Ti FP, $TiH_2$(con.) and HTC-$TiH_2$. **F**, $NH_3$ yields and FEs for titanium hydride at different concentrations of $NO_3^-$(without iR compensation). **g**, $NH_3$ yields and FEs for titanium hydride against various work potentials in 1 M KOH solution with 0.1 M $NO_3^-$(without iR compensation). **h**, $NH_3$ yield rate and FE of Cu NWs-SR catalysts for stable 15 cycles in the electrolyte containing 0.1 M $NO_3^-$ at -0.7 V (vs. RHE). **i**, Isotope experiment on hydrogen source in ammonia. **g**, Isotope experiment on nitrogen source in ammonia.

Guided by this principle, we engineered a model HTC electrocatalyst by transforming titanium fiber paper via controlled acid reaction (Fig. 4a, Supplementary Fig. 66 to 67). The resulting $TiH_2$ exhibits the characteristic porous, defect-rich microstructure of an HTC, featuring microcracks, etching trenches, and dislocation networks (Fig. 4b to 4d), in stark contrast to the dense morphology of conventional $TiH_2$ (Supplementary Fig. 68). This is the deliberate architectural outcome of orchestrated corrosion and embrittlement.

Electrocatalytic evaluation confirms the functional superiority of this design. HTC-$TiH_2$ achieves a current density of 1.07 A $cm^{-2}$ at -0.7 V vs. RHE (Fig. 4e), outperforming pristine Ti and conventional $TiH_2$ by factors of 2.8 and 2, respectively. It also maintains high Faradaic efficiency over a wide potential range and under varying nitrate concentrations (Fig. 4f,g), demonstrating sustained activity where conventional hydrides typically fail. The HTC structure shows excellent long-term stability (Fig. 4h). The preservation of the titanium hydride phase, as confirmed by XRD, coupled with the retention of a high density of defects as evidenced by TEM/SEM analysis, collectively demonstrates the outstanding structural and catalytic stability of the sample throughout the $NO_3^-$RR process(Supplementary Figs 69).

To gain deeper insight into the mechanism behind the performance enhancement, we conducted systematic isotope tracing experiments to probe the hydrogen source (Fig. 4i). When pure Ti was used as the catalyst, the $^1H$ NMR spectrum detected ammonia signals originating solely from hydrogen in the electrolyte, indicating that its catalytic process relies entirely on external hydrogen sources. In contrast, when HTC-$TiH_2$ was tested as the catalyst in a $D_2O$-based electrolyte, the $^1H$ NMR spectrum clearly revealed characteristic $NH_4^+$ peaks derived from lattice hydrogen. Conversely, when titanium deuteride was employed as the catalyst in an $H_2O$ system, distinct $ND_4^+$ signals originating from lattice deuterium could be detected. These results collectively confirm the existence of a dynamic and reversible exchange cycle between lattice hydrogen and solution hydrogen in HTC-$TiH_2$, and demonstrate that the HTC structure greatly facilitates this exchange process. More importantly, under identical $H_2O$ testing conditions, the $ND_4^+$ signal intensity generated by HTC-$TiD_2$ was significantly stronger than that from conventional $TiD_2$, directly proving that the HTC architecture substantially enhances the reactivity and mobility of $H^-$ within the material. These findings indicate that the HTC structure enables efficient capture, lattice stabilization, and rapid conversion of hydrogen species ($H^-/H^+$): lattice $H^-$ migrates rapidly via the cage channels to the surface to participate in nitrate hydrogenation, while protons from the solution are promptly captured and stabilized in the lattice, thereby enabling sustained catalytic cycling and structural stability (Supplementary Fig. 69). Furthermore, nitrogen source isotope labeling experiments using $^{14}NO_3^-$ and $15NO_3^-$ confirmed that the nitrogen in the produced ammonia originates entirely from nitrate, effectively ruling out the possibility of nitrogen contamination from other sources (Fig. 4j).

In summary, the exceptional performance stems directly from the multifunctional design of the engineered hydrogen-trapping cage (HTC). The cage's defect-rich scaffold enables efficient proton capture, the local stress field ($\Delta P_{eq}$) and electron localization ensure hydrogen stabilization within the lattice, and its 3D interconnected channels facilitate ultrafast hydrogen transport and conversion. This work thus demonstrates that the HTC is a functional materials platform in which the synthesis pathway inherently constructs the architecture for high-performance hydrogen management.

## Conclusion

This work establishes the paradigm of transforming acidic corrosion and hydrogen embrittlement into a functional microstructure, the in-situ engineered hydrogen-trapping cage (HTC). Central to this approach is the quantitative criterion $|\Delta P_{eq}| > \Delta P_{ph}$, which deciphers the stabilization mechanism and enables predictive synthesis, as demonstrated by the targeted preparation of challenging hydrides such as LiH and NaH. This criterion underpins our cage-engineering platform. The power of this platform is exemplified by a cage-rich titanium hydride electrocatalyst, which achieves a high current density of 1.07 $A \cdot cm^{-2}$ with stable efficiency, a direct result of the HTC's integrated capabilities for hydrogen capture, stabilization, and on-demand conversion within a single architecture. By transcending classical reaction models, we have created an integrated system that unifies these three key steps.

This "destruction-to-construction" methodology provides a practical, efficient route to hydrides and opens avenues for next-generation functional materials where in-situ defect engineering is crucial. More broadly, the principle of harnessing failure mechanisms to build functional microstructures offers a transformative design philosophy that could inspire innovative strategies across materials chemistry, paving the way toward a more sustainable and integrated energy future.

## References and Notes

1 Dresselhaus, M. S. & Thomas, I. L. Alternative energy technologies. *Nature* **414**, 332-337 (2001).

2 Turner, J. A. Sustainable Hydrogen Production. *Science* **305**, 972-974 (2004).

3 Griessen, R., Strohfeldt, N. & Giessen, H. Thermodynamics of the hybrid interaction of hydrogen with palladium nanoparticles. *Nature Materials* **15**, 311-317 (2016).

4 Zhang, X. *et al.* Atomic reconstruction for realizing stable solar-driven reversible hydrogen storage of magnesium hydride. *Nature Communications* **15**, 2815 (2024).

5 Allendorf, M. D. *et al.* An assessment of strategies for the development of solid-state adsorbents for vehicular hydrogen storage. *Energy & Environmental Science* **11**, 2784-2812 (2018).

6 Pasini, J. M. *et al.* Metal hydride material requirements for automotive hydrogen storage systems. *International journal of hydrogen energy* **38**, 9755-9765 (2013).

7 Bhatia, S. K. & Myers, A. L. Optimum Conditions for Adsorptive Storage. *Langmuir* **22**, 1688-1700 (2006).

8 Witman, M. *et al.* Extracting an Empirical Intermetallic Hydride Design Principle from Limited Data via Interpretable Machine Learning. *The Journal of Physical Chemistry Letters* **11**, 40-47 (2020).
9 Guo, X. *et al.* Direct, Nonoxidative Conversion of Methane to Ethylene, Aromatics, and Hydrogen. *Science* **344**, 616-619 (2014).
10 Yuan, Y. *et al.* Steam methane reforming using a regenerable antenna–reactor plasmonic photocatalyst. *Nature Catalysis* **7**, 1339-1349 (2024).
11 Li, A. *et al.* Atomically dispersed hexavalent iridium oxide from $MnO_2$ reduction for oxygen evolution catalysis. *Science* **384**, 666-670 (2024).
12 Chu, S. & Majumdar, A. Opportunities and challenges for a sustainable energy future. *Nature* **488**, 294-303 (2012).
13 Allendorf, M. D. *et al.* Challenges to developing materials for the transport and storage of hydrogen. *Nature Chemistry* **14**, 1214-1223 (2022).
14 Johnson, W. H. On Some Remarkable Changes Produced in Iron and Steel by the Action of Hydrogen and Acids. *Nature* **11**, 393-393 (1875).
15 Jiang, S. *et al.* Structurally complex phase engineering enables hydrogen-tolerant Al alloys. *Nature* (2025).
16 Westlake, D. GENERALIZED MODEL FOR HYDROGEN EMBRITTLEMENT. (Argonne National Lab., Ill., 1969).
17 Forcey, K., Ross, D. & Earwaker, L. Investigation of the effectiveness of oxidised fecralloy as a containment for tritium in fusion reactors. *Zeitschrift für Physikalische Chemie* **143**, 213-228 (1985).
18 Qiao, L.-j. *et al.* The concentration of hydrogen at crack tip of austenitic stainless steel after stress corrosion and polarization. *Scripta metallurgica* **22**, 627-630 (1988).
19 El-Meligi, A. A. Hydrogen production by aluminum corrosion in hydrochloric acid and using inhibitors to control hydrogen evolution. *International Journal of Hydrogen Energy* **36**, 10600-10607 (2011).
20 Li, Z., Higuchi, E., Liu, B. & Suda, S. Electrochemical properties and characteristics of a fluorinated AB2-alloy. *Journal of alloys and compounds* **293**, 593-600 (1999).
21 Ajay, P., Velkannan, V., Ram Kumar, P. & Kottaisamy, M. Enhancing corrosion resistance of mild steel with a hybrid surface coating: Film-forming zinc titanate/oleic acid composites. *Ceramics International* **51**, 751-763 (2025).
22 Caldona, E. B., Smith Jr, D. W. & Wipf, D. O. Surface electroanalytical approaches to organic polymeric coatings. *Polymer International* **70**, 927-937 (2021).
23 Xu, J. *et al.* Multi-physics instrument: Total scattering neutron time-of-flight diffractometer at China Spallation Neutron Source. *Nuclear Instruments and Methods in Physics Research Section A: Accelerators, Spectrometers, Detectors and Associated Equipment* **1013**, 165642 (2021).
24 Guo, X. *et al.* Self-accelerated corrosion of nuclear waste forms at material interfaces. *Nature Materials* **19**, 310-316 (2020).
25 Bard, A. J., Parsons, R. & Jordan, J. *Standard potentials in aqueous solution*. (Routledge, 2017).
26 Kuzovnikov, M. A., Tkacz, M., Meng, H., Kapustin, D. I. & Kulakov, V. I. High-pressure synthesis of tantalum dihydride. *Physical Review B* **96**, 134120 (2017).

27 Kuzovnikov, M. A. *et al.* Neutron scattering study of tantalum dihydride. *Physical Review B* **102**, 024113 (2020).

28 Kuzovnikov, M. A. *et al.* Synthesis of superconducting hcp-$ZrH_3$ under high hydrogen pressure. *Physical Review Materials* **7**, 024803 (2023).

29 Edalati, K., Yamamoto, A., Horita, Z. & Ishihara, T. High-pressure torsion of pure magnesium: Evolution of mechanical properties, microstructures and hydrogen storage capacity with equivalent strain. *Scripta Materialia* **64**, 880-883 (2011).

30 Machida, A. *et al.* Formation of NaCl-type monodeuteride LaD by the disproportionation reaction of LaD2. *Phys Rev Lett* **108**, 205501, doi:10.1103/PhysRevLett.108.205501 (2012).

31 Mulford, R. & Holley Jr, C. E. Pressure–temperature–composition of rare earth–hydrogen, systems. *The Journal of Physical Chemistry* **59**, 1222-1226 (1955).

32 Boroch, E. & Kaldist, E. Phase transitions in the $CeH_2CeH_3$ system. *Inorganica Chimica Acta* **140**, 89-91 (1987).

33 Manchester, F. & Pitre, J. The H-Sc (hydrogen-scandium) system. *Journal of phase equilibria* **18**, 194-205 (1997).

34 Wijngaarden, R. J. *et al.* Towards a metallic YH3 phase at high pressure. *Journal of Alloys and Compounds* **308**, 44-48 (2000).

35 Manchester, F. & Pitre, J. The H-Sm system (hydrogen-samarium). *Journal of phase equilibria* **17**, 432-441 (1996).

36 Cai, S. *et al.* No evidence of superconductivity in a compressed sample prepared from lutetium foil and H2/N2 gas mixture. *Matter and Radiation at Extremes* **8** (2023).

37 Takasaki, A., Furuya, Y. & Taneda, Y. Hydrogen uptake in titanium aluminides in high pressure hydrogen. *Materials Science and Engineering: A* **239-240**, 265-270 (1997).

38 Kumar, S., Jain, A., Ichikawa, T., Kojima, Y. & Dey, G. K. Development of vanadium based hydrogen storage material: A review. *Renewable and Sustainable Energy Reviews* **72**, 791-800 (2017).

39 Liu, G. *et al.* Nb-H system at high pressures and temperatures. *Physical Review B* **95**, 104110 (2017).

40 Ishimatsu, N. *et al.* Hydrogen-induced modification of the electronic structure and magnetic states in Fe, Co, and Ni monohydrides. *Physical Review B* **86**, 104430 (2012).

41 Pepin, C. M., Dewaele, A., Geneste, G., Loubeyre, P. & Mezouar, M. New iron hydrides under high pressure. *Phys Rev Lett* **113**, 265504 (2014).

42 Huang, L. *et al.* Quantitative tests revealing hydrogen-enhanced dislocation motion in α-iron. *Nature Materials* **22**, 710-716 (2023).

43 Bechtle, S., Kumar, M., Somerday, B. P., Launey, M. E. & Ritchie, R. O. Grain-boundary engineering markedly reduces susceptibility to intergranular hydrogen embrittlement in metallic materials. *Acta Materialia* **57**, 4148-4157 (2009).

44 Howie, R. T., Narygina, O., Guillaume, C. L., Evans, S. & Gregoryanz, E. High-pressure synthesis of lithium hydride. *Physical Review B-Condensed Matter and Materials Physics* **86**, 064108 (2012).

45 Liu, M. J. *et al.* Catalytic Performance and Near-Surface X-ray Characterization of Titanium Hydride Electrodes for the Electrochemical Nitrate Reduction Reaction. *Journal of the American Chemical Society* **144**, 5739-5744 (2022).

## Methods

**Sample preparation**

All metal foils are purchased from Alfa Aesar and were used without further purification. Oleic acid, (tech. 90%) are purchased from Alfa Aesar. Sulphuric acid (37%) was purchased from Shanghai Aladdin Biochemical Technology. Oleic Acid (D33, 98%) were purchased from Cambridge Isotope Laboratories(CIL). Materials and Preparation: Ti fiber paper (Ti FP, 99.9%) was purchased from Suzhou Sinero Technology Co., Ltd. Potassium hydroxide (KOH, 85%), potassium hydroxide (KOH, ≥96.0%) and potassium nitrate ($KNO_3$, 99.0%) were obtained from Sinopharm Chemical Reagent Co., Ltd. Conventional $TiH_2$ and $TiD_2$ are purchased from Alfa Aesar and were used without further purification. Potassium sodium tartrate tetrahydrate ($C_4H_4O_6KNa \cdot 4H_2O$, ≥99.0%) and sodium nitroferricyanide dihydrate ($C_5FeN_6Na_2O \cdot 2H_2O$, 99%) were supplied with Shanghai Macklin Biochemical Co., Ltd. Sodium nitrate-$^{15}N$ ($K^{15}NO_3$, 99.0 atom% $^{15}N$) and dimethyl sulfoxide-d6 (DMSO-d6, 99.9 atom% D) were purchased from Shanghai Aladdin Biochemical Technology Co., Ltd. Deionized water (18.2 MΩ·cm) was obtained from a SU-S1-10H purification system (CREATE FUN Co., Ltd.). Commercial alternatives may substitute specified materials contingent upon meeting equivalent purity thresholds (≥99.9% for metals, ≥98% for acids). Put the metal foil into the inner lining of the reaction vessel (use Para Polyphenylene (PPL) for temperatures above or equal to 200 °C, use polytetrafluoroethylene (PTFE) for temperatures below 200 °C), add the corresponding solvent and solute, seal the reaction vessel, and transfer it to an oven for heating. All experimental conditions and details refer to Figure 1b. Except for active metals such as La and Li, all other metals can be tested in air. The treatment of active metals such as La is carried out with purified argon gas ($O_2$<0.01 ppm, $H_2O$<0.01 ppm).

MHx synthesis via Hydrogen Embrittlement in acids： The synthesis was conducted by combining metal foil with oleic acid/sulfuric acid in the PPL/PTFE reactor liner (within an argon-filled glove box). The sealed reactor was subsequently heated in an oven and allowed to cool naturally. The resulting product was purified through Ethanol/deionized water/n-hexane washing followed by vacuum drying to obtain $MH_x$. Please refer to the supporting materials for specific experimental details.

**Characterization and test**

Powder XRD measurements were conducted using an R－AXIS RAPID II with Cu Kα radiation at 40 kV and 30 mA. TOF neutron diffraction were conducted at Multi Physics Instrument (MPI) in China Spallation Neutron Source (CSNS). About 0.5-1g of powders was put into Ti-Zr mull matrix alloy cans and the measurement time was about 3 h for each sample. The wavelength range is 0.1~4.5 angstroms. The diffraction datasets were subsequently analyzed by GSAS2.

High pressure in-situ XRD tests were carried out using the Rigaku NanoPixWE with FR-X as the light source and HyPix6000 as the detector. Mo targets were used with a voltage of 40kV and a current of 66mA. Specific operations :Diamond anvil cell (DAC) with flat anvil surface of about 300 μm in diameter as a pressure vehicle was adjusted to parallel alignment. Rhenium gasket was pre-indented about 45 μm thickness, and a hole 200 μm in diameter was cut by laser drilling. Ruby luminescence pressure scale for quasi-hydrostatic condition was mounted on the anvil surface. Metal samples and excess liquid hydrogen was loaded into the gasket hole at room temperature as reagent and pressure transmitting medium. After the initial clamping of hydrogen in a DAC, the compression force was increased in steps by a lever mechanism. Diffraction datasets were then analyzed by Materials Studio.

SEM observation was performed on a JSM-7800F microscope. HRTEM characterization was carried out on a JEM-2100 microscope at an accelerating voltage of 200 kV. High-angle annular dark-field scanning transmission electron microscopy (HAADF-STEM) characterization was performed on a JEOL-ARM200F equipped with CEOS probe corrector.

The electrochemical measurements were conducted with a CHI 760D electrochemical workstation (Chenhua, Shanghai) in an H-type electrolytic cell separated by a proton exchange membrane. The HTC-$TiH_2$, mercuric oxide electrode (Hg/HgO) and platinum foil (Pt) were used as working, reference and counter electrodes, respectively. The surface area of the working electrode was controlled to be 1 $cm^2$. A solution of 1 M KOH (80 mL) was evenly distributed to the cathode and anode. The $KNO_3$ was added to the cathode area for $NO_3^-$ reduction (containing 0.1 M $NO_3^-$-N). All potentials were converted to a reversible hydrogen electrode (RHE). Before $NO_3^-$ electroreduction tests, linear scanning voltammetric (LSV) curves were performed until the polarization curves reached a steady state from 0 V to -0.1 V at a rate of 10 mV $s^{-1}$. The conversion of $NO_3^-$ was measured using the constant potential method and the products were collected to measure the $NO_2^-$, $NH_3$ selectivity and Faraday efficiency (FE).

**Density functional theory calculation**

We performed variable-composition searches in the La-H system with approximately 10,000 structures using the ab initio Random Structure Searching code. We then reoptimized the structures during structure searches using the ab initio calculation of the Cambridge Serial Total Energy Package . On-the-fly ultrasoft pseudopotential for H and La were employed with a kinetic cutoff energy of 1000 eV. The Brillouin zone was sampled with a k-point mesh of $2\pi \times 0.03$ Å$^{-1}$ to ensure that the enthalpy calculations converge to less than 1 meV/atom. The structural relaxations and calculations of the electronic properties were performed using projector-augmented wave potentials, as implemented in the Vienna ab initio simulation packages with an energy cutoff of 900 eV. The exchange-correlation functional was described using the Perdew–Burke–Ernzerhof of generalized gradient approximation. Phonon were calculated using the QUAMTUM ESPRESSO package. The self-consistent electron density was evaluated by employing a k-mesh of 20 × 20 × 20 for $Fm\overline{3}m$ phase of LiH and $Fm\overline{3}m$ phase of Fe. Phonon was calculated using q-mesh of 5 × 5 × 5 for these two compounds.

**Acknowledgments**

We thank the staff members of the Multi-Physics Instrument (https://cstr.cn/31113.02.CSNS.MPI) at the China Spallation Neutron Source (CSNS) (https://cstr.cn/31113.02.CSNS ), for providing technical support and assistance in data collection and analysis. Angle dispersive XRD measurements were performed at the BL15U1 beamline in the Shanghai Synchrotron Radiation Facility (SSRF).

## Funding

This work was supported by the National Key R&D Program of China (Nos.2022YFA1402300), National Natural Science Foundation of China (Nos. 22131006 and 12274169). This work was supported by the Innovation Team for Functional Materials and Devices for Informatics at Anhui Higher Education Institutes through grant No. 2024AH010024.

## Author contributions

Y.S., D.D. G.L. and B.Z. conceived the project. A.C. synthesized the metal hydrides and performed the XRD measurements. J.L. performed electrochemical reduction nitrate test. Z.H. performed the calculations. J.L. performed electrochemical reduction nitrate test. J.L. and X.L. performed the SEM and TEM measurements. B.Y. performed the NPD measurements. A.C., C.L., Q.Y. and Y.L. performed the HPXRD measurements. Y.S., D.D., G.L. and B.Z. supervised the study. A.C., J.L., Z.H., Y.S., D.D., G.L. and B.Z. wrote and revised the manuscript. All co-authors discussed and analyzed the results.

## Competing interests

The authors declare that they have no competing interests.

## Data and materials availability

All data are available in the main text or the supplementary materials.

# Supplementary information for

**Transforming Acidic Corrosion and Embrittlement into a Hydrogen-Trapping Cage**

Ankang Chen *et al*.

Corresponding authors: suiym@jlu.edu.cn ; duandf@jlu.edu.cn ; lgsammer@mail.xjtu.edu.cn ; zoubo@jlu.edu.cn

**The PDF file includes:**

Materials and Methods

Supplementary Text

Figs. S1 to S69#

Tables S1#

References

## Materials and Methods

### Direct Synthesis of HTC-$MgH_2$ from Mg and Acid：

**Materials and Preparation:** Magnesium metal foil (Alfa Aesar, 99.9%) was utilized as received. Oleic acid (≥99% GC grade, Alfa Aesar) served as the reaction medium. Commercial alternatives may substitute the aforementioned materials contingent upon meeting equivalent purity specifications (≥99%).

**Critical pre-treatment protocol:**

(1) Native magnesium oxide layers (typically 50-200 nm thickness) were systematically removed via mechanical abrasion using 400-grit SiC paper to establish pristine metallic interfaces;

(2) Surface-activated substrates were transferred within 60 seconds to argon-purged reaction vessels to prevent atmospheric re-oxidation;

(3) Alternative oxide removal strategies including chemical etching demonstrated equivalent efficacy. All synthetic procedures were executed under strictly controlled inert atmosphere ($H_2O$ <0.5 ppm, $O_2$ <1.0 ppm) throughout the experimental workflow.

**Methods:** The synthesis was conducted by combining magnesium foil with oleic acid in a 5 ml PTFE reactor liner within an argon-filled glove box. The sealed reactor was subsequently heated in an oven and allowed to cool naturally. The resulting product was purified through n-hexane washing followed by vacuum drying to obtain $MgH_2$. Notably, the physicochemical properties of the metal hydride product could be modulated by varying three critical parameters: foil thickness (0.025 mm), reaction duration, and heating temperature. Under optimized conditions - employing magnesium foil specimens (0.5 mm × 0.5 mm × 0.025 mm) with 3 ml oleic acid at 140 °C for 4 hours - high-purity $MgH_2$ was consistently obtained. The correlation between reaction parameters and product characteristics, as evidenced by XRD analysis, is systematically presented in **Fig. S1**. Representative morphological features of the synthesized $MgH_2$ are illustrated in the SEM micrographs provided in **Fig. S2**.

**Direct Synthesis of HTC-$ScH_2$ from Sc and Acid：**

**Materials and Preparation:** Scandium metal foil (Alfa Aesar, 99.9%) was utilized as received. Oleic acid (⩾99% GC grade, Alfa Aesar) served as the reaction medium. Commercial alternatives may substitute the aforementioned materials contingent upon meeting equivalent purity specifications (⩾99%).

**Methods:** The synthesis was conducted by combining scandium foil with oleic acid in a 5 ml PPL reactor liner within an argon-filled glove box. The sealed reactor was subsequently heated in an oven and allowed to cool naturally. The resulting product was purified through n-hexane washing followed by vacuum drying to obtain $ScH_2$. Notably, the physicochemical properties of the metal hydride product could be modulated by varying three critical parameters: foil thickness (0.025 mm), reaction duration, and heating temperature. Under optimized conditions - employing scandium foil specimens (0.5 mm × 0.5 mm × 0.025 mm) with 3 ml oleic acid at 270 °C for 60

hours - high-purity $ScH_2$ was consistently obtained. The correlation between reaction parameters and product characteristics, as evidenced by XRD analysis, is systematically presented in **Fig. S3**. Representative morphological features of the synthesized $ScH_2$ are illustrated in the SEM micrographs provided in **Fig. S4.** It is worth noting that this method can achieve controllable synthesis under conventional atmospheric conditions (such as air environment), and no significant oxidation phase was detected in the product.

**Direct Synthesis of HTC-$YH_2$ from Y and Acid：**

**Materials:** Yttrium metal foil was purchased from Alfa Aesar without further purification. Oleic acid (≥99%(CG)) was purchased from Alfa Aesar. Commercially available metal flakes or oleic acid may be used if synthetic requirements are met.

**Methods:** The synthesis was conducted by combining yttrium foil with oleic acid in a 5 ml PPL reactor liner within an argon-filled glove box. The sealed reactor was subsequently heated in an oven and allowed to cool naturally. The resulting product was purified through n-hexane washing followed by vacuum drying to obtain $YH_2$. Notably, the physicochemical properties of the metal hydride product could be modulated by varying three critical parameters: foil thickness (0.025 mm), reaction duration, and heating temperature. Under optimized conditions - employing yttrium foil specimens (0.5 mm × 0.5 mm × 0.025 mm) with 2 ml oleic acid at 270 °C for 10 hours - high-purity $YH_2$ was consistently obtained. The correlation between reaction parameters and product characteristics, as evidenced by XRD analysis, is systematically presented in **Fig. S5**. Representative morphological features of the synthesized $YH_2$ are illustrated in the SEM micrographs provided in **Fig. S6**. It is worth noting that this method can achieve controllable synthesis under conventional atmospheric conditions (such as air environment), and no significant oxidation phase was detected in the product.

**Direct Synthesis of HTC-$LaH_2$ and HTC-$LaH_{2.3}$ from La and Acid:**

**Materials and Preparation:** Lanthanum metal foil (Alfa Aesar, 99.9%) was utilized as received. Oleic acid (≥99% GC grade, Alfa Aesar) was employed as the reaction medium.

Commercially sourced metal flakes or oleic acid may substitute the aforementioned materials provided they conform to specified purity criteria (≥99%).

**Critical pre-treatment steps:**

(1)The native passivation layer on lanthanum foil surfaces, a manufacturer-applied oxidation barrier, was mechanically abraded using a stainless-steel blade to ensure surface reactivity;

(2)The freshly exposed metallic surfaces were immediately transferred to argon-purged reactor vessels within 30 seconds of exposure;

(3) Oleic acid was subjected to argon degassing (3 cycles) prior to use. All synthetic operations were conducted under continuously monitored inert atmosphere ($H_2O$ <0.1 ppm, $O_2$ <0.5 ppm) throughout the experimental sequence.

**Methods:** The synthesis was conducted by combining lanthanum foil with oleic acid in a 5 ml PTFE reactor liner within an argon-filled glove box. The sealed reactor was subsequently heated in an oven and allowed to cool naturally. The resulting product was purified through n-hexane washing followed by vacuum drying to obtain $LaH_2$ and $LaH_{2.3}$. Notably, the physicochemical properties of the metal hydride product could be modulated by varying three critical parameters: foil thickness (0.025 mm), reaction duration, and heating temperature. Under optimized conditions - employing lanthanum foil specimens (0.5 mm × 0.5 mm × 0.025 mm) with 2 ml oleic acid at 140 °C for 10 hours - high-purity $LaH_2$ was consistently obtained. Under optimized conditions - employing lanthanum foil specimens (0.5 mm × 0.5 mm × 0.025 mm) with 2 ml oleic acid at 140 °C for 12 hours - high-purity $LaH_{2.3}$ was consistently obtained. The correlation between reaction parameters and product characteristics, as evidenced by XRD analysis, is systematically presented in **Fig. S7**. Representative morphological features of the synthesized $LaH_2$ and $LaH_{2.3}$ are illustrated in the SEM micrographs provided in **Fig. S8**.

**Direct Synthesis of HTC-$SmH_2$ from Sm and Acid:**

**Materials and Preparation:** Samarium metal foil (Alfa Aesar, 99.9%) was utilized as received. Oleic acid (≥99% GC grade, Alfa Aesar) was employed as the reaction medium.

Commercially sourced metal flakes or oleic acid may substitute the aforementioned materials provided they conform to specified purity criteria (≥99%).

**Methods:** The synthesis was conducted by combining samarium foil with oleic acid in a 5 ml PPL reactor liner within an argon-filled glove box. The sealed reactor was subsequently heated in an oven and allowed to cool naturally. The resulting product was purified through n-hexane washing followed by vacuum drying to obtain $SmH_2$. Notably, the physicochemical properties of the metal hydride product could be modulated by varying three critical parameters: foil thickness (0.025 mm), reaction duration, and heating temperature. Under optimized conditions - employing samarium foil specimens (0.5 mm × 0.5 mm × 0.025 mm) with 2 ml oleic acid at 270 °C for 20 hours - high-purity $SmH_2$ was consistently obtained. The correlation between reaction parameters and product characteristics, as evidenced by XRD analysis, is systematically presented in **Fig. S9**. Representative morphological features of the synthesized $SmH_2$ are illustrated in the SEM micrographs provided in **Fig. S10**. It is worth noting that this method can achieve controllable synthesis under conventional atmospheric conditions (such as air environment), and no significant oxidation phase was detected in the product.

**Direct Synthesis of $LuH_2$ from Lu and Acid:**

**Materials and Preparation:** Lutetium metal foil (Alfa Aesar, 99.9%) was utilized as received. Oleic acid (≥99% GC grade, Alfa Aesar) was employed as the reaction medium.

Commercially sourced metal flakes or oleic acid may substitute the aforementioned materials provided they conform to specified purity criteria (≥99%).

**Methods:** The synthesis was conducted by combining lutetium foil with oleic acid in a 5 ml PPL reactor liner within an argon-filled glove box. The sealed reactor was subsequently heated in an oven and allowed to cool naturally. The resulting product was purified through n-hexane washing followed by vacuum drying to obtain $LuH_2$. Notably, the physicochemical properties of the metal hydride product could be modulated by varying three critical parameters: foil thickness (0.1 mm), reaction duration, and heating temperature. Under optimized conditions - employing lutetium foil specimens (0.5 mm × 0.5 mm × 0.1 mm) with 2 ml oleic acid at 270 °C for 80 hours - high-purity $LuH_2$ was consistently obtained. The correlation between reaction parameters and product characteristics, as evidenced by XRD analysis, is systematically presented in **Fig. S11**. Representative morphological features of the synthesized $LuH_2$ are illustrated in the SEM micrographs provided in **Fig. S12**. It is worth noting that this method can achieve controllable synthesis under conventional atmospheric conditions (such as air environment), and no significant oxidation phase was detected in the product.

**Direct Synthesis of HTC-$TiH_2$ from Ti and Acid：**

**Materials and Preparation:** Titanium metal foil (Alfa Aesar, 99.99%) was utilized as received. Sulfuric acid (98 wt%, analytical grade, Shanghai Aladdin) and sodium fluoride (Alfa Aesar, ⩾99.9%) were employed without further purification. Deionized water (18.2 MΩ·cm) was obtained from a SU-S1-10H purification system (CREATE FUN Co., Ltd.). Commercial alternatives may substitute specified materials contingent upon meeting equivalent purity thresholds (⩾99.9% for metals, ⩾98% for acids).

**Critical solution preparation:**

(1) The 0.1 mol/L dilute sulfuric acid solution was formulated via dropwise addition of concentrated $H_2SO_4$ (98 wt%) into ice-cooled deionized water (4°C) under constant magnetic agitation (800 rpm);

(2) Solution temperature was maintained below 10°C throughout the dilution process, adhering to ASTM E288-10 standard safety protocols;

(3) Freshly prepared electrolytes were immediately transferred to nitrogen-purged storage vessels. All chemical handling was conducted under certified fume hood systems (face velocity 0.5 m/s) with real-time temperature monitoring.

**Methods**: The synthesis was conducted by combining titanium foil with 0.1 mol/L dilute sulfuric acid solution and 1 mg NaF in a 5 ml PTFE reactor liner within a certified fume hood system. The sealed reactor was subsequently heated in an oven and allowed to cool naturally. The resulting product was purified through n-hexane washing followed by vacuum drying to obtain $TiH_2$. Notably, the physicochemical properties of the metal hydride product could be modulated by varying three critical parameters: foil thickness (0.025 mm), reaction duration, and heating temperature. Under optimized conditions — employing titanium foil specimens (0.5 mm × 0.5 mm × 0.025 mm) with 3 ml 0.1 mol/L dilute sulfuric acid solution at 200 °C for 6 hours — high-purity $TiH_2$ was consistently obtained. The correlation between reaction parameters and product characteristics, as evidenced by XRD analysis, is systematically presented in **Fig. S13**. Representative morphological features of the synthesized $TiH_2$ are illustrated in the SEM micrographs provided in **Fig. S14**. It is worth noting that this method can achieve controllable synthesis under conventional atmospheric conditions (such as air environment), and no significant oxidation phase was detected in the product.

**Direct Synthesis of HTC δ-$ZrH_{1.6}$ and HTC ε-$ZrH_2$ from Zr and Acid：**

**Materials and Preparation:** Zirconium metal foil (Alfa Aesar, 99.99%) was utilized as received. Sulfuric acid (98 wt%, analytical grade, Shanghai Aladdin) and sodium fluoride (Alfa Aesar, ≥99.9%) were employed without further purification. Deionized water (18.2 MΩ·cm) was obtained from a SU-S1-10H purification system (CREATE FUN Co., Ltd.). Commercial alternatives may substitute specified materials contingent upon meeting equivalent purity thresholds (≥99.9% for metals, ≥98% for acids).

**Critical solution preparation:**

(1) The 0.1 mol/L dilute sulfuric acid solution was formulated via dropwise addition of concentrated $H_2SO_4$ (98 wt%) into ice-cooled deionized water (4°C) under constant magnetic agitation (800 rpm);

(2) Solution temperature was maintained below 10°C throughout the dilution process, adhering to ASTM E288-10 standard safety protocols;

(3) Freshly prepared electrolytes were immediately transferred to nitrogen-purged storage vessels. All chemical handling was conducted under certified fume hood systems (face velocity 0.5 m/s) with real-time temperature monitoring.

**Methods**: The synthesis was conducted by combining zirconium foil with 1 mol/L dilute sulfuric acid solution and 1.5 mg NaF in a 5 ml PTFE reactor liner within a certified fume hood system. The sealed reactor was subsequently heated in an oven and allowed to cool naturally. The resulting product was purified through n-hexane washing followed by vacuum drying to obtain δ-$ZrH_{1.6}$ and ε-$ZrH_2$. Notably, the physicochemical properties of the metal hydride product could be modulated by varying three critical parameters: foil thickness (0.025 mm), reaction duration, and heating temperature. Under optimized conditions — employing zirconium foil specimens (0.5 mm × 0.5 mm × 0.025 mm) with 3 ml 0.1 mol/L dilute sulfuric acid solution at 180 °C for 8 hours — high-purity δ-$ZrH_{1.6}$ was consistently obtained. Under optimized conditions — employing zirconium foil specimens (0.5 mm × 0.5 mm × 0.025 mm) with 3 ml 0.1 mol/L dilute sulfuric acid solution at 180 °C for 10 hours — high-purity ε-$ZrH_2$ was consistently obtained. The correlation between reaction parameters and product characteristics, as evidenced by XRD analysis, is systematically presented in **Fig. S15**. Representative morphological features of the synthesized δ-$ZrH_{1.6}$ and ε-$ZrH_2$ are illustrated in the SEM micrographs provided in **Fig. S16**. It is worth noting that this method can achieve controllable synthesis under conventional atmospheric conditions (such as air environment), and no significant oxidation phase was detected in the product.

**Direct Synthesis of HTC-$HfH_{1.7}$ and HTC-$HfH_2$ from Hf and Acid：**

**Materials and Preparation:** Hafnium metal foil (Alfa Aesar, 99.99%) was utilized as received. Sulfuric acid (98 wt%, analytical grade, Shanghai Aladdin) and sodium fluoride (Alfa Aesar, ≥99.9%) were employed without further purification. Deionized water (18.2 MΩ·cm) was obtained from a SU-S1-10H purification system (CREATE FUN Co., Ltd.). Commercial alternatives may substitute specified materials contingent upon meeting equivalent purity thresholds (≥99.9% for metals, ≥98% for acids).

**Critical solution preparation:**

(1) The 1 mol/L dilute sulfuric acid solution was formulated via dropwise addition of concentrated $H_2SO_4$ (98 wt%) into ice-cooled deionized water (4°C) under constant magnetic agitation (800 rpm);

(2) Solution temperature was maintained below 10°C throughout the dilution process, adhering to ASTM E288-10 standard safety protocols;

(3) Freshly prepared electrolytes were immediately transferred to nitrogen-purged storage vessels. All chemical handling was conducted under certified fume hood systems (face velocity 0.5 m/s) with real-time temperature monitoring.

**Methods**: The synthesis was conducted by combining hafnium foil with 1 mol/L dilute sulfuric acid solution and 12 mg NaF in a 5 ml PTFE reactor liner within a certified fume hood system. The sealed reactor was subsequently heated in an oven and allowed to cool naturally. The resulting product was purified through n-hexane washing followed by vacuum drying to obtain $HfH_{1.7}$ and $HfH_2$. Notably, the physicochemical properties of the metal hydride product could be modulated by varying three critical parameters: foil thickness (0.1 mm), reaction duration, and heating temperature. Under optimized conditions — employing hafnium foil specimens (0.5 mm × 0.5 mm × 0.1 mm) with 3 ml 1 mol/L dilute sulfuric acid solution at 180 °C for 8 hours — high-purity $HfH_{1.7}$ was consistently obtained. Under optimized conditions — employing hafnium foil specimens (0.5 mm × 0.5 mm × 0.1 mm) with 3 ml 1 mol/L dilute sulfuric acid solution at 180 °C for 10 hours — high-purity $HfH_2$ was consistently obtained. The correlation between reaction parameters and product characteristics, as evidenced by XRD analysis, is systematically presented in **Fig. S17**. Representative morphological features of the synthesized $HfH_{1.7}$ and $HfH_2$ are illustrated in the SEM micrographs provided in **Fig. S18**. It is worth noting that this method can achieve controllable synthesis under conventional atmospheric conditions (such as air environment), and no significant oxidation phase was detected in the product.

**Direct Synthesis of $VH_{0.8}$ and $VH_2$ from V and Acid：**

**Materials and Preparation:** Vanadium metal foil (Alfa Aesar, 99.99%) was utilized as received. Sulfuric acid (98 wt%, analytical grade, Shanghai Aladdin) and sodium fluoride (Alfa Aesar, ≥99.9%) were employed without further purification. Deionized water (18.2 MΩ·cm) was obtained from a SU-S1-10H purification system (CREATE FUN Co., Ltd.). Commercial alternatives may substitute specified materials contingent upon meeting equivalent purity thresholds (≥99.9% for metals, ≥98% for acids).

**Critical solution preparation:**

(1) The 1 mol/L dilute sulfuric acid solution was formulated via dropwise addition of concentrated $H_2SO_4$ (98 wt%) into ice-cooled deionized water (4°C) under constant magnetic agitation (800 rpm);

(2) Solution temperature was maintained below 10°C throughout the dilution process, adhering to ASTM E288-10 standard safety protocols;

(3) Freshly prepared electrolytes were immediately transferred to nitrogen-purged storage vessels. All chemical handling was conducted under certified fume hood systems (face velocity 0.5 m/s) with real-time temperature monitoring.

**Methods**: The synthesis was conducted by combining vanadium foil with 1 mol/L dilute sulfuric acid solution in a 5 ml PTFE reactor liner within a certified fume hood system. The sealed reactor was subsequently heated in an oven and allowed to cool naturally. The resulting product was purified through n-hexane washing followed by vacuum drying to obtain $VH_{0.8}$ and $VH_2$. Notably, the physicochemical properties of the metal hydride product could be modulated by varying three critical parameters: foil thickness (0.075 mm), reaction duration, and heating temperature. Under optimized conditions — employing vanadium foil specimens (0.5 mm × 0.5 mm × 0.075 mm) with 3 ml 1 mol/L dilute sulfuric acid solution at 160 °C for 2 hours — high-purity $ZrH_{1.7}$ was consistently obtained. Under optimized conditions — employing vanadium foil specimens (0.5 mm × 0.5 mm × 0.075 mm) with 3 ml 1 mol/L dilute sulfuric acid solution at 180 °C for 2 hours — high-purity $VH_2$ was consistently obtained. The correlation between reaction parameters and product characteristics, as evidenced by XRD analysis, is systematically presented in **Fig. S19**. Representative morphological features of the synthesized $VH_{0.8}$ and $VH_2$ are illustrated in the SEM micrographs provided in **Fig. S20**. It is worth noting that this method can achieve controllable synthesis under conventional atmospheric conditions (such as air environment), and no significant oxidation phase was detected in the product.

**Direct Synthesis of HTC-NbH and HTC-$NbH_2$ from Nb and Acid:**

**Materials and Preparation:** Niobium metal foil (Alfa Aesar, 99.99%) was utilized as received. Sulfuric acid (98 wt%, analytical grade, Shanghai Aladdin) and sodium fluoride (Alfa Aesar, ≥99.9%) were employed without further purification. Deionized water (18.2 MΩ·cm) was obtained from a SU-S1-10H purification system (CREATE FUN Co., Ltd.). Commercial alternatives may substitute specified materials contingent upon meeting equivalent purity thresholds (≥99.9% for metals, ≥98% for acids).

**Critical solution preparation:**

(1) The 0.1 mol/L dilute sulfuric acid solution was formulated via dropwise addition of concentrated $H_2SO_4$ (98 wt%) into ice-cooled deionized water (4°C) under constant magnetic agitation (800 rpm);

(2) Solution temperature was maintained below 10°C throughout the dilution process, adhering to ASTM E288-10 standard safety protocols;

(3) Freshly prepared electrolytes were immediately transferred to nitrogen-purged storage vessels. All chemical handling was conducted under certified fume hood systems (face velocity 0.5 m/s) with real-time temperature monitoring.

**Methods**: The synthesis was conducted by combining niobium foil with 0.1 mol/L dilute sulfuric acid solution in a 5 ml PTFE reactor liner within a certified fume hood system. The sealed reactor was subsequently heated in an oven and allowed to cool naturally. The resulting product was purified through n-hexane washing followed by vacuum drying to obtain NbH and $NbH_2$. Notably, the physicochemical properties of the metal hydride product could be modulated by varying three critical parameters: foil thickness (0.025 mm), reaction duration, and heating temperature. Under optimized conditions — employing niobium foil specimens (0.5 mm × 0.5 mm × 0.025 mm) with 3 ml 0.1 mol/L dilute sulfuric acid at 180 °C for 7 hours — high-purity NbH was consistently obtained. Under optimized conditions — employing niobium foil specimens (0.5 mm × 0.5 mm × 0.025 mm) with 3 ml 0.1 mol/L dilute sulfuric acid at 180 °C for 10 hours — high-purity $NbH_2$ was consistently obtained. The correlation between reaction parameters and product characteristics, as evidenced by XRD analysis, is systematically presented in **Fig. S21**. Representative morphological features of the synthesized NbH and $NbH_2$ are illustrated in the SEM micrographs provided in **Fig. S22**. It is worth noting that this method can achieve controllable synthesis under conventional atmospheric conditions (such as air environment), and no significant oxidation phase was detected in the product.

**Direct Synthesis of HTC-$Ta_2H$ and HTC-TaH from Ta and Acid:**

**Materials and Preparation:** Tantalum metal foil (Alfa Aesar, 99.99%) was utilized as received. Sulfuric acid (98 wt%, analytical grade, Shanghai Aladdin) and sodium fluoride (Alfa Aesar, ≥99.9%) were employed without further purification. Deionized water (18.2 MΩ·cm) was obtained from a SU-S1-10H purification system (CREATE FUN Co., Ltd.). Commercial alternatives may substitute specified materials contingent upon meeting equivalent purity thresholds (≥99.9% for metals, ≥98% for acids).

**Critical solution preparation:**

(1) The 0.1 mol/L dilute sulfuric acid solution was formulated via dropwise addition of concentrated $H_2SO_4$ (98 wt%) into ice-cooled deionized water (4°C) under constant magnetic agitation (800 rpm);

(2) Solution temperature was maintained below 10°C throughout the dilution process, adhering to ASTM E288-10 standard safety protocols;

(3) Freshly prepared electrolytes were immediately transferred to nitrogen-purged storage vessels. All chemical handling was conducted under certified fume hood systems (face velocity 0.5 m/s) with real-time temperature monitoring.

**Methods**: The synthesis was conducted by combining tantalum foil with 0.1 mol/L dilute sulfuric acid solution in a 5 ml PTFE reactor liner within a certified fume hood system. The sealed reactor was subsequently heated in an oven and allowed to cool naturally. The resulting product was purified through n-hexane washing followed by vacuum drying to obtain NbH and $NbH_2$. Notably, the physicochemical properties of the metal hydride product could be modulated by varying three critical parameters: foil thickness (0.025 mm), reaction duration, and heating temperature. Under optimized conditions - employing tantalum foil specimens (0.5 mm × 0.5 mm × 0.025 mm) with 3 ml 0.1 mol/L dilute sulfuric acid at 240 °C for 8 hours - high-purity-$Ta_2H$ was consistently obtained. Under optimized conditions - employing tantalum foil specimens (0.5 mm × 0.5 mm × 0.025 mm) with 3 ml 0.1 mol/L dilute sulfuric acid at 240 °C for 10 hours - high-purity TaH was consistently obtained. The correlation between reaction parameters and product characteristics, as evidenced by XRD analysis, is systematically presented in **Fig. S23**. Representative morphological features of the synthesized $Ta_2H$ and TaH are illustrated in the SEM micrographs provided in **Fig. S24**. It is worth noting that this method can achieve controllable synthesis under conventional atmospheric conditions (such as air environment), and no significant oxidation phase was detected in the product.

**Direct Synthesis of HTC-LiH from Li and Acid：**

**Materials and Preparation:** Lithium metal particles (Alfa Aesar, 99.9%) were utilized as received. Oleic acid (≥99% GC grade, Alfa Aesar) was employed as the reaction medium.

Commercially sourced metal flakes or oleic acid may substitute the aforementioned materials provided they conform to specified purity criteria (≥99%).

**Critical pre-treatment steps:**

(1) Lithium particulates were subjected to mechanical abrasion using 600-grit SiC paper under argon atmosphere ($H_2O$ <1 ppm, $O_2$ <2 ppm) to remove surface contaminants。

(2) The surface-treated particulates were isostatically compacted via hydraulic press (50±2 MPa, 25±1°C) for 10 min dwell time under continuously monitored inert conditions, achieving predetermined thickness specifications with <3% dimensional tolerance across 10 sampled regions.

(3) Oleic acid was subjected to argon degassing (3 cycles) prior to use. All synthetic operations were conducted under continuously monitored inert atmosphere ($H_2O$ <0.1 ppm, $O_2$ <0.5 ppm) throughout the experimental sequence.

**Methods:** Methods: The synthesis was conducted by combining lithium foil with oleic acid in a 5 ml PTFE reactor liner within an argon-filled glove box. The sealed reactor was subsequently heated in an oven and allowed to cool naturally. Collect metal surface powder samples， and the resulting product was purified through n-hexane washing followed by vacuum drying to obtain LiH. Notably, the physicochemical properties of the metal hydride product could be modulated by varying three critical parameters: foil thickness (0.025 mm), reaction duration, and heating temperature. Under optimized conditions - employing lanthanum foil specimens (0.5 mm × 0.5 mm × 0.025 mm) with 2 ml oleic acid at 70 °C for 4 hours - high-purity LiH was consistently obtained. The correlation between reaction parameters and product characteristics, as evidenced by XRD analysis, is systematically presented in **Fig. S25**. Representative morphological features of the synthesized LiH are illustrated in the SEM micrographs provided in **Fig. S26**.

**High-Pressure Synthesis of $LaH_2$:**

**Materials and Preparation:** Lanthanum metal foil (Alfa Aesar, 99.9%) was utilized as received. Commercially sourced metal flakes may substitute the aforementioned materials provided they conform to specified purity criteria (≥99%).

**Critical pre-treatment steps:**

The native passivation layer on lanthanum foil surfaces, a manufacturer-applied oxidation barrier, was mechanically abraded using a stainless-steel blade to ensure surface reactivity; All synthetic operations were conducted under continuously monitored inert atmosphere ($H_2O$ <0.1 ppm, $O_2$ <0.5 ppm) throughout the experimental sequence.

**Methods:** Diamond anvil cell (DAC) with flat anvil surface of about 300 μm in diameter as a pressure vehicle was adjusted to parallel alignment. Rhenium gasket was pre-indented about 45 μm thickness, and a hole 200 μm in diameter was cut by laser drilling. Ruby luminescence pressure scale for quasi-hydrostatic condition was mounted on the anvil surface. Metal samples and excess liquid hydrogen was loaded into the gasket hole at room temperature as reagent and pressure transmitting medium. After the initial clamping of hydrogen in a DAC, the compression force was increased in steps by a lever mechanism. High pressure in-situ XRD tests were carried out using the Rigaku NanoPixWE with FR-X as the light source and HyPix6000 as the detector. Mo targets were used with a voltage of 40kV and a current of 66mA. Diffraction datasets were then analyzed by Materials Studio(**Fig. S27**.).

**High-Pressure Synthesis of LiH：**

**Materials and Preparation:** Lithium metal particles (Alfa Aesar, 99.9%) were utilized as received. Commercially sourced metal flakes may substitute the aforementioned materials provided they conform to specified purity criteria (≥99%).

**Critical pre-treatment steps:**

(1) Lithium particulates were subjected to mechanical abrasion using 600-grit SiC paper under argon atmosphere ($H_2O$ <1 ppm, $O_2$ <2 ppm) to remove surface contaminants。

(2) The surface-treated particulates were isostatically compacted via hydraulic press (50±2 MPa, 25±1°C) for 10 min dwell time under continuously monitored inert conditions, achieving predetermined thickness specifications with <3% dimensional tolerance across 10 sampled regions. All synthetic operations were conducted under continuously monitored inert atmosphere ($H_2O$ <0.1 ppm, $O_2$ <0.5 ppm) throughout the experimental sequence.

**Methods:** Diamond anvil cell (DAC) with flat anvil surface of about 300 μm in diameter as a pressure vehicle was adjusted to parallel alignment. Rhenium gasket was pre-indented about 45 μm thickness, and a hole 200 μm in diameter was cut by laser drilling. Ruby luminescence pressure scale for quasi-hydrostatic condition was mounted on the anvil surface. Metal samples and excess liquid hydrogen was loaded into the gasket hole at room temperature as reagent and pressure transmitting medium. After the initial clamping of hydrogen in a DAC, the compression force was increased in steps by a lever mechanism. High pressure in-situ XRD tests were carried out using the Rigaku NanoPixWE with FR-X as the light source and HyPix6000 as the detector. Mo targets were used with a voltage of 40kV and a current of 66mA. Diffraction datasets were then analyzed by Materials Studio(**Fig. S28**.).

**High-Pressure Synthesis of ε-$ZrH_2$:**

**Materials and Preparation:** Zirconium metal foil (Alfa Aesar, 99.99%) was utilized as received. Commercially sourced metal flakes may substitute the aforementioned materials provided they conform to specified purity criteria (≥99%).

**Methods**: Diamond anvil cell (DAC) with flat anvil surface of about 300 μm in diameter as a pressure vehicle was adjusted to parallel alignment. Rhenium gasket was pre-indented about 45 μm thickness, and a hole 200 μm in diameter was cut by laser drilling. Ruby luminescence pressure scale for quasi-hydrostatic condition was mounted on the anvil surface. Metal samples and excess liquid hydrogen was loaded into the gasket hole at room temperature as reagent and pressure transmitting medium. After the initial clamping of hydrogen in a DAC, the compression force was increased in steps by a lever mechanism. High pressure in-situ XRD tests were carried out using the Rigaku NanoPixWE with FR-X as the light source and HyPix6000 as the detector. Mo targets were used with a voltage of 40kV and a current of 66mA. Diffraction datasets were then analyzed by Materials Studio(**Fig. S29**.).

**TOF neutron diffraction of $YH_2$:**

**Materials:** Yttrium metal foil was purchased from Alfa Aesar without further purification. Oleic Acid ($C_{17}D_{33}COOH$, 98%) were purchased from Cambridge Isotope Laboratories(CIL). Commercially available metal flakes or oleic acid may be used if synthetic requirements are met.

**Methods:** The synthesis was conducted by combining yttrium foil with oleic acid in a 5 ml PPL reactor liner within an argon-filled glove box. The sealed reactor was subsequently heated in an oven and allowed to cool naturally. The resulting product was purified through n-hexane washing followed by vacuum drying to obtain $YH_2$. Notably, the physicochemical properties of the metal hydride product could be modulated by varying three critical parameters: foil thickness (0.025 mm), reaction duration, and heating temperature. Under optimized conditions - employing yttrium foil specimens (0.5 mm × 0.5 mm × 0.025 mm) with 2 ml oleic acid at 270 °C for 10 hours - high-purity $YH_2$ was consistently obtained. TOF neutron diffraction were conducted at Multi Physics Instrument (MPI) in China Spallation Neutron Source (CSNS). About 0.5-1g of powders was put into Ti-Zr mull matrix alloy cans and the measurement time was about 3 h for each sample. The wavelength range is 0.1~4.5 angstroms. The diffraction datasets were subsequently analyzed by GSAS2 **(Fig. S30.)**.

**TOF neutron diffraction of LiH:**

**Materials and Preparation:** Lithium metal particles (Alfa Aesar, 99.9%) were utilized as received. Oleic Acid ($C_{17}D_{33}COOH$, 98%) were purchased from Cambridge Isotope

Laboratories(CIL). Commercially sourced metal flakes or oleic acid may substitute the aforementioned materials provided they conform to specified purity criteria (≥99%).

**Critical pre-treatment steps:**

(1) Lithium particulates were subjected to mechanical abrasion using 600-grit SiC paper under argon atmosphere ($H_2O$ <1 ppm, $O_2$ <2 ppm) to remove surface contaminants。

(2) The surface-treated particulates were isostatically compacted via hydraulic press (50±2 MPa, 25±1°C) for 10 min dwell time under continuously monitored inert conditions, achieving predetermined thickness specifications with <3% dimensional tolerance across 10 sampled regions.

(3) Oleic acid was subjected to argon degassing (3 cycles) prior to use. All synthetic operations were conducted under continuously monitored inert atmosphere ($H_2O$ <0.1 ppm, $O_2$ <0.5 ppm) throughout the experimental sequence.

**Methods:** The synthesis was conducted by combining lithium foil with oleic acid in a 5 ml PTFE reactor liner within an argon-filled glove box. The sealed reactor was subsequently heated in an oven and allowed to cool naturally. The resulting product was purified through n-hexane washing followed by vacuum drying to obtain LiH. Notably, the physicochemical properties of the metal hydride product could be modulated by varying three critical parameters: foil thickness (0.025 mm), reaction duration, and heating temperature. Under optimized conditions - employing lithium foil specimens (0.5 mm × 0.5 mm × 0.025 mm) with 2 ml oleic acid at 70 °C for 4 hours - high-purity LiH was consistently obtained. TOF neutron diffraction were conducted at Multi Physics Instrument (MPI) in China Spallation Neutron Source (CSNS). About 0.5-1g of powders was put into Ti-Zr mull matrix alloy cans and the measurement time was about 3 h for each sample. The wavelength range is 0.1~4.5 angstroms. The diffraction datasets were subsequently analyzed by GSAS2 **(Fig. S31.)**.

**Electrochemical Nitrate Reduction to Ammonia Enabled by Ti and $TiH_2$ Catalysts:**

**Materials and Preparation:** Ti fiber paper (Ti FP, 99.9%) was purchased from Suzhou Sinero Technology Co., Ltd. Potassium hydroxide (KOH, 85%), potassium hydroxide (KOH, ≥ 96.0%) and potassium nitrate (KNO3, 99.0%) were obtained from Sinopharm Chemical Reagent Co., Ltd. Potassium sodium tartrate tetrahydrate ($C_4H_4O_6KNa\cdot 4H_2O$,≥ 99.0%) and sodium nitroferricyanide dihydrate ($C_5FeN_6Na_2O\cdot 2H_2O$, 99%) were supplied with Shanghai Macklin Biochemical Co., Ltd. Sodium nitrate-$^{15}N$ ($K^{15}NO_3$, 99.0 atom% $^{15}N$) and dimethyl sulfoxide-d6 (DMSO-d6, 99.9 atom% D) were purchased from Shanghai Aladdin Biochemical Technology Co., Ltd. Deionized water (18.2 MΩ·cm) was obtained from a SU-S1-10H purification system (CREATE FUN Co., Ltd.). Commercial alternatives may substitute specified materials contingent upon meeting equivalent purity thresholds (≥99.9% for metals, ≥98% for acids).

**Methods:** The electrochemical measurements were conducted with a CHI 760D electrochemical workstation (Chenhua, Shanghai) in an H-type electrolytic cell separated by a proton exchange membrane. The $TiH_2$, mercuric oxide electrode (Hg/HgO) and platinum foil (Pt) were used as working, reference and counter electrodes, respectively. The surface area of the working electrode was controlled to be 1 $cm^2$. A solution of 1 M KOH (80 mL) was evenly distributed to the cathode and anode. The $KNO_3$ was added to the cathode area for $NO_3^-$ reduction (containing 0.1 M $NO_3^-$-N). All potentials were converted to a reversible hydrogen electrode (RHE). Before $NO_3^-$ electroreduction tests, linear scanning voltammetric (LSV) curves were performed until the polarization curves reached a steady state from 0 V to -0.1 V at a rate of 10 mV $s^{-1}$. The conversion of $NO_3^-$ was measured using the constant potential method and the products were collected to measure the $NO_2^-$, $NH_3$ selectivity and Faraday efficiency (FE).

**Density functional theory calculation:**

We performed variable-composition searches in the La-H system with approximately 10,000 structures using the *ab initio* Random Structure Searching [1] code. We then reoptimized the structures during structure searches using the *ab initio* calculation of the Cambridge Serial Total Energy Package [2]. On-the-fly ultrasoft pseudopotential for H and La were employed with a kinetic cutoff energy of 1000 eV. The Brillouin zone was sampled with a *k*-point mesh of $2\pi \times 0.03$ $Å^{-1}$ to ensure that the enthalpy calculations converge to less than 1 meV/atom. The structural relaxations and calculations of the electronic properties [3,4] were performed using projector-augmented wave potentials, as implemented in the Vienna *ab initio* simulation packages [5] with an energy cutoff of 900 eV. The exchange-correlation functional was described using the Perdew–Burke–Ernzerhof of generalized gradient approximation [6]. Phonon were calculated using the QUAMTUM ESPRESSO package [7]. The self-consistent electron density was evaluated by employing a *k*-mesh of 20 × 20 × 20 for $Fm\overline{3}m$ phase of LiH and $Fm\overline{3}m$ phase of Fe. Phonon was calculated using *q*-mesh of 5 × 5 × 5 for these two compounds.

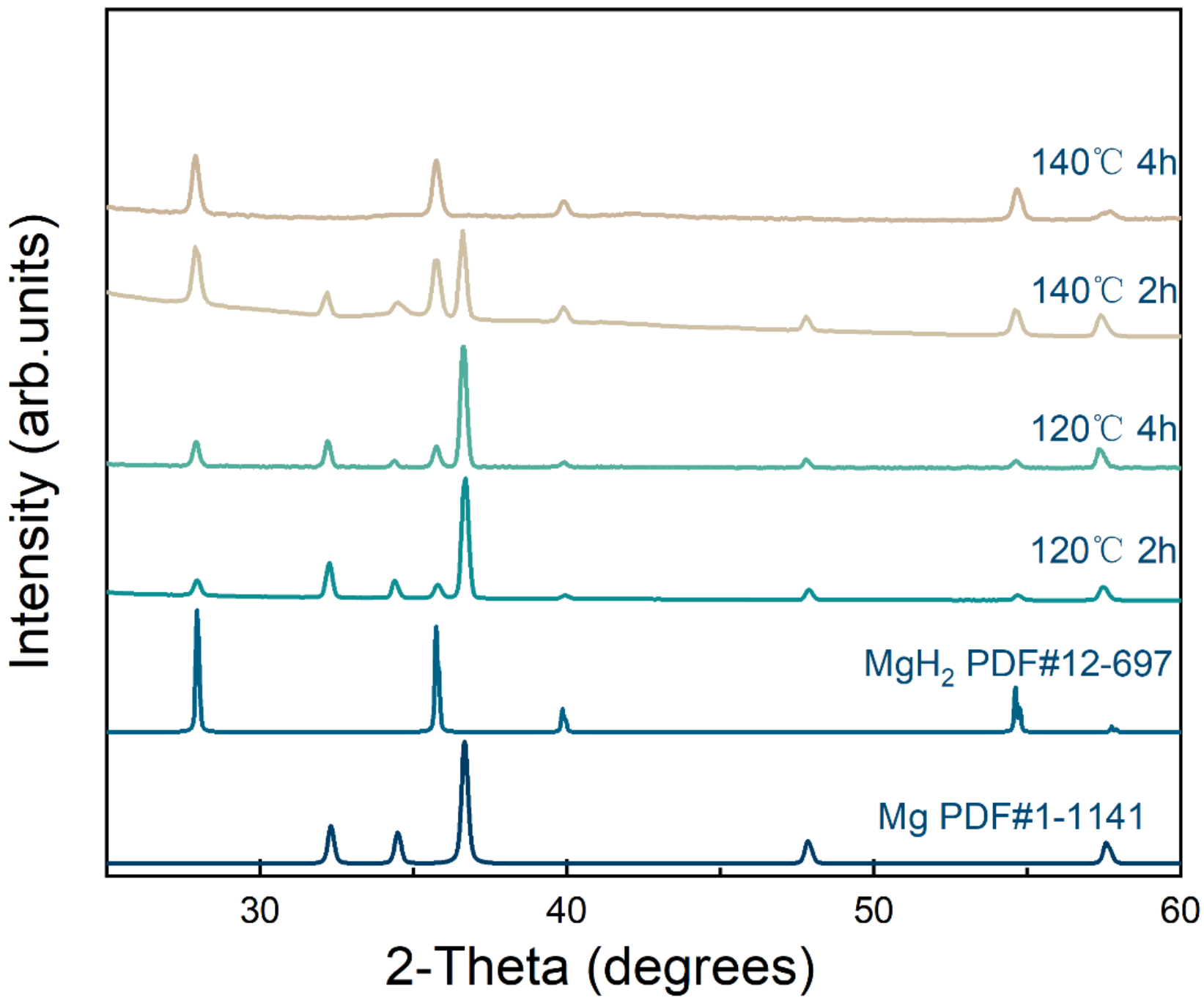

**Fig. S1. XRD phase of $MgH_2$ analysis under varying synthetic conditions.** The diffraction pattern shows that the product obtained at 120 °C for a short duration (2 hours) exhibits predominant metallic magnesium characteristics, indicating incomplete hydrogenation. By raising the temperature to 140 ° C and extending the reaction time, a complete phase transition towards crystalline $MgH_2$ was achieved. This phase transition is attributed to the HE phenomenon in acidic solutions, where hydrogen evolution corrosion increases the specific surface area of the metal and generates initial cracks. The crack propagation induced by HE promotes accelerated hydrogen diffusion pathways, ultimately promoting hydride formation (**Figure S2**).

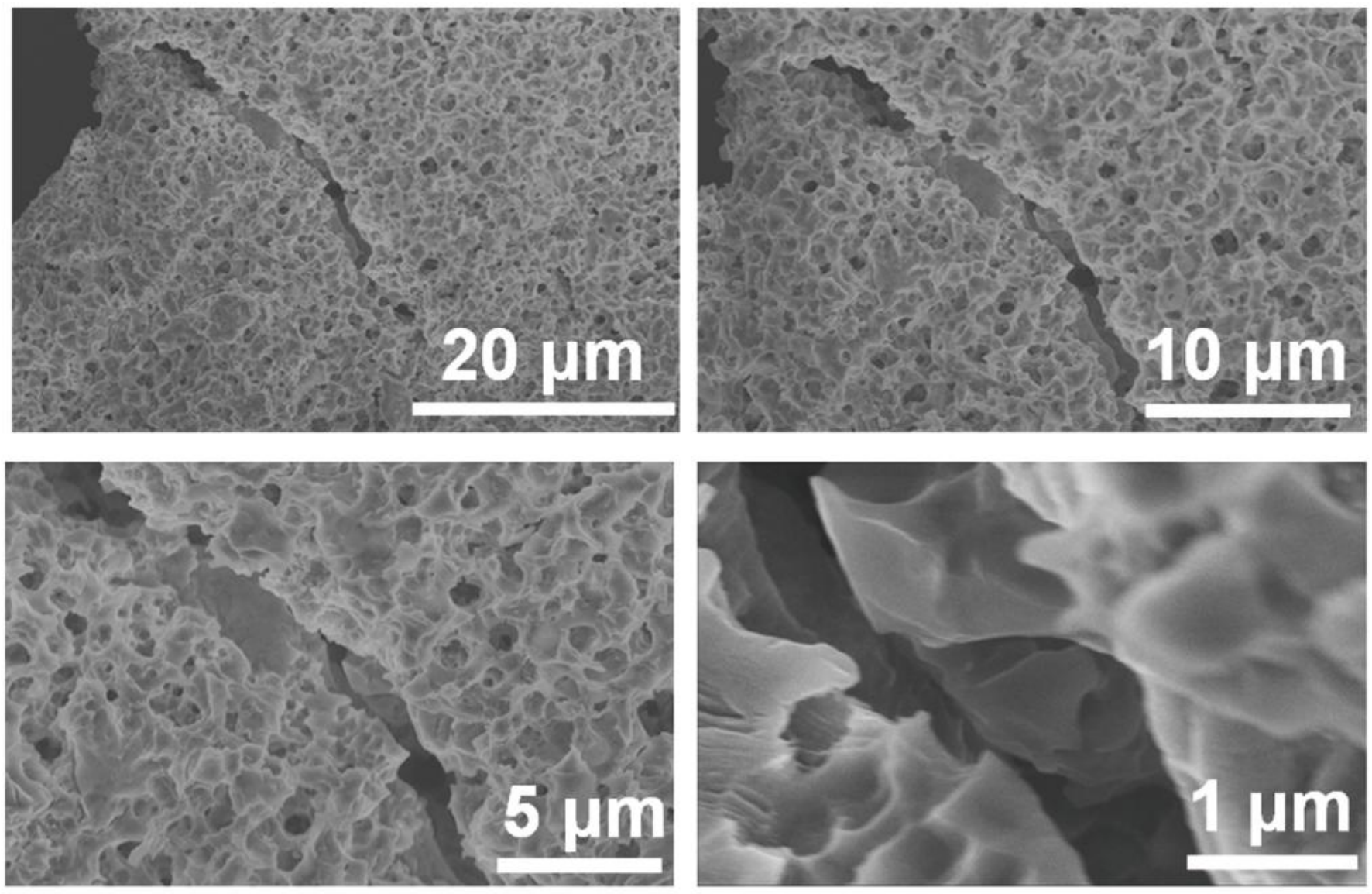

**Fig. S2. Representative SEM images of HTC-$MgH_2$.** The SEM micrographs reveal characteristic surface degradation patterns with pronounced corrosive morphology and micrometer-scale crack propagation (1 μm in length), demonstrating synergistic damage mechanisms arising from hydrogen evolution corrosion coupled with hydrogen embrittlement effects.

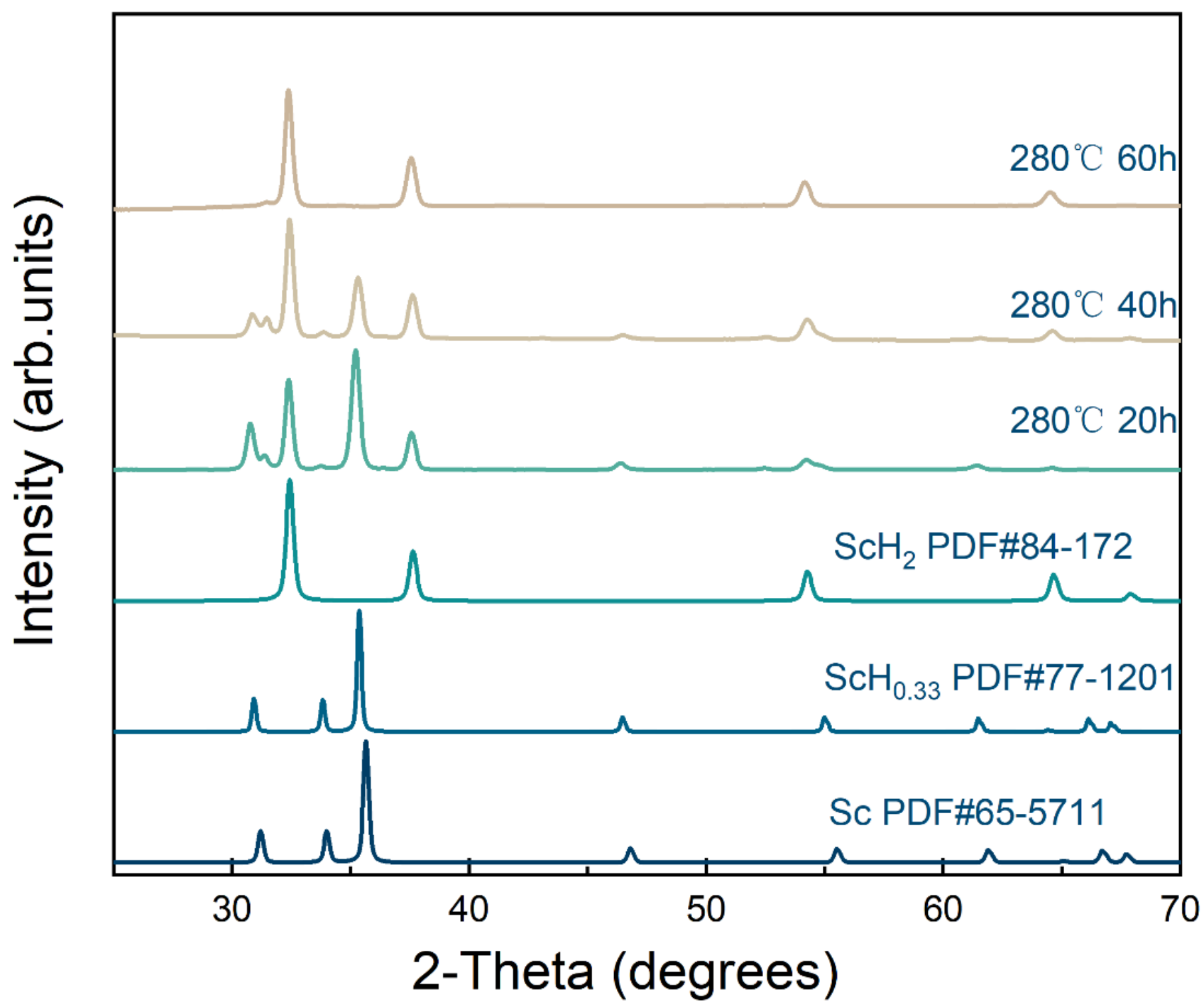

**Fig. S3. XRD phase of HTC-$ScH_2$ analysis under varying synthetic conditions.** The diffraction pattern shows that the product obtained at 280 °C for a short duration (20 h) exhibits the main $ScH_{0.33}$ features, indicating incomplete hydrogenation. By raising the temperature to 280 ° C and extending the reaction time (60 h), a complete phase transition towards crystalline $ScH_2$ was achieved. This phase transition is attributed to the HE phenomenon in acidic solutions, where hydrogen evolution corrosion increases the specific surface area of the metal and generates initial cracks. The crack propagation

induced by HE promotes accelerated hydrogen diffusion pathways, ultimately promoting hydride formation **(Figure S4**).

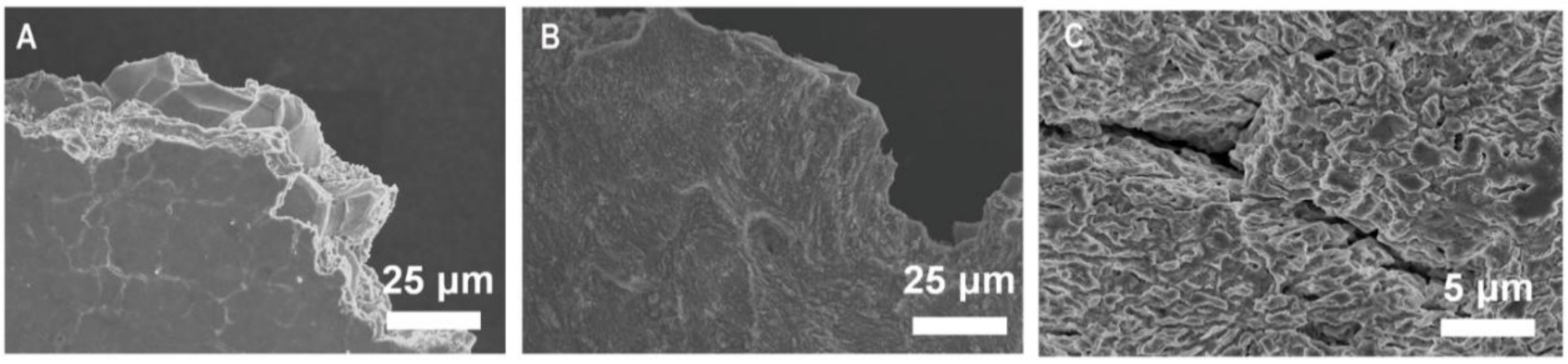

**Fig. S4. Representative SEM images of HTC-$ScH_2$.** (**A**) Initial stage: The metal surface is relatively smooth and flat. (**B**) Intermediate stage of reaction (280 ℃, 40 hours): Significant hydrogen evolution corrosion marks appear on the metal surface, and microcracks (<200 nm width) appear. (**C**) End of reaction (280 ℃, 60 hours): The surface of the sample is severely corroded and cracks with a width of 1.2 ± 0.3 µm appear. This indicates that HE in acid solution provides H through hydrogen evolution corrosion and generates initial cracks. Crack propagation accelerates hydrogen diffusion and promotes the formation of $ScH_2$. The morphological evolution is consistent with the XRD phase transition shown in **Figure S3**.

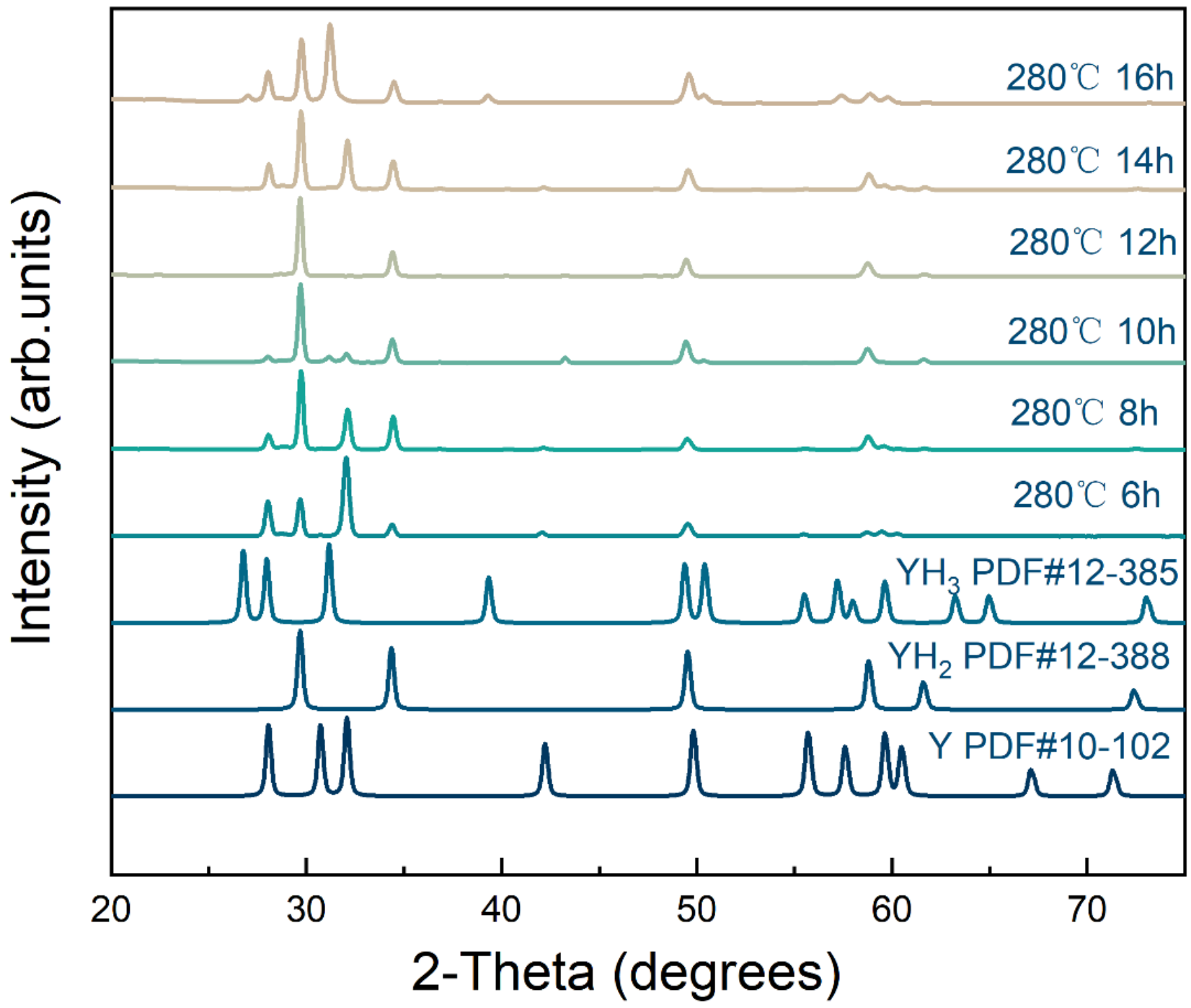

**Fig. S5**. **XRD phase of HTC-$YH_2$ analysis under varying synthetic conditions.** The diffraction pattern shows that the product obtained at 280 °C for a short period of time (6 hours) exhibits the main characteristics of Y, indicating incomplete hydrogenation. By raising the temperature to 280 ° C and extending the reaction time (12 hours), a complete phase transition towards crystalline $YH_2$ was achieved. This phase transition is attributed to the HE phenomenon in acidic solutions, where hydrogen evolution corrosion increases the specific surface area of the metal and generates initial cracks.

The crack propagation caused by HE accelerates the hydrogen diffusion pathway, ultimately promoting the formation of hydrides (**Figure S6**).

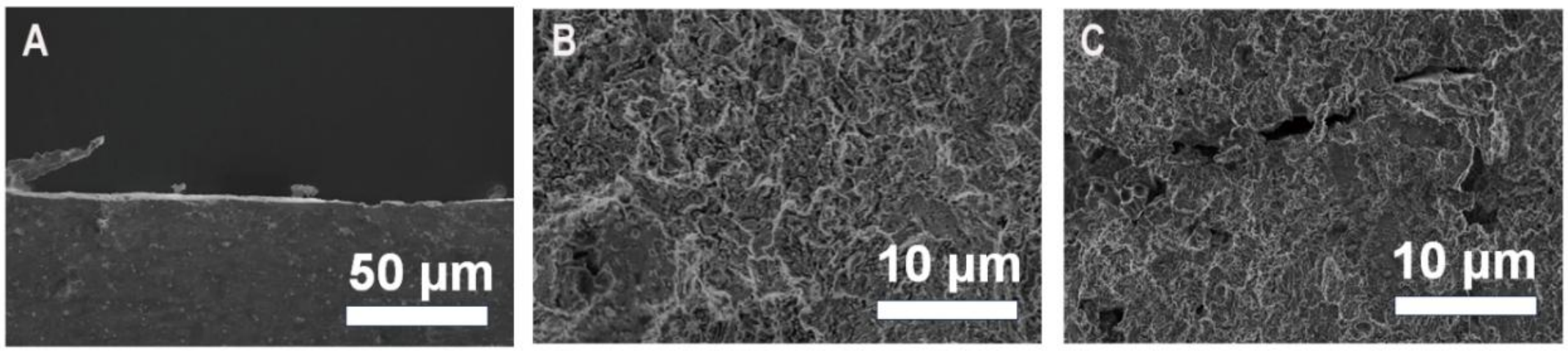

**Fig. S6. Representative SEM images of HTC-$YH_2$.** (**A**) Initial stage: The metal surface is relatively smooth and flat. (**B**) Intermediate stage of reaction (280 °C, 8 hours): Significant hydrogen evolution corrosion marks appear on the metal surface, and microcracks (0.5-1 μm wide) appear. (**C**) End of reaction (280 °C, 10 hours): The surface of the sample is severely corroded, and cracks with a width of 1-2 μm appear. This indicates that HE in acidic solutions provides H through hydrogen evolution corrosion and generates initial cracks. Crack propagation accelerates the diffusion of hydrogen and promotes the formation of $YH_2$. The morphological evolution is consistent with the XRD phase transition shown in **Figure S5**.

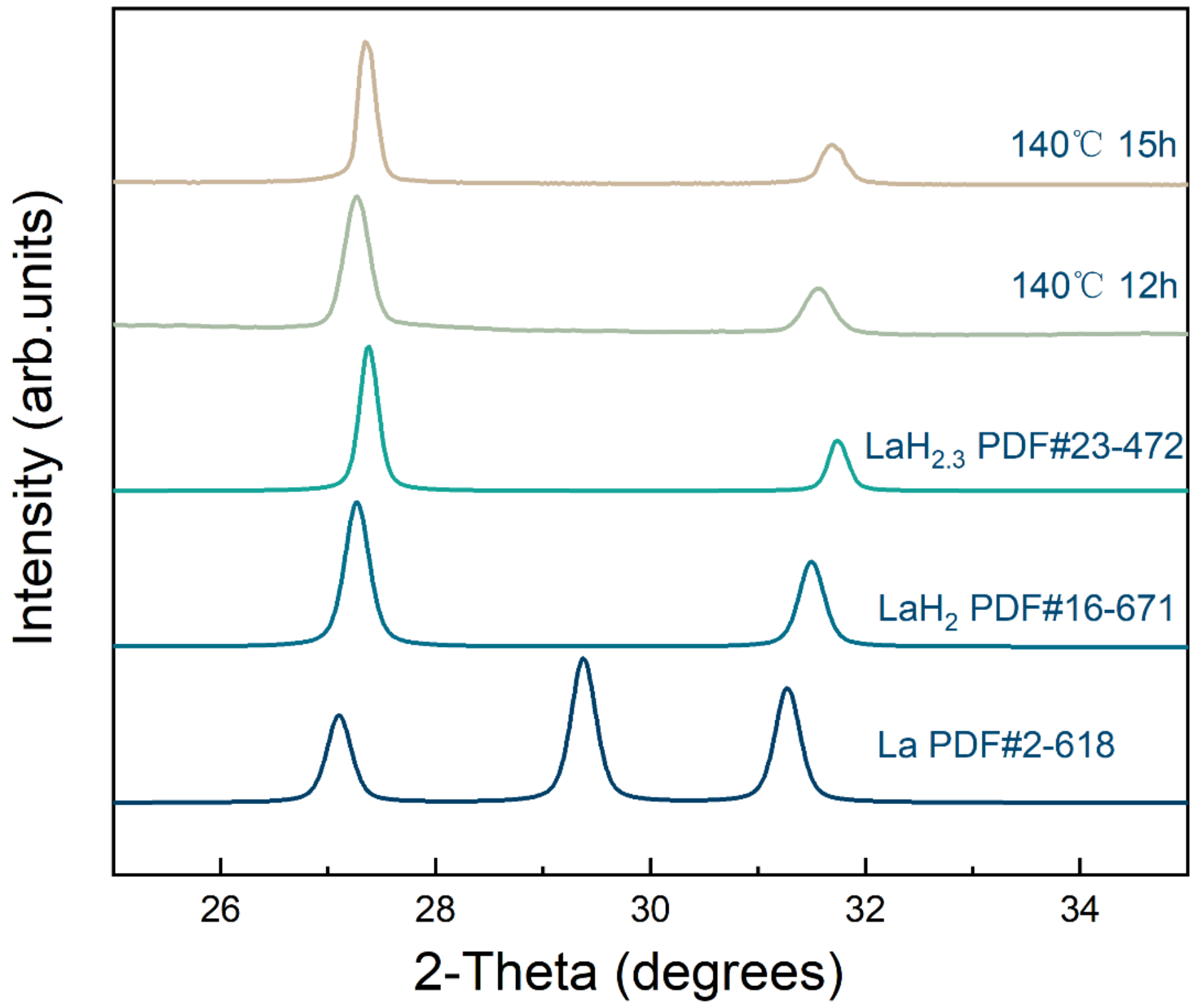

**Fig. S7. XRD phase analysis under varying synthetic conditions.** When the reaction conditions are (140 °C 20h), the XRD diffraction peak is mainly $LaH_2$, and there are no other diffraction peaks, indicating that the product is metallic $LaH_2$ with high purity at this time; When the reaction proceeds to (140 °C 15h), the XRD pattern shows that the XRD diffraction peak is mainly $LaH_{2.3}$, and there are no other diffraction peaks, indicating that the product is metallic $LaH_{2.3}$ with high purity at this time.

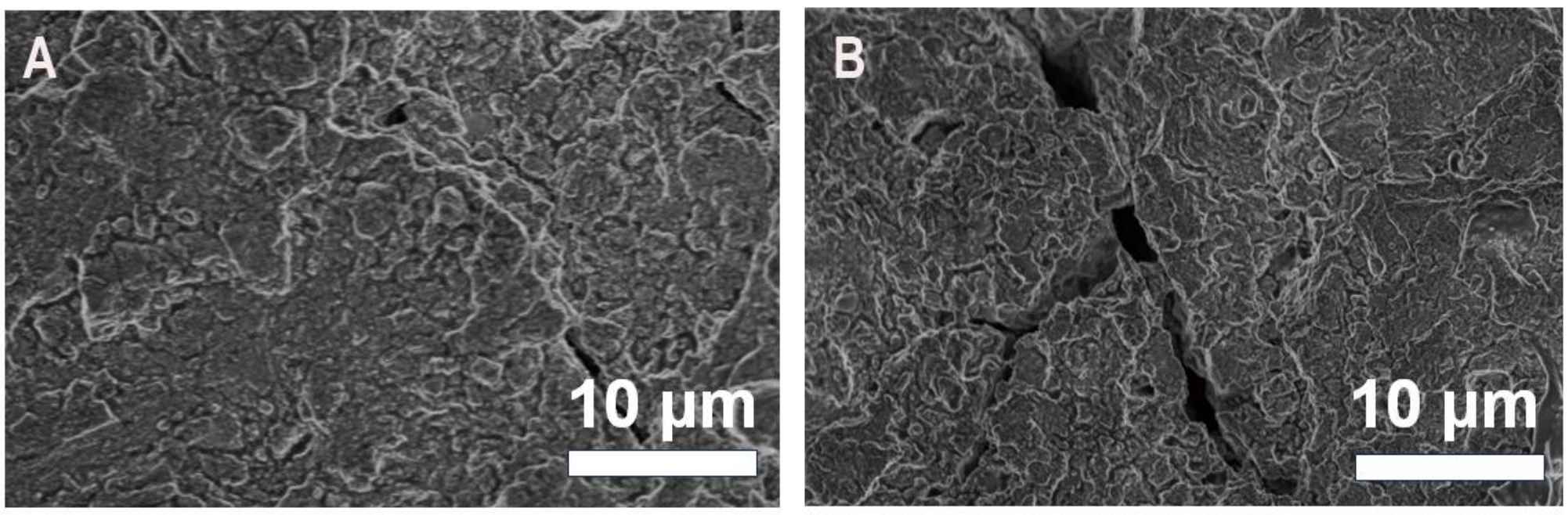

**Fig. S8. Representative SEM images of HTC-$LaH_2$ and HTC-$LaH_{2.3}$.** Fig. S8. Representative SEM images of $LaH_2$ and $LaH_{2.3}$. Characterization by scanning electron microscopy (SEM) revealed that $LaH_2$ and $LaH_{2.3}$ the surface of the sample exhibits significant acid etching characteristics and microcrack structures. Quantitative analysis shows that the surface crack width of $LaH_2$ is distributed in the range of 0.5-1 μm, while $LaH_{2.3}$ Surface cracks propagate to 1-2 μm, accompanied by more pronounced local corrosion pit morphology. This indicates that HE in acidic solutions provides H through hydrogen evolution corrosion and generates initial cracks. Crack propagation accelerates the diffusion of hydrogen and promotes the formation of $LaH_2$ and $LaH_{2.3}$. The morphological evolution is consistent with the XRD phase transition shown in **Figure S7**.

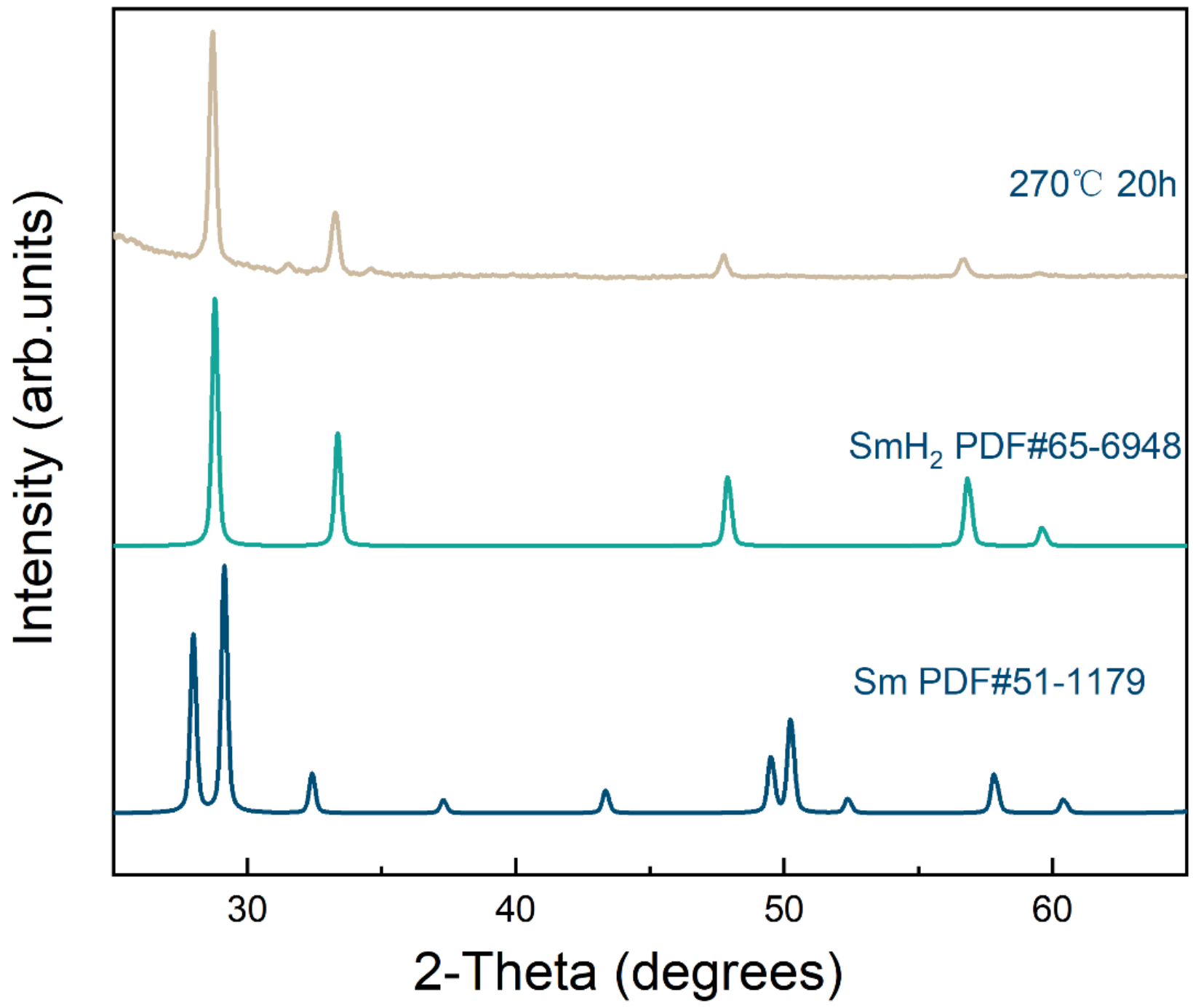

**Fig. S9. XRD phase analysis under varying synthetic conditions.** When the reaction conditions are (270 °C 40h), the XRD diffraction peak is mainly $SmH_2$, and there are no other diffraction peaks, indicating that the product is metallic $SmH_2$ with high purity.

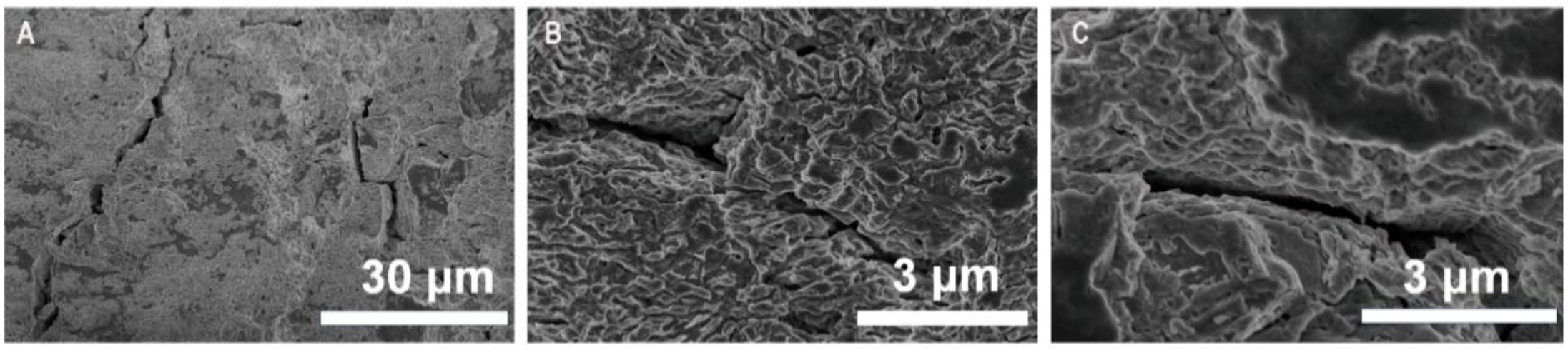

**Fig. S10. Representative SEM images of** HTC-**$SmH_2$.** The surface of $SmH_2$ has severe acid corrosion marks, and the 0.15 μm crack has formed on the surface, which is caused by the combined effects of hydrogen evolution corrosion and hydrogen embrittlement. This indicates that HE in acidic solutions provides H through hydrogen evolution corrosion and generates initial cracks. Crack propagation accelerates the diffusion of hydrogen and promotes the formation of $SmH_2$.

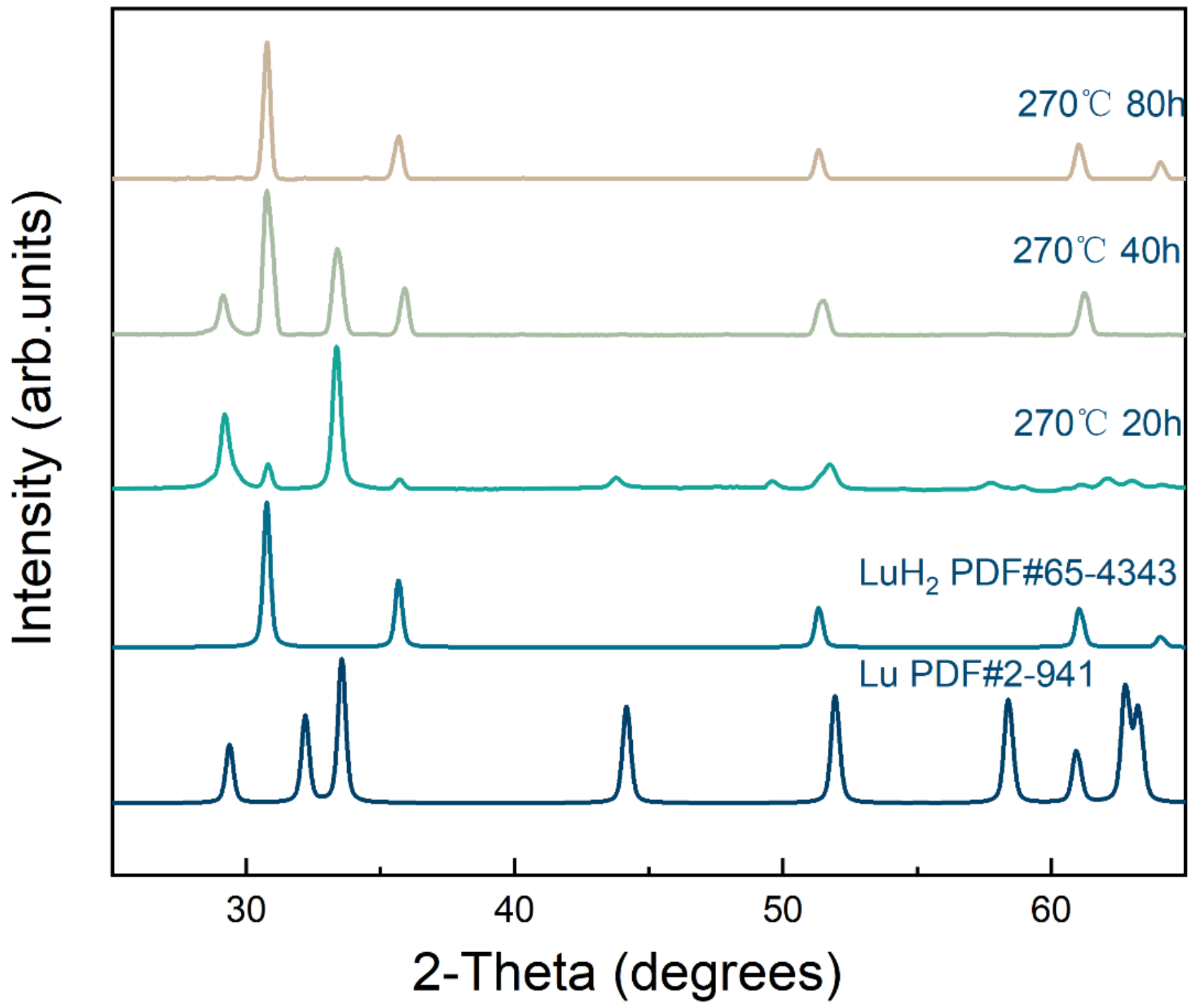

**Fig. S11. XRD phase analysis under varying synthetic conditions.** The diffraction pattern shows that the product obtained at 280 °C for a short period of time (20 hours) exhibits the main characteristics of Lu, indicating incomplete hydrogenation. By raising the temperature to 280 ° C and extending the reaction time (80 hours), a complete phase transition towards crystalline $LuH_2$ was achieved. This phase transition is attributed to the HE phenomenon in acidic solutions, where hydrogen evolution corrosion increases the specific surface area of the metal and generates initial cracks. The crack propagation caused by HE accelerates the hydrogen diffusion pathway, ultimately promoting the formation of hydrides (**Figure S12**).

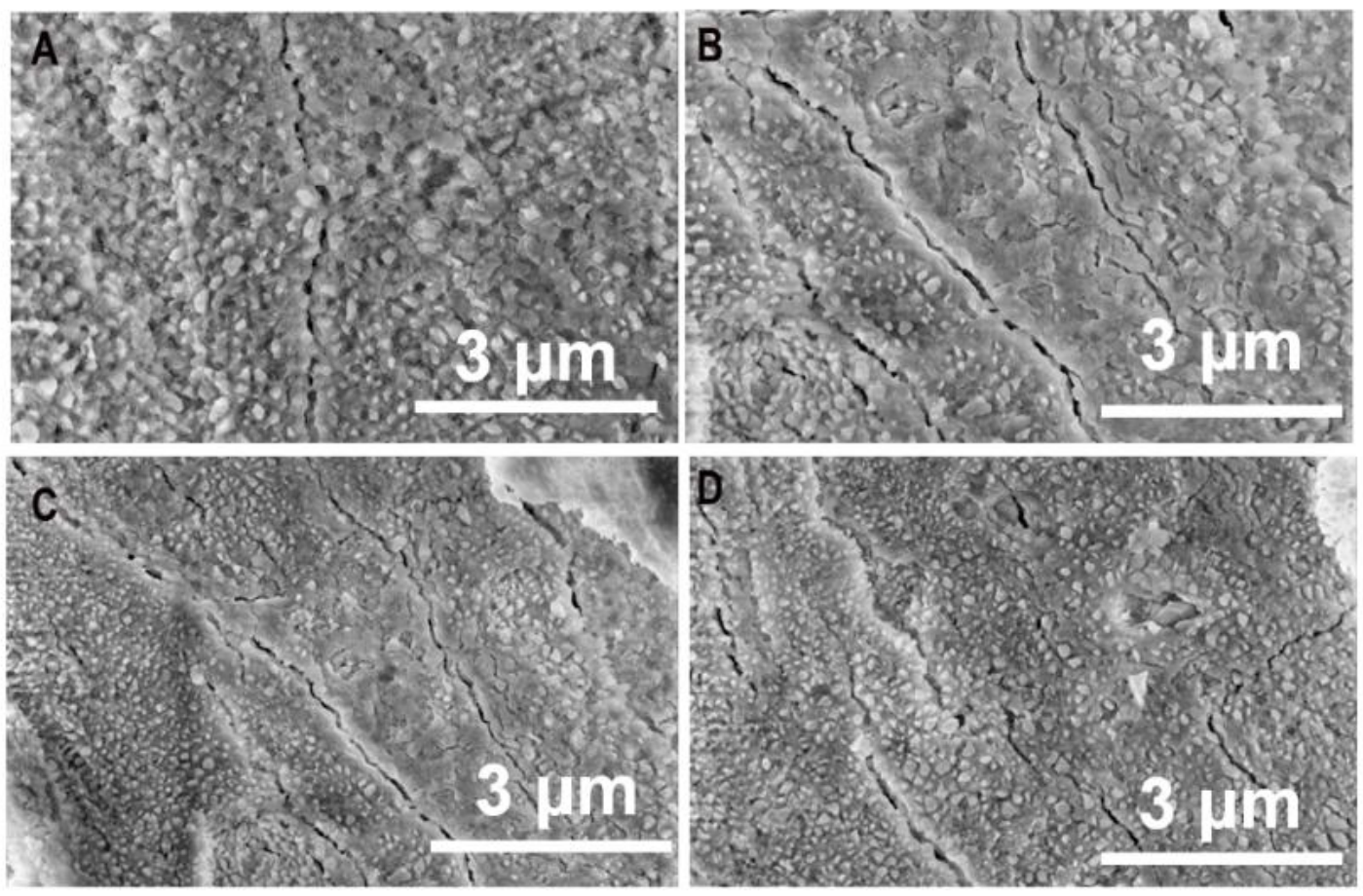

**Fig. S10. Representative SEM images of HTC-$LuH_2$.** The surface of $LuH_2$ has severe acid corrosion marks, and the 0.2 μm crack has formed on the surface. This indicates that HE in acidic solutions provides H through hydrogen evolution corrosion and generates initial cracks. Crack propagation accelerates the diffusion of hydrogen and promotes the formation of $LuH_2$.

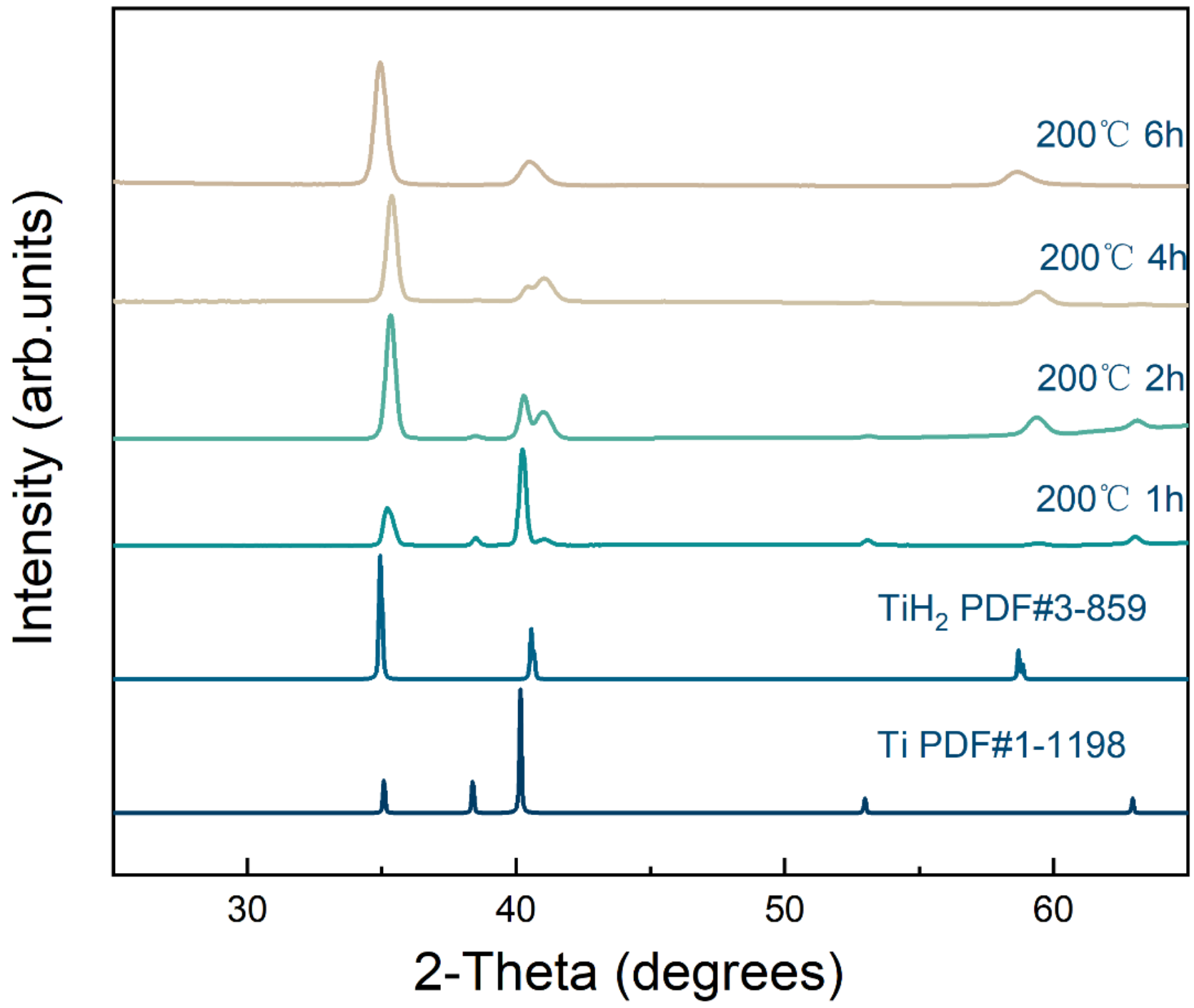

**Fig. S13. XRD phase analysis under varying synthetic conditions.** The diffraction pattern shows that the product obtained at 200 °C for a short period of time (1 hour) exhibits the main characteristics of Ti, indicating incomplete hydrogenation. By extending the reaction time (6 hours), a complete phase transition towards crystalline $TiH_2$ was achieved. This phase transition is attributed to the HE phenomenon in acidic solutions, where hydrogen evolution corrosion increases the specific surface area of the metal and generates initial cracks. The crack propagation caused by HE accelerates the hydrogen diffusion pathway, ultimately promoting the formation of hydrides (**Figure S14**).

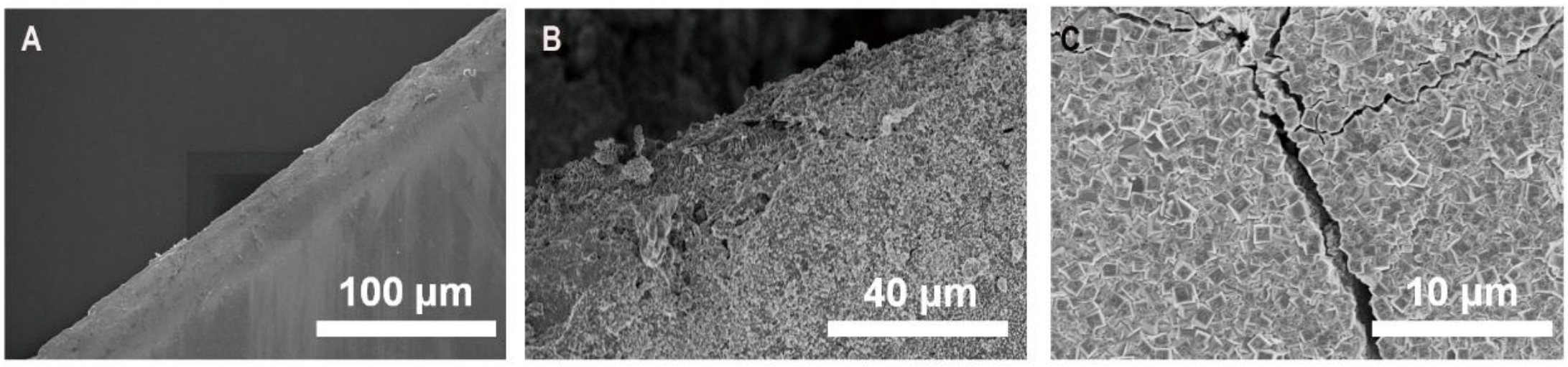

**Fig. S14. Representative SEM images of HTC-$TiH_2$.** (**A**) Initial stage: The metal surface is relatively smooth and flat. (**B**) Intermediate stage of reaction (200 °C, 1 hours): There are obvious hydrogen evolution corrosion marks on the metal surface, and microcracks appear. (**C**) End of reaction (200 °C, 6 hours): The surface of the sample is severely corroded, and cracks with a width of 1 μm appear. This indicates that HE in acidic solutions provides H through hydrogen evolution corrosion and generates initial cracks. Crack propagation accelerates the diffusion of hydrogen and promotes the formation of $TiH_2$. The morphological evolution is consistent with the XRD phase transition shown in **Figure S13**.

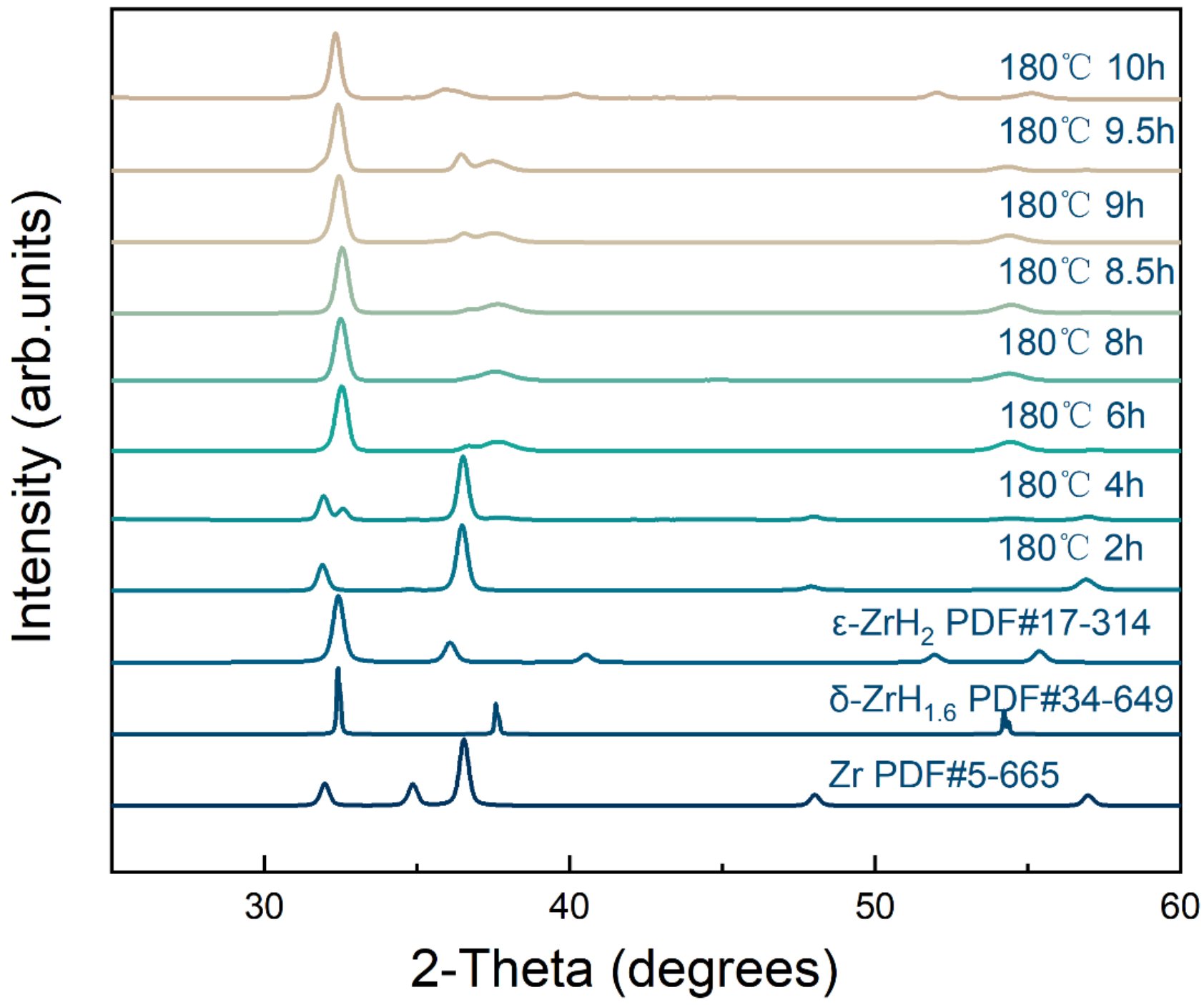

**Fig. S15. XRD phase analysis under varying synthetic conditions.** The diffraction pattern shows that the product obtained at 180 °C for a short period of time (2 hours) exhibits the main characteristics of Zr, indicating incomplete hydrogenation. By extending the reaction time (6 hours), a complete phase transition towards crystalline **δ-**$ZrH_{1.6}$ was achieved. By extending the reaction time (10 hours), a complete phase transition towards crystalline **ε-**$ZrH_2$ was achieved. This phase transition is attributed to the HE phenomenon in acidic solutions, where hydrogen evolution corrosion increases the specific surface area of the metal and generates initial cracks. The crack propagation caused by HE accelerates the hydrogen diffusion pathway, ultimately promoting the formation of hydrides (**Figure S16**).

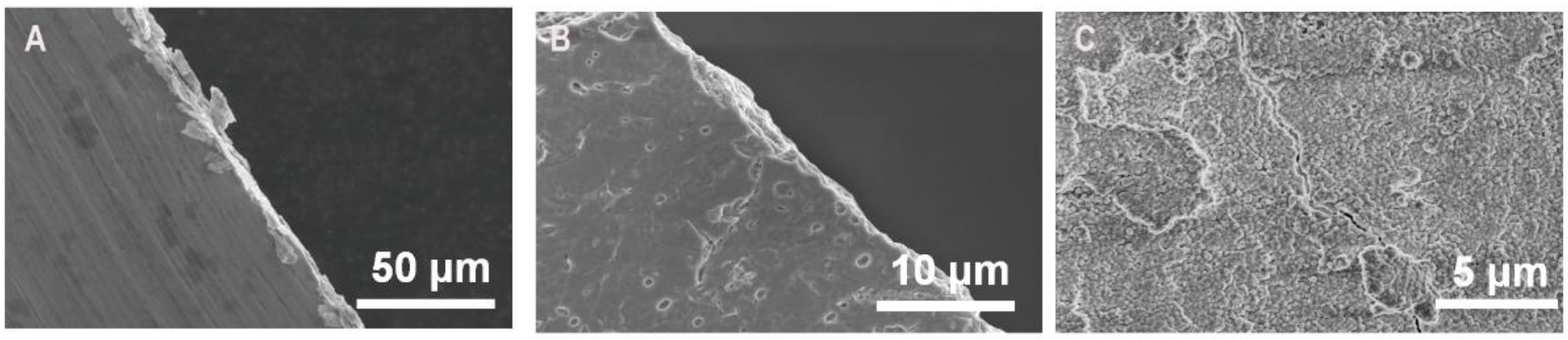

**Fig. S16. Representative SEM images of HTC δ-$ZrH_{1.6}$ and HTC ε-$ZrH_2$.** (**A**) Initial stage: The metal surface is relatively smooth and flat. (**B**) δ-$ZrH_{1.6}$ (180 °C, 8 hours): There are obvious hydrogen evolution corrosion marks on the metal surface, and 0.5 μm microcracks appear. (**C**) ε-$ZrH_2$ (180 °C, 10 hours): The surface of the sample is severely corroded, and cracks with a width of 1 μm appear. This indicates that HE in acidic solutions provides H through hydrogen evolution corrosion and generates initial cracks. Crack propagation accelerates the diffusion of hydrogen and promotes the formation of δ-$ZrH_{1.6}$ and ε-$ZrH_2$. The morphological evolution is consistent with the XRD phase transition shown in **Figure S15**.

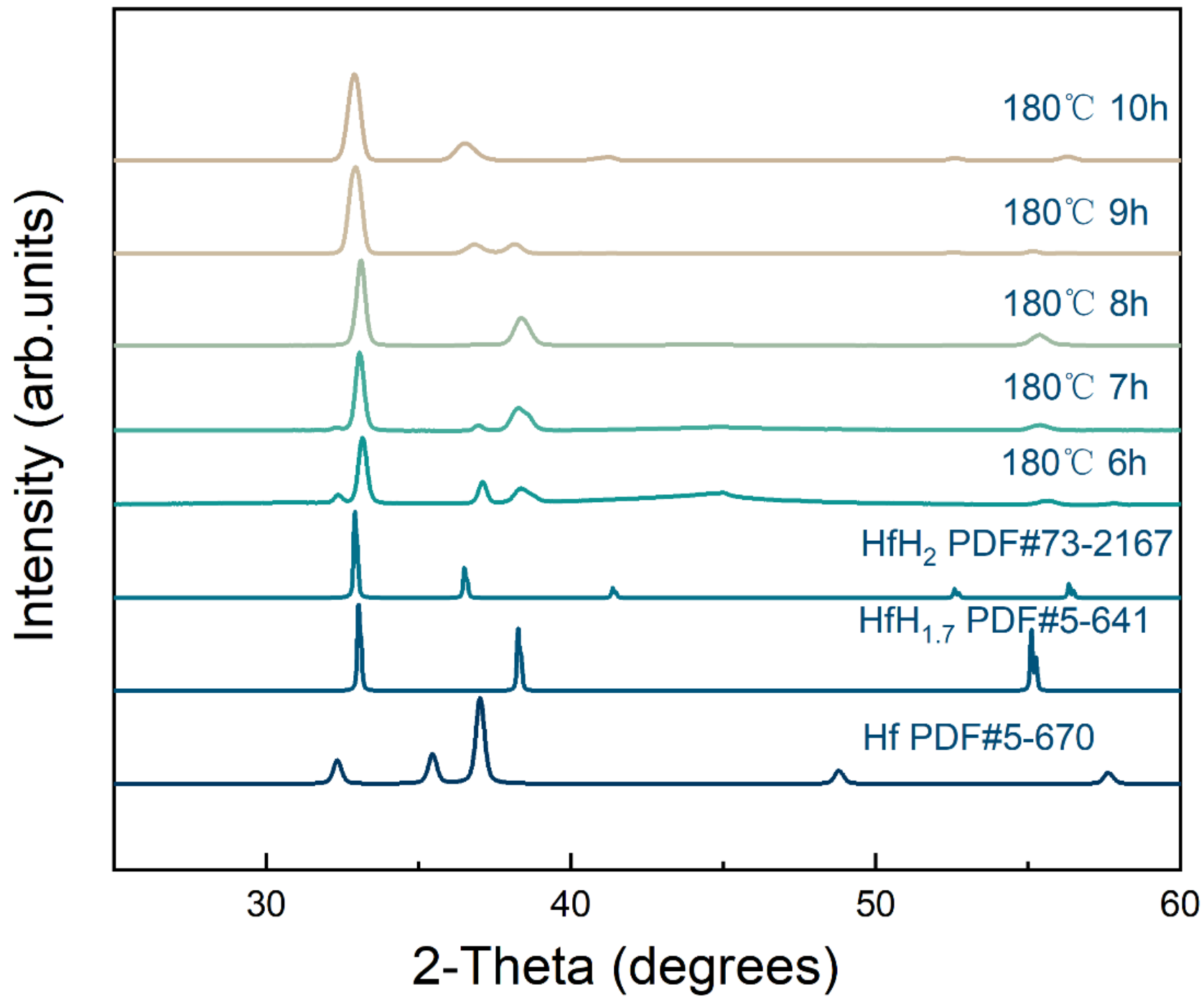

**Fig. S17. XRD phase analysis under varying synthetic conditions.** The diffraction pattern shows that the product obtained at 180 °C for a short period of time (6 hours) exhibits the main characteristics of Hf, indicating incomplete hydrogenation. By extending the reaction time (8 hours), a complete phase transition towards crystalline $HfH_{1.7}$ was achieved. By extending the reaction time (10 hours), a complete phase transition towards crystalline $HfH_2$ was achieved. This phase transition is attributed to the HE phenomenon in acidic solutions, where hydrogen evolution corrosion increases the specific surface area of the metal and generates initial cracks. The crack propagation

caused by HE accelerates the hydrogen diffusion pathway, ultimately promoting the formation of hydrides (**Figure S18**).

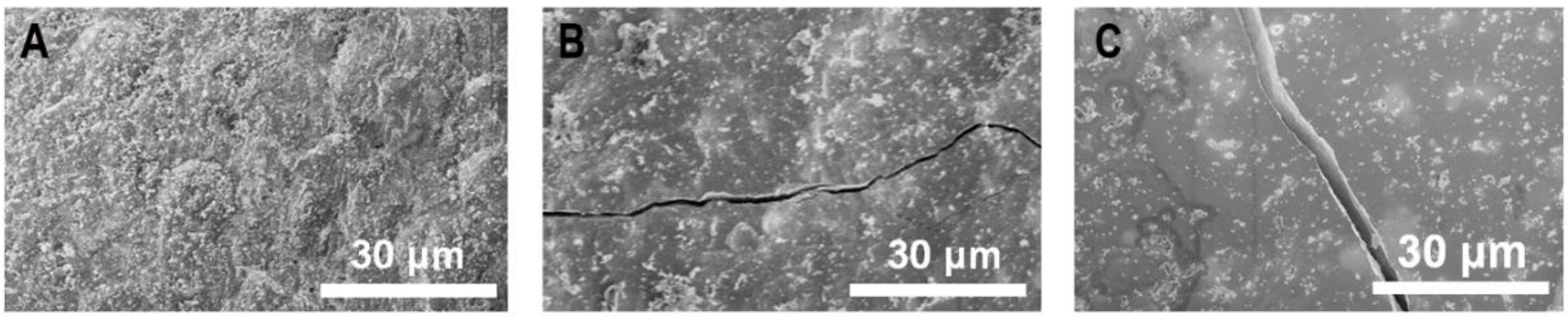

**Fig. S18. Representative SEM images of HTC-$HfH_{1.7}$ and HTC-$HfH_2$.** (A) Initial stage: The metal surface is relatively smooth and flat. (B) $HfH_{1.7}$ (180 ℃, 8 hours): There are obvious traces of hydrogen evolution corrosion on the metal surface, and 1.5 μm microcracks appear. (C) $HfH_2$ (180 ℃, 10 hours): The sample surface is severely corroded and cracks with a width of 4 μm appear. This indicates that HE in acidic solutions provides H through hydrogen evolution corrosion and generates initial cracks. Crack propagation accelerates the diffusion of hydrogen and promotes the formation of $HfH_{1.7}$ and $HfH_2$. The morphological evolution is consistent with the XRD phase transition shown in **Figure S17**.

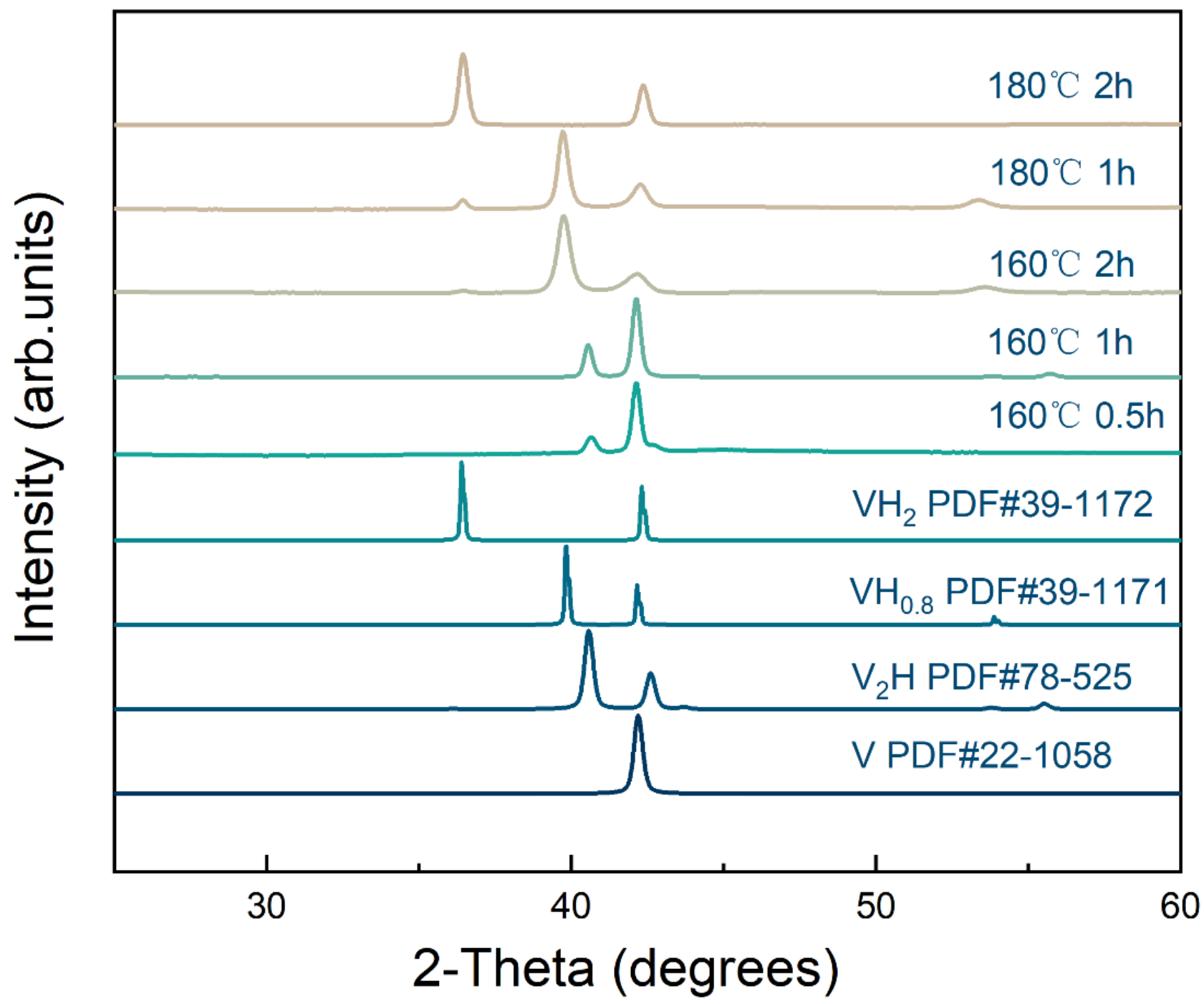

**Fig. S19. XRD phase analysis under varying synthetic conditions.** The diffraction pattern shows that the product obtained at 180 °C for a short period of time (6 hours) exhibits the main characteristic of V, indicating incomplete hydrogenation. By extending the reaction time (8 hours), a complete phase transition towards crystalline $VH_{0.8}$ was achieved. By extending the reaction time (10 hours), a complete phase transition towards crystalline $VH_2$ was achieved. This phase transition is attributed to the HE phenomenon in acidic solutions, where hydrogen evolution corrosion increases the specific surface area of the metal and generates initial cracks. The crack propagation

caused by HE accelerates the hydrogen diffusion pathway, ultimately promoting the formation of hydrides (**Figure S20**).

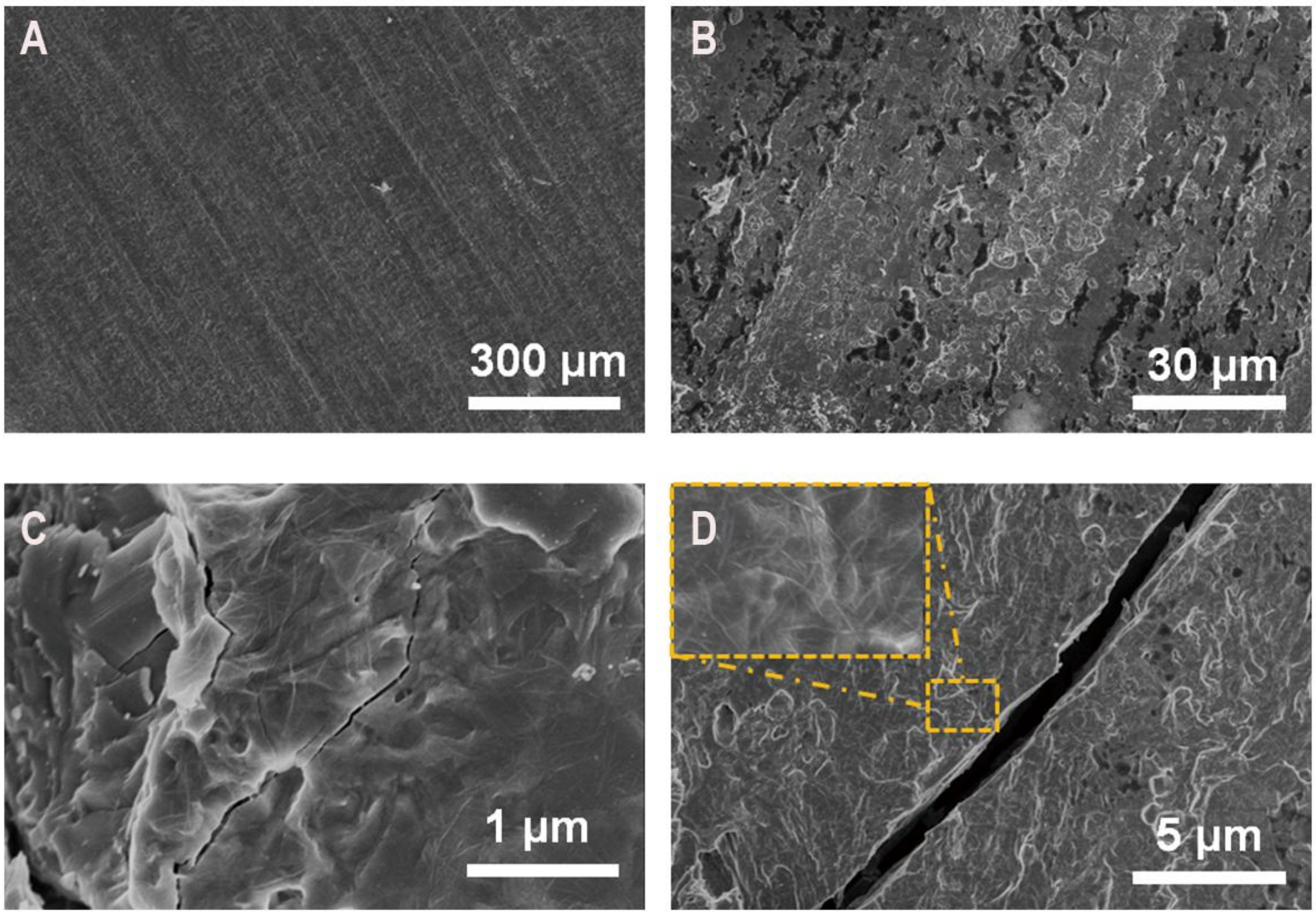

**Fig. S20. Representative SEM images of HTC-$VH_{0.8}$ and HTC-$VH_2$.** (**A**) Initial stage: The metal surface is relatively smooth and flat. (**B**) Reacting at 180 ℃ for 6 hours, hydrogen evolution corrosion marks appeared on the metal surface (**C**) $VH_{0.8}$ (180 ℃, 8 hours): There were obvious hydrogen evolution corrosion marks on the metal surface, with micro cracks of 0.06 μm. (**D**) $VH_2$ (180 ℃, 10 hours): The sample surface is severely corroded and cracks with a width of 1.2 μm appear. This indicates that HE in acidic solutions provides H through hydrogen evolution corrosion and generates initial cracks. The crack propagation accelerated the diffusion of hydrogen

and promoted the formation of $VH_{0.8}$ and $VH_2$. The morphological evolution is consistent with the XRD phase transition shown in Figure S15.

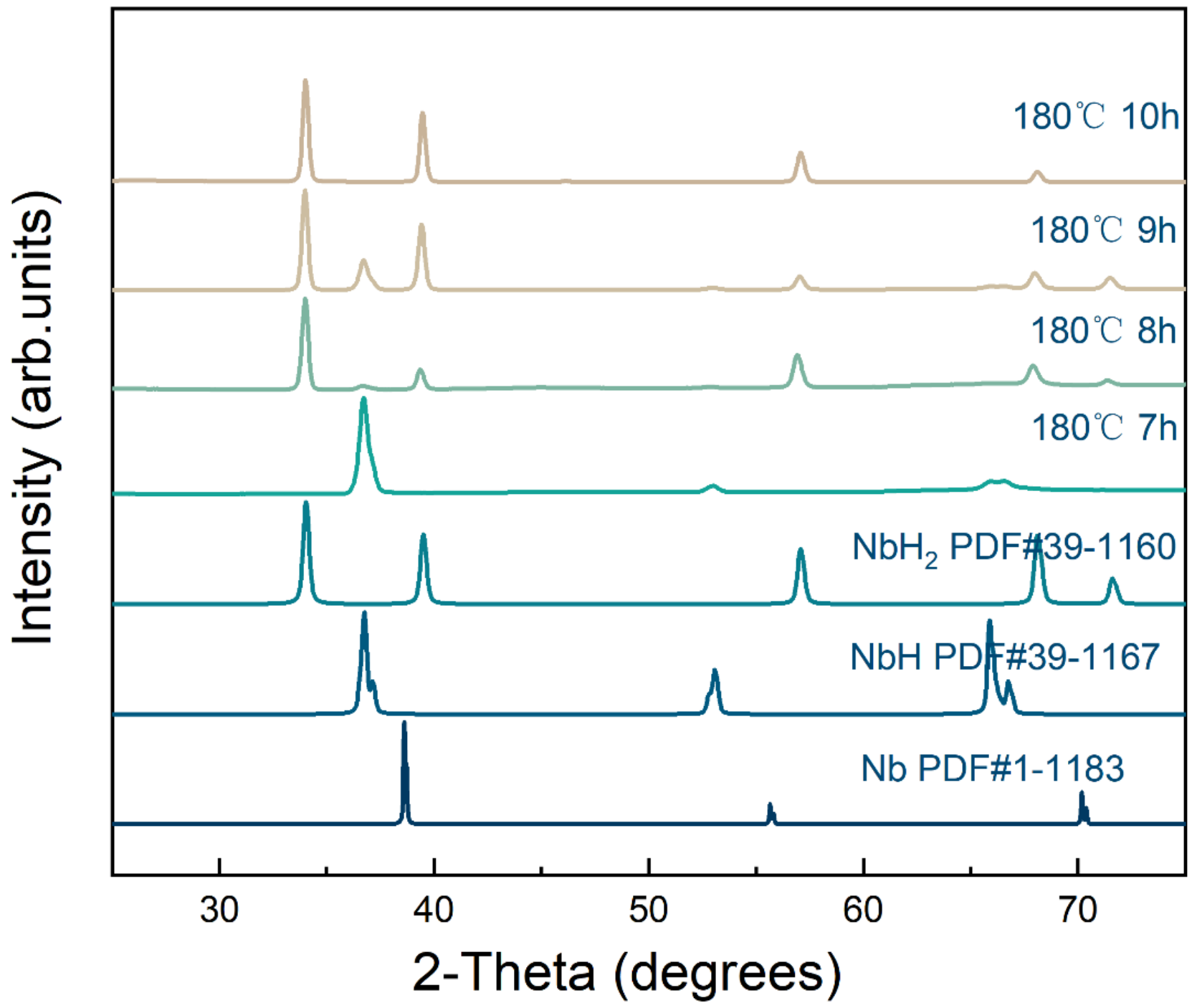

**Fig. S21. XRD phase analysis under varying synthetic conditions.** The diffraction pattern shows that a complete phase transition of NbH was achieved at 180 °C for 7 hours. By extending the time interval (10 hours), a complete phase transition towards crystalline $NbH_2$ was achieved. This phase transition is attributed to the HE phenomenon in acidic solutions, where hydrogen evolution corrosion increases the specific surface area of the metal and generates initial cracks. The crack propagation caused by HE accelerates the hydrogen diffusion pathway, ultimately promoting the formation of hydrides (**Figure S22**).

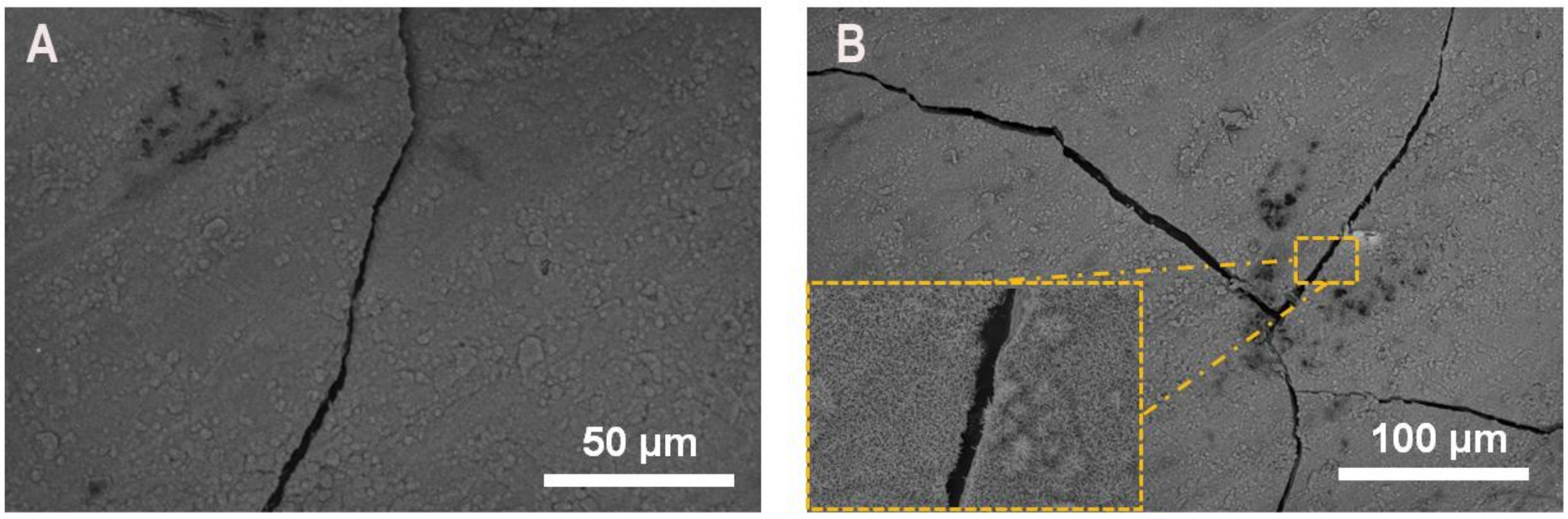

**Fig. S22. Representative SEM images of HTC-NbH and HTC-$NbH_2$.** (**A**) NbH (180 ℃, 7 hours): There are obvious traces of hydrogen evolution corrosion on the metal surface, with microcracks of 2.5 μm. (**B**) $NbH_2$ (180 ℃, 10 hours): The sample surface is severely corroded and cracks with a width of 5 μm appear. This indicates that HE in acidic solutions provides H through hydrogen evolution corrosion and generates initial cracks. Crack propagation accelerates the diffusion of hydrogen and promotes the formation of NbH and $NbH_2$. The morphological evolution is consistent with the XRD phase transition shown in **Figure S21**.

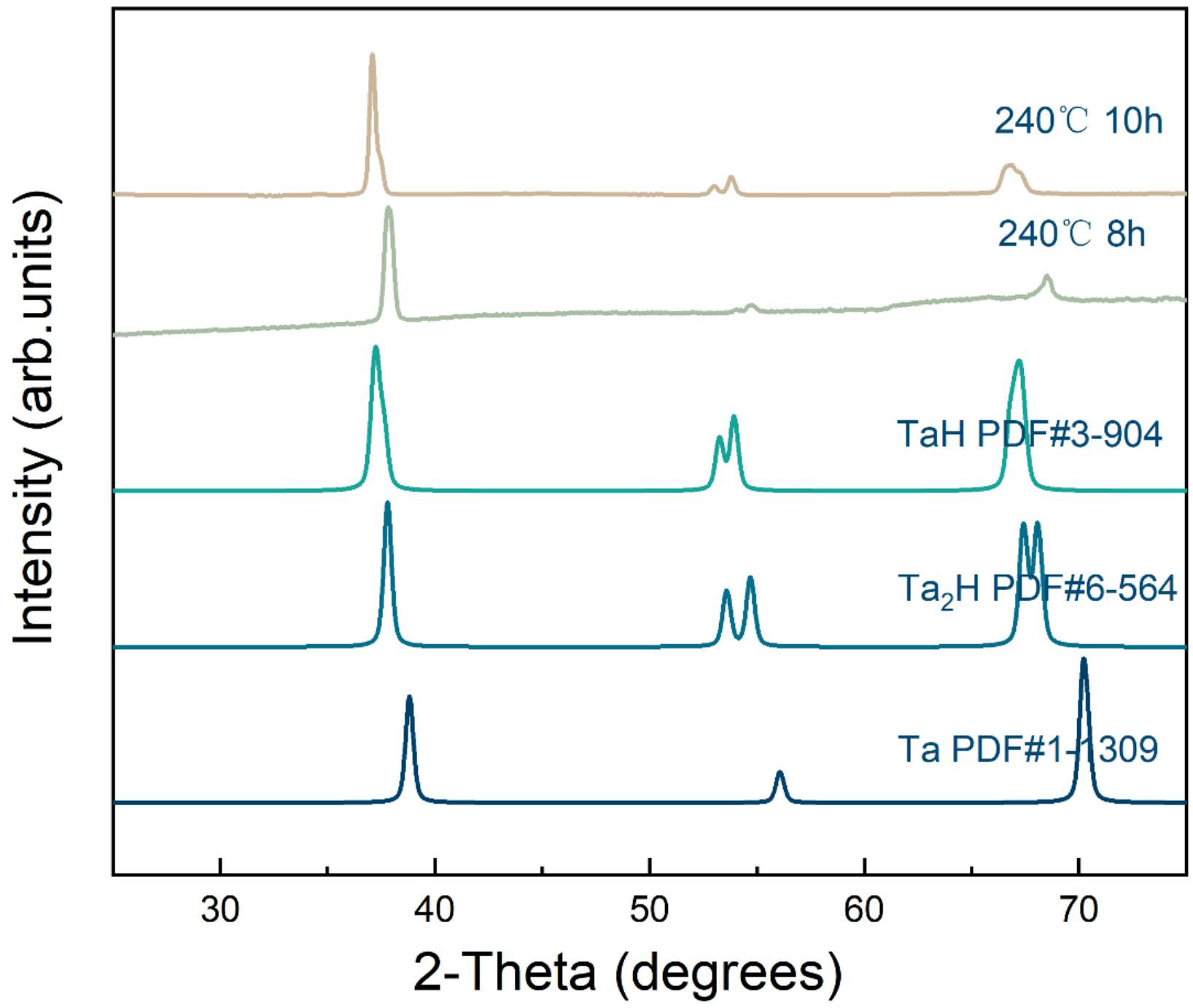

**Fig. S23. XRD phase analysis under varying synthetic conditions.** The diffraction pattern shows that $Ta_2H$ was fully formed after 8 hours at 240 °C. By extending the time (10 hours), a complete phase transition towards crystalline TaH was achieved. This phase transition is attributed to the HE phenomenon in acidic solutions, where hydrogen evolution corrosion increases the specific surface area of the metal and generates initial cracks. The crack propagation caused by HE accelerates the hydrogen diffusion pathway, ultimately promoting the formation of hydrides (**Figure S24**).

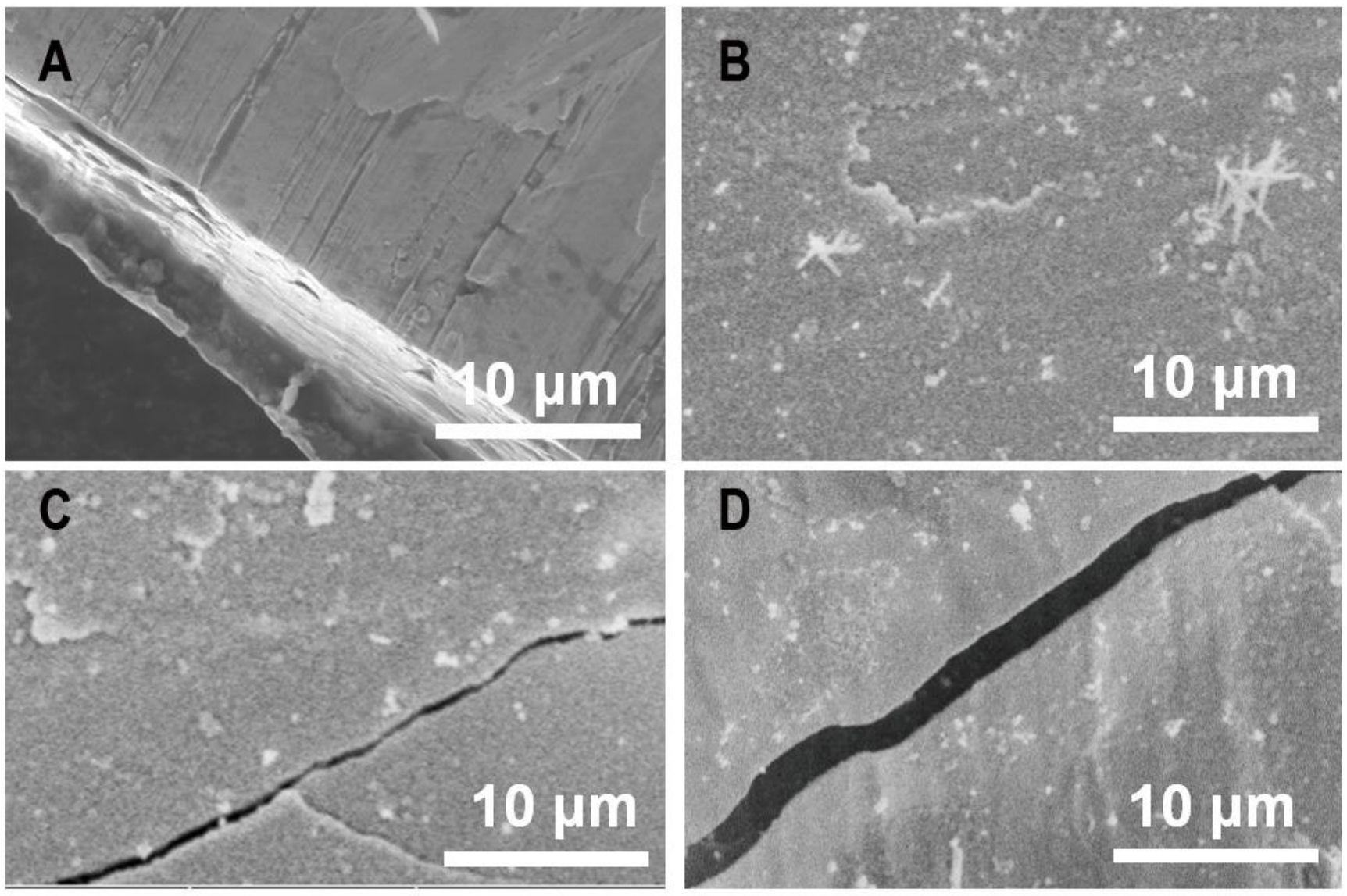

**Fig. S24. Representative SEM images of HTC-$Ta_2H$ and TaH.** (**A**) Initial stage: The metal surface is relatively smooth and flat. (**B**) Reacting at 240 °C for 2 hours, hydrogen evolution corrosion marks appeared on the metal surface (**C**) $Ta_2H$ (240 °C, 8 hours): There were obvious hydrogen evolution corrosion marks on the metal surface, with micro cracks of 0.5 μm. (**D**) TaH (240 °C, 10 hours): The sample surface is severely corroded and cracks with a width of 2 μm appear. This phase transition is attributed to the HE phenomenon in acidic solutions, where hydrogen evolution corrosion increases the specific surface area of the metal and generates initial cracks. The crack propagation caused by HE accelerates the hydrogen diffusion pathway, ultimately promoting the formation of hydrides (**Figure S23**).

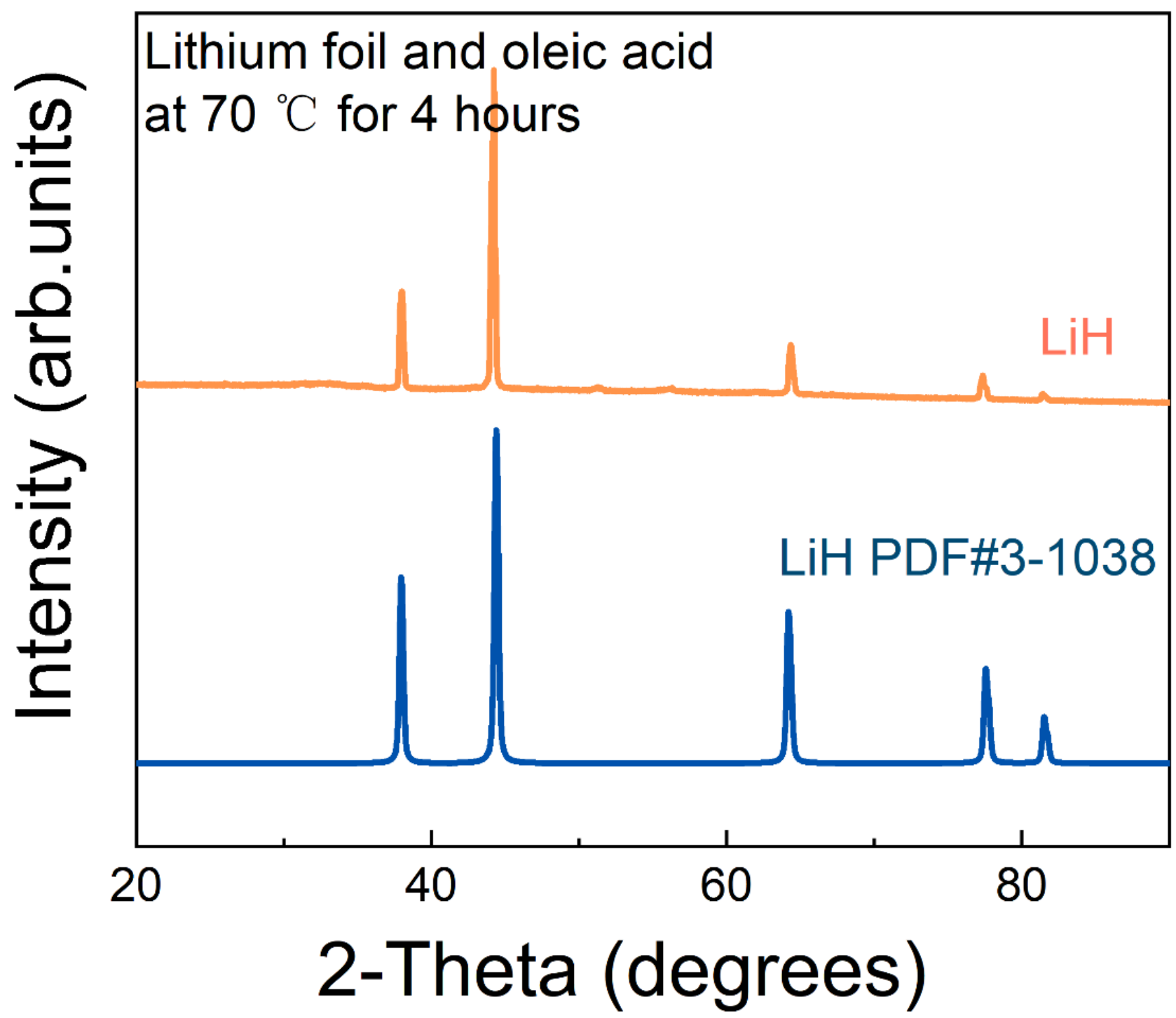

**Fig. S25. XRD phase analysis under varying synthetic conditions.** Treated Li foil reacted with oleic acid at 70°C for 4 h; after washing and drying, XRD under inert atmosphere matches LiH standard pattern.

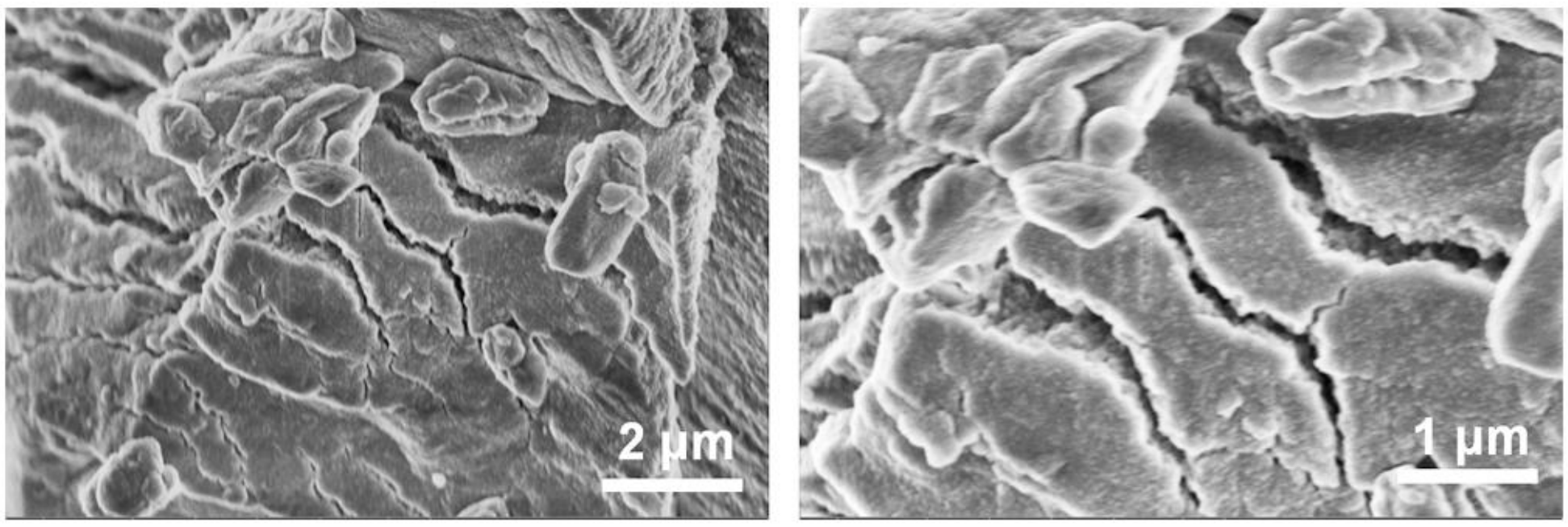

**Fig. S26. Representative SEM images of HTC-LiH.**

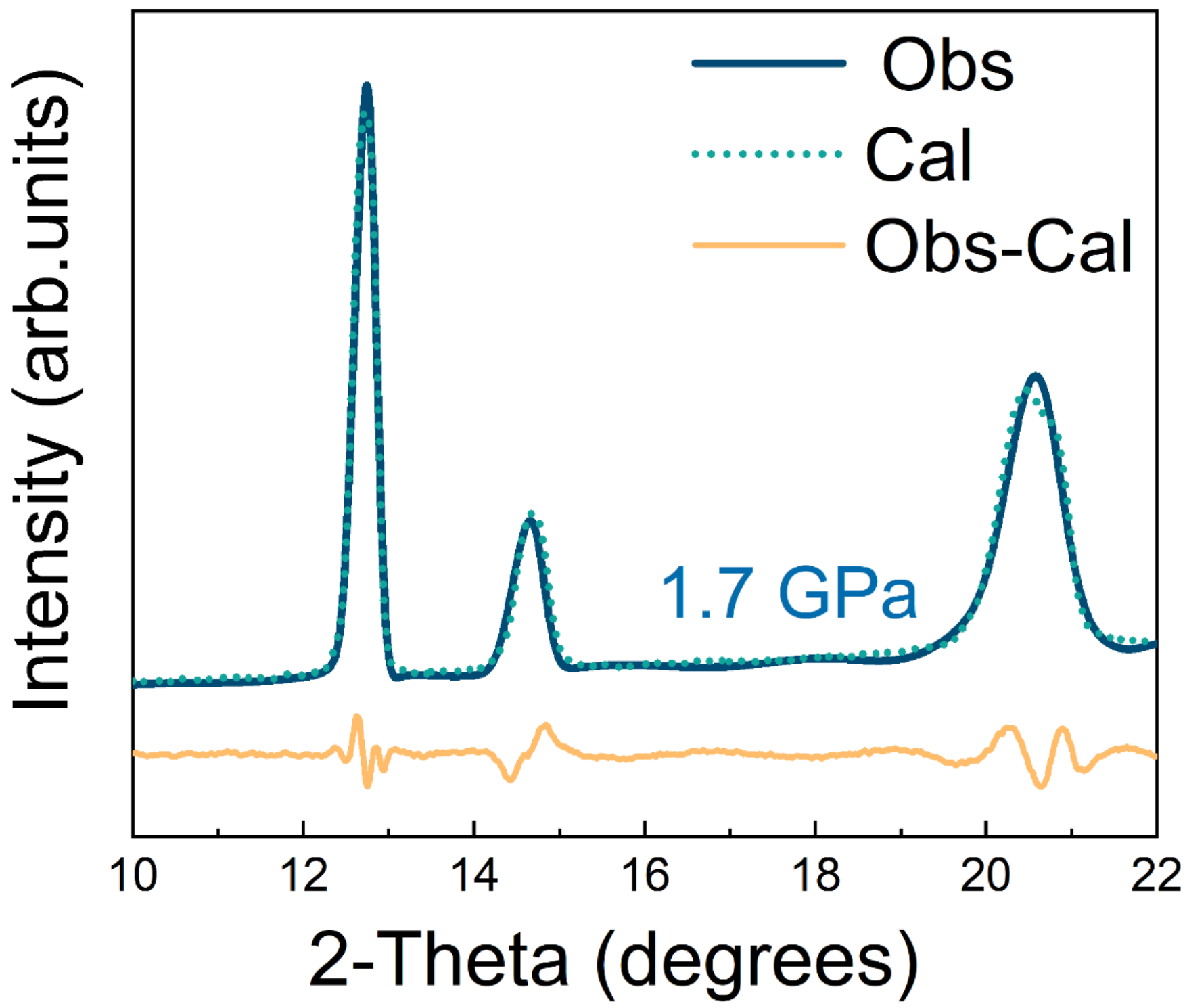

**Fig. S27. XRD refinement pattern of $LaH_2$ at 1.7 GPa.** DAC synthesis of $LaH_2$ from La and $H_2$ at 1.7 GPa demonstrates that its thermodynamic pressure $\Delta P_{th} \leq 1.7$ GPa.

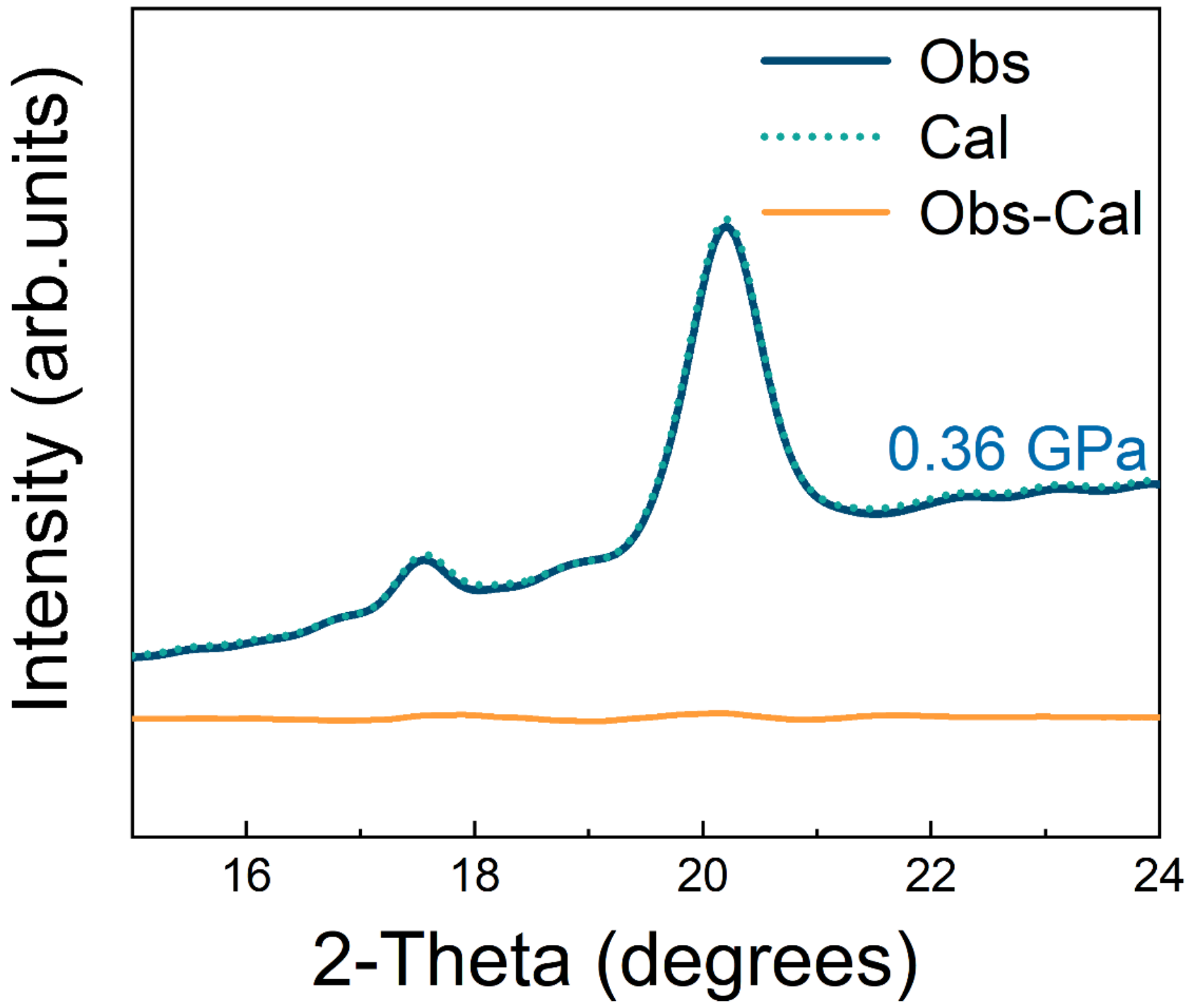

**Fig. S28. XRD refinement pattern of LiH at 0.36 GPa.** DAC synthesis of LiH from Li and $H_2$ at 0.36 GPa demonstrates that its thermodynamic pressure $\Delta P_{th} \leq 1.7$ GPa.

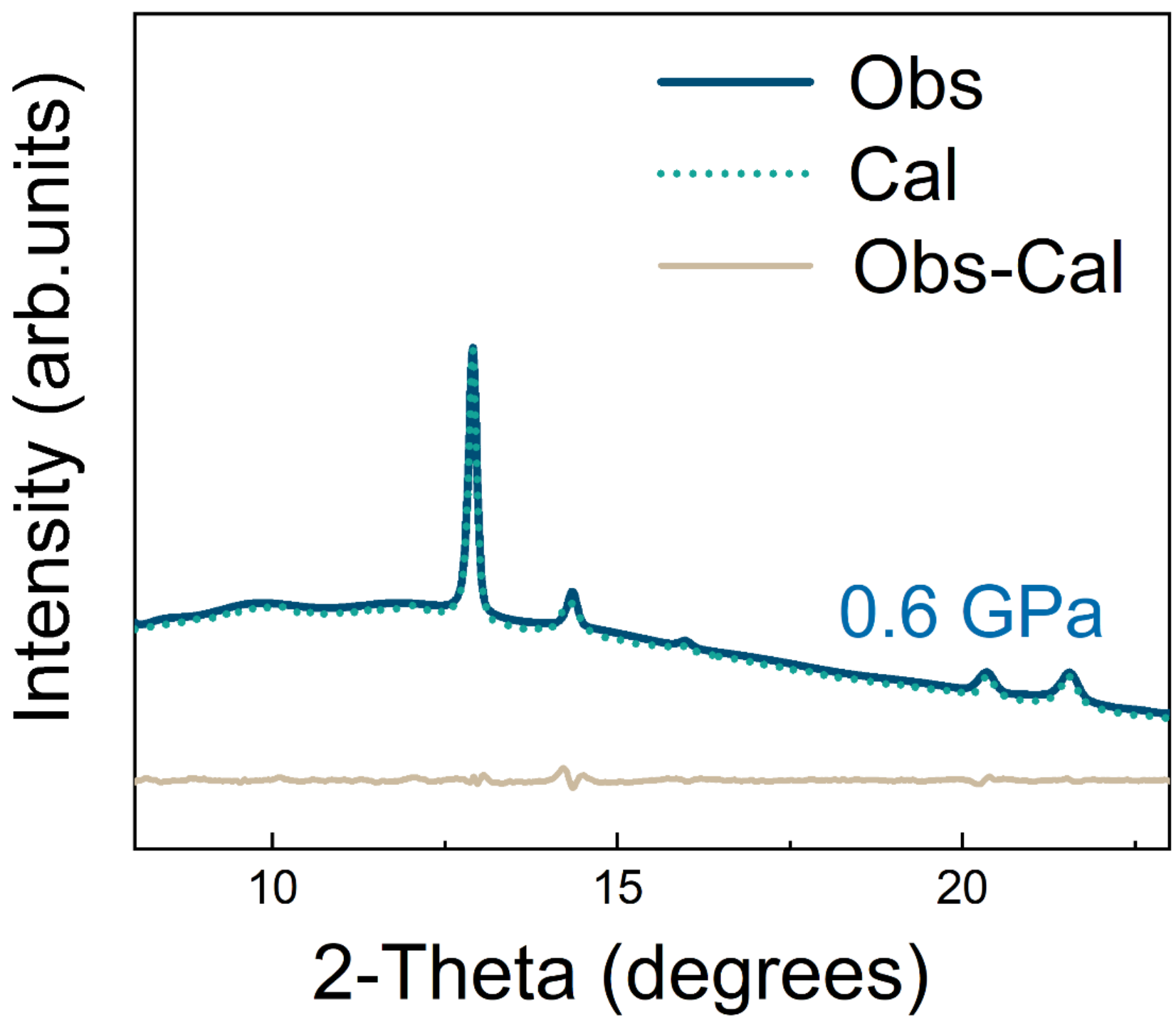

**Fig. S29. XRD refinement pattern of $ZrH_2$ at 0.6 GPa.** DAC synthesis of ε-$ZrH_2$ from Zr and $H_2$ at 0.6 GPa demonstrates that its thermodynamic pressure $\Delta P_{th} \leq 0.6$ GPa.

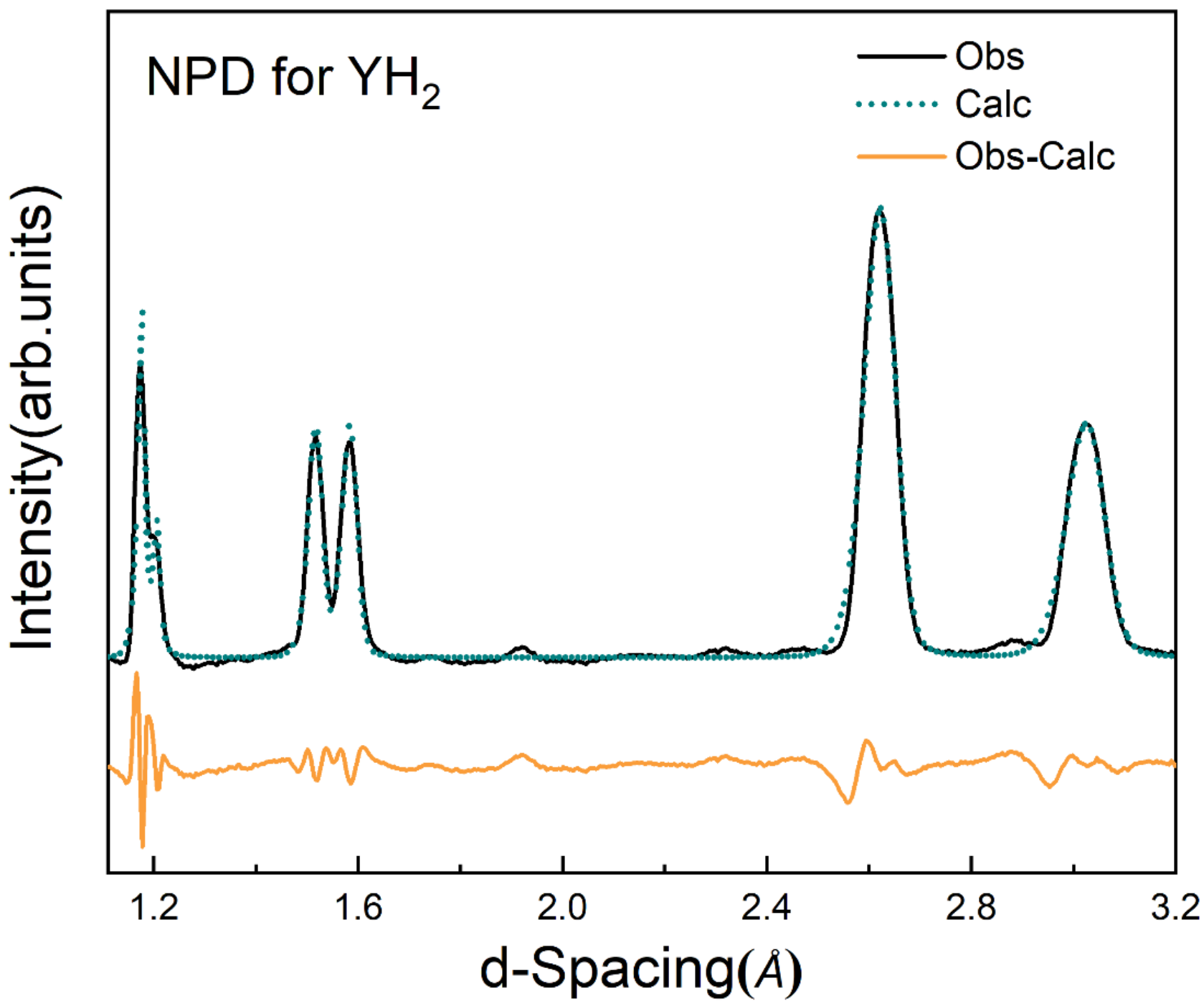

**Fig. S30. NPD refinement pattern of $YH_2$.** Neutron powder diffraction identifies $YH_2$ from Y foil reaction with oleic acid-$d_{33}$ ($C_{17}D_{33}COOH$), confirming deuterium source as the carboxy group.

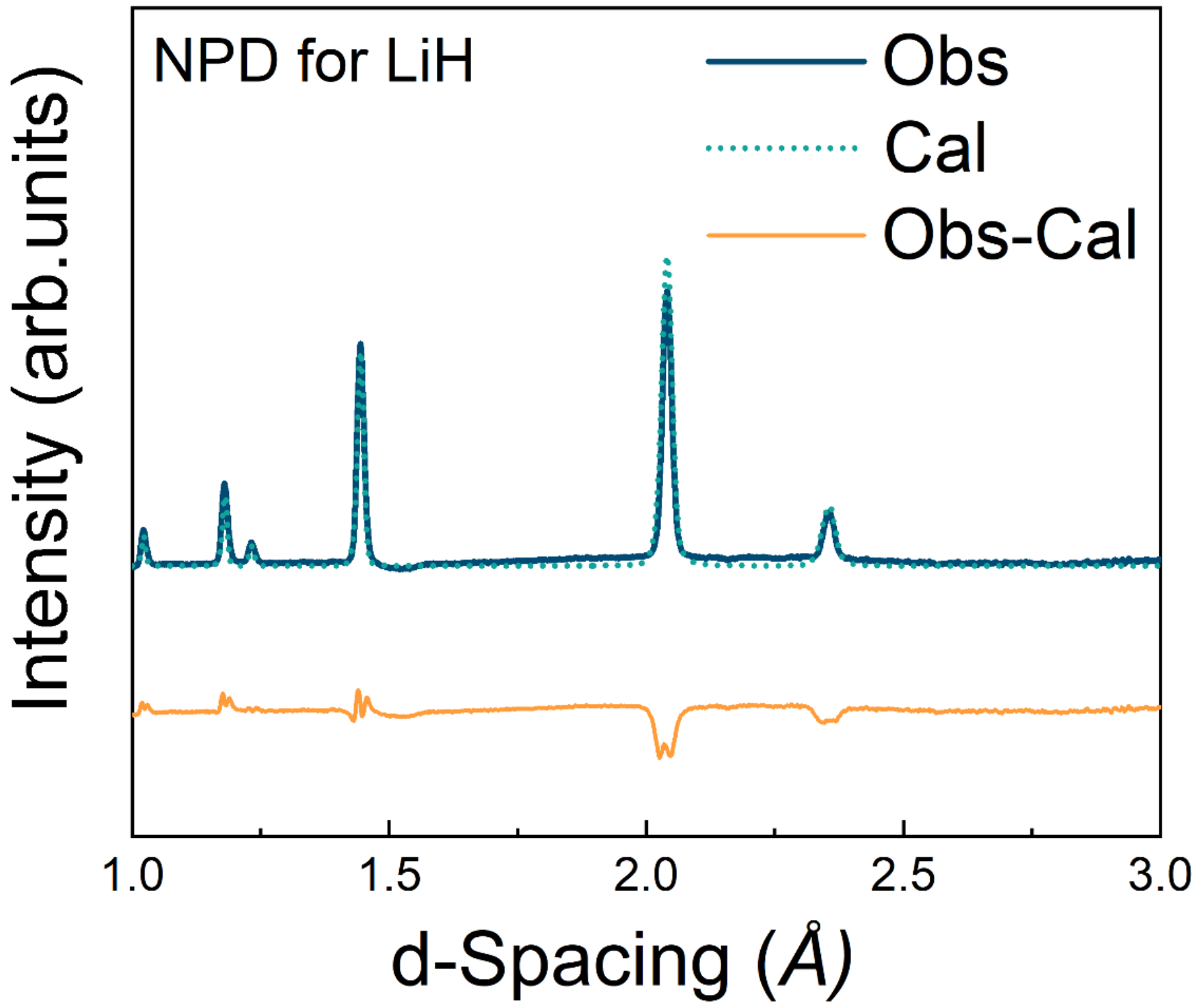

**Fig. S31. NPD refinement pattern of LiH.** Neutron powder diffraction identifies LiH from Li foil reaction with oleic acid-$d_{33}$ ($C_{17}D_{33}COOH$), confirming deuterium source as the carboxy group.

**Supplementary Text**

**Acid-Mediated Mechanisms in Hydride Synthesis**

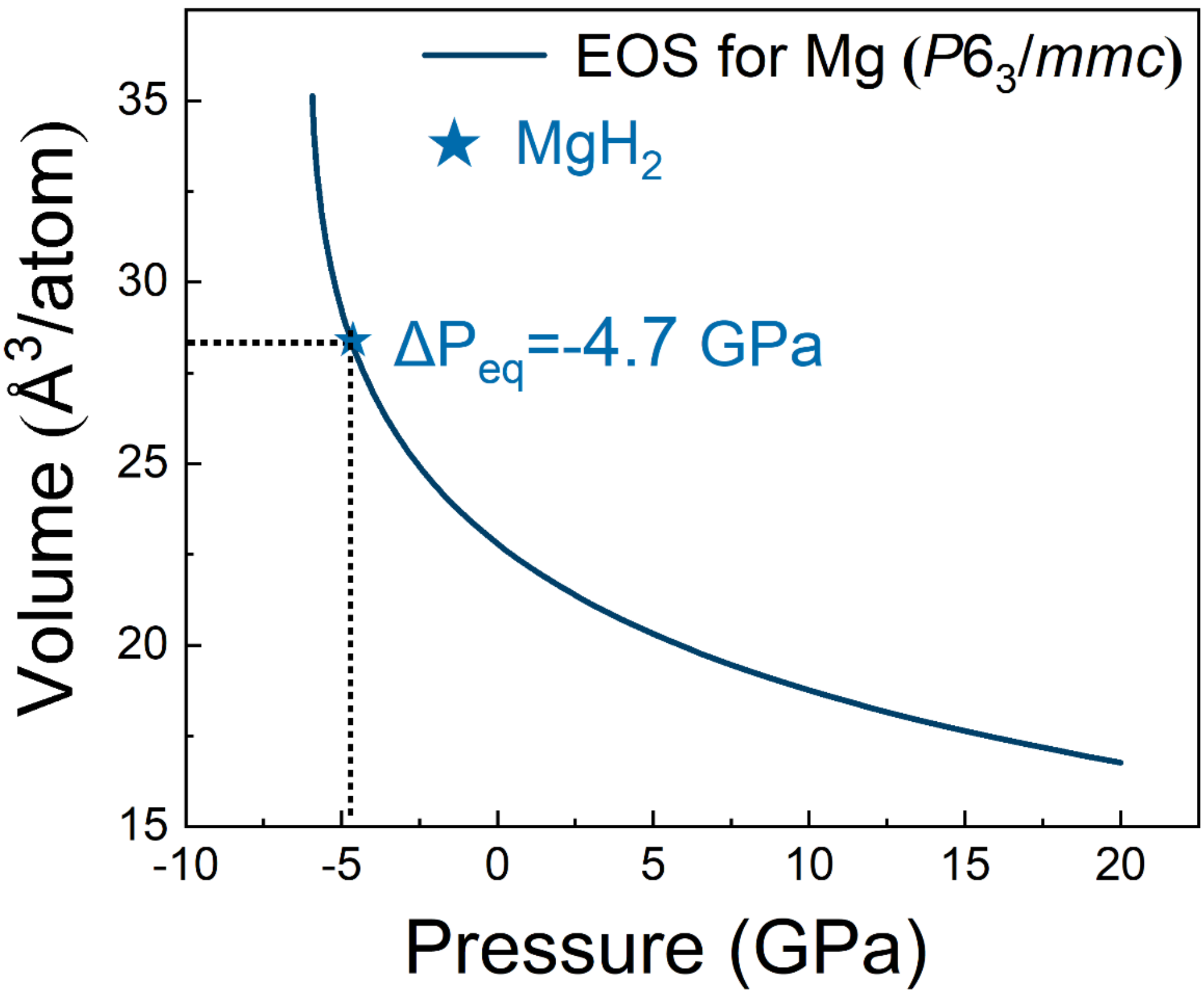

**Fig. S30. Eos for Mg (*P*6$_3$*/mmc*) and $\Delta P_{eq}$ of $MgH_2$.**

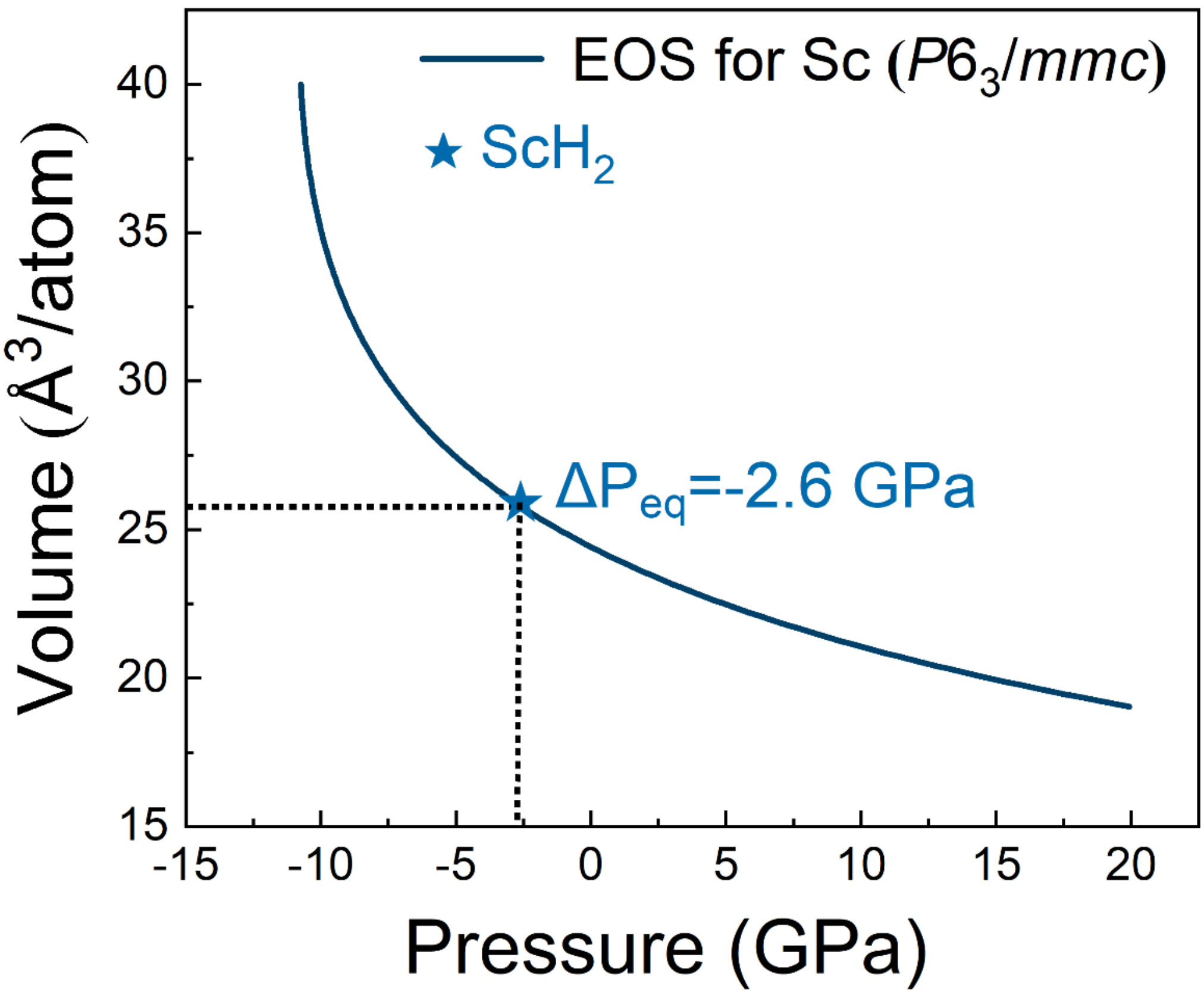

**Fig. S31. Eos for Sc (*P*6₃/*mmc*) and ΔP$_{eq}$ of ScH$_2$.**

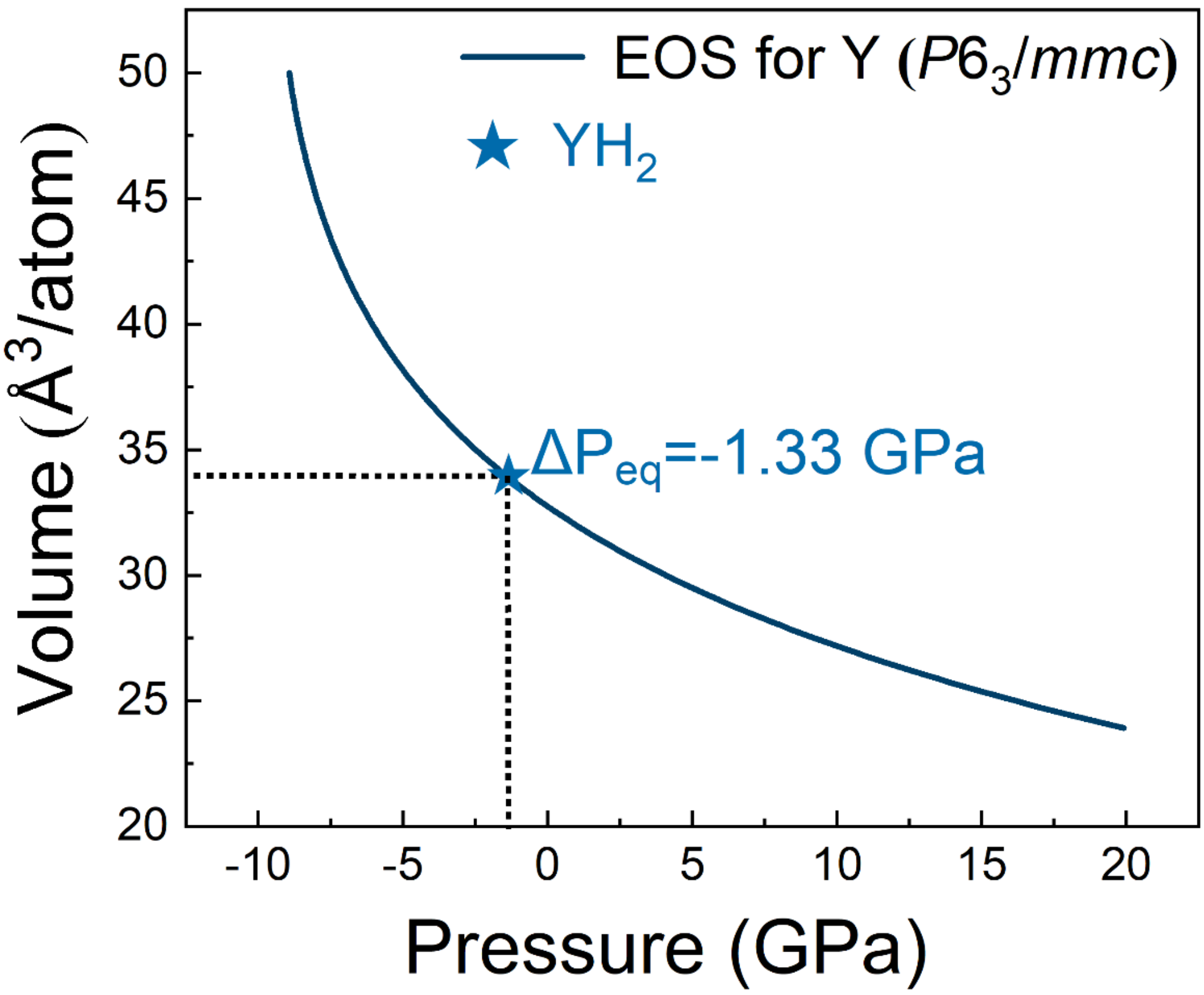

**Fig. S32. Eos for Y (*P*$6_3$*/mmc*) and $\Delta P_{eq}$ of $YH_2$.**

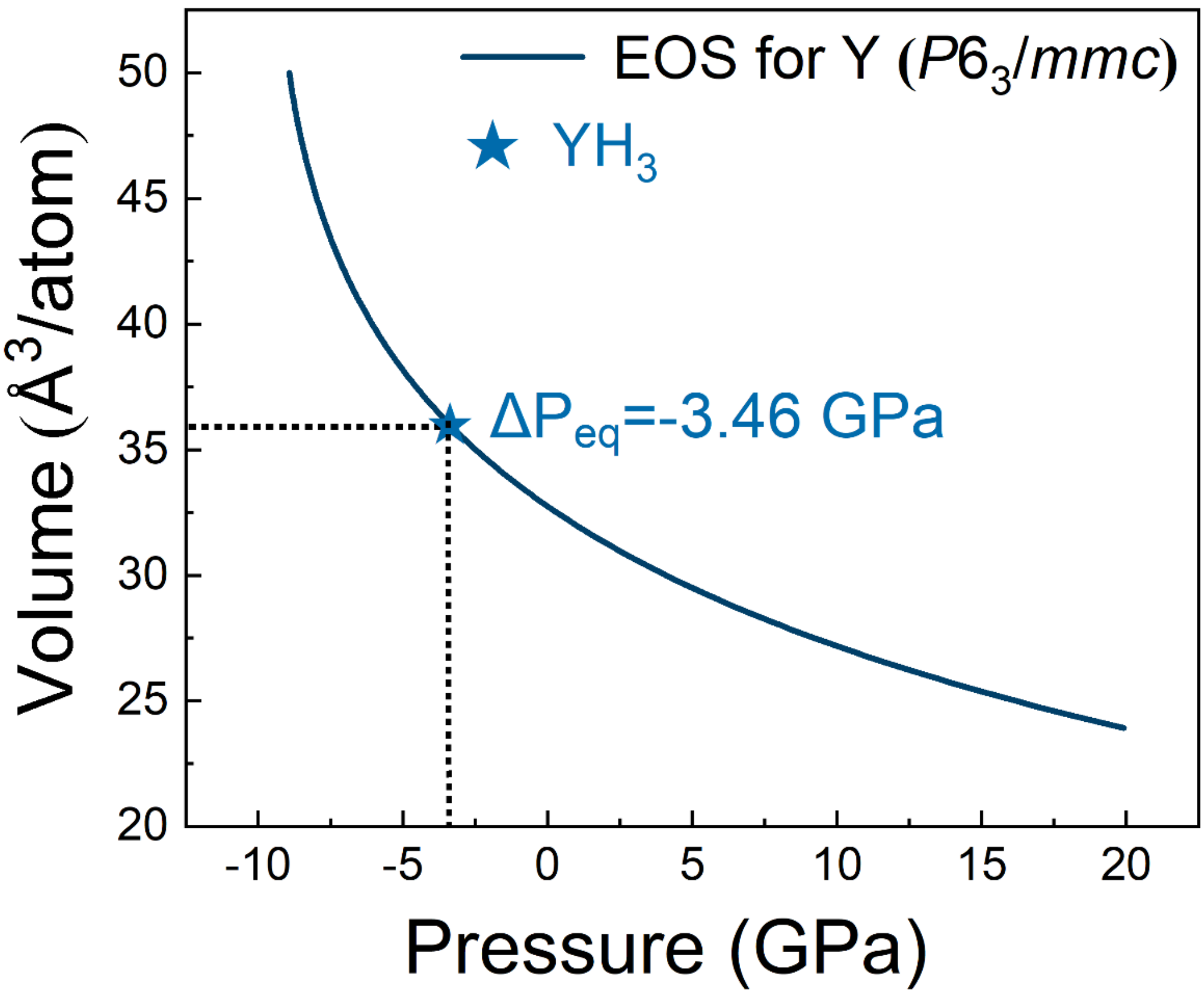

**Fig. S33. Eos for Y (*P*6$_3$*/mmc*) and Δ$P_{eq}$ of $YH_3$.**

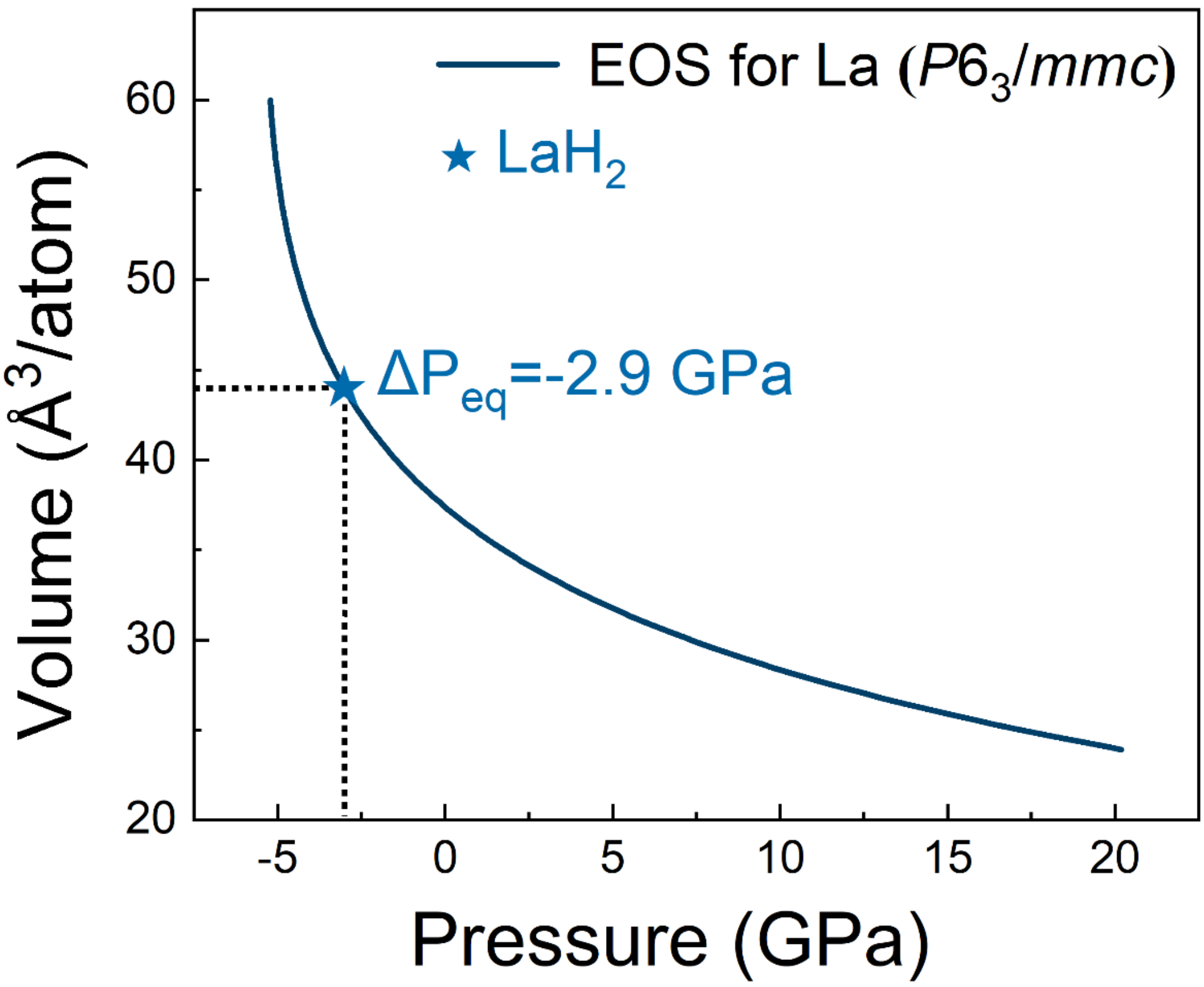

**Fig. S34. Eos for La (*P*6$_3$*/mmc*) and $\Delta P_{eq}$ of $LaH_2$.**

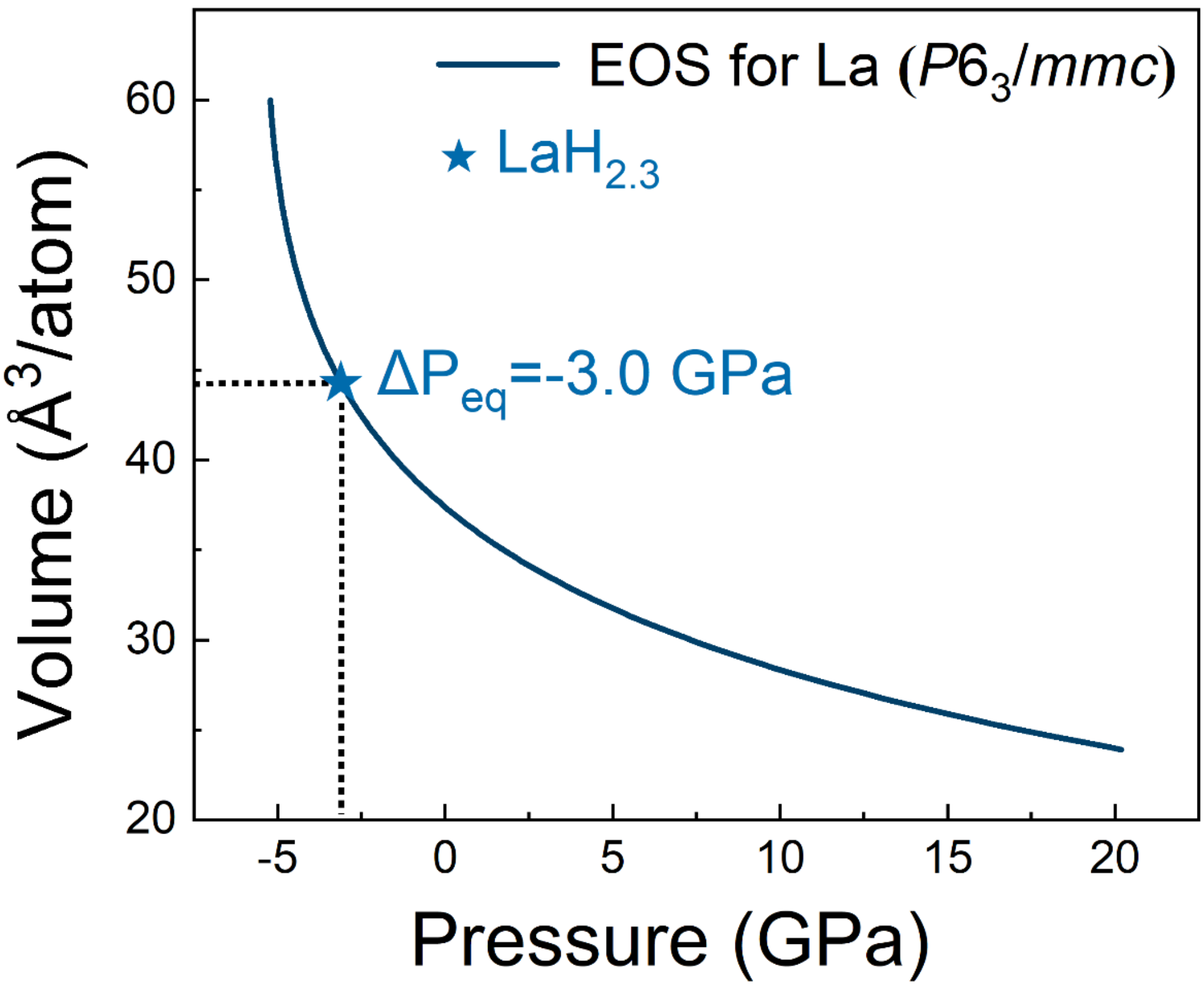

**Fig. S35. Eos for La (*P*6$_3$*/mmc*) and ΔP$_{eq}$ of LaH$_{2.3}$.**

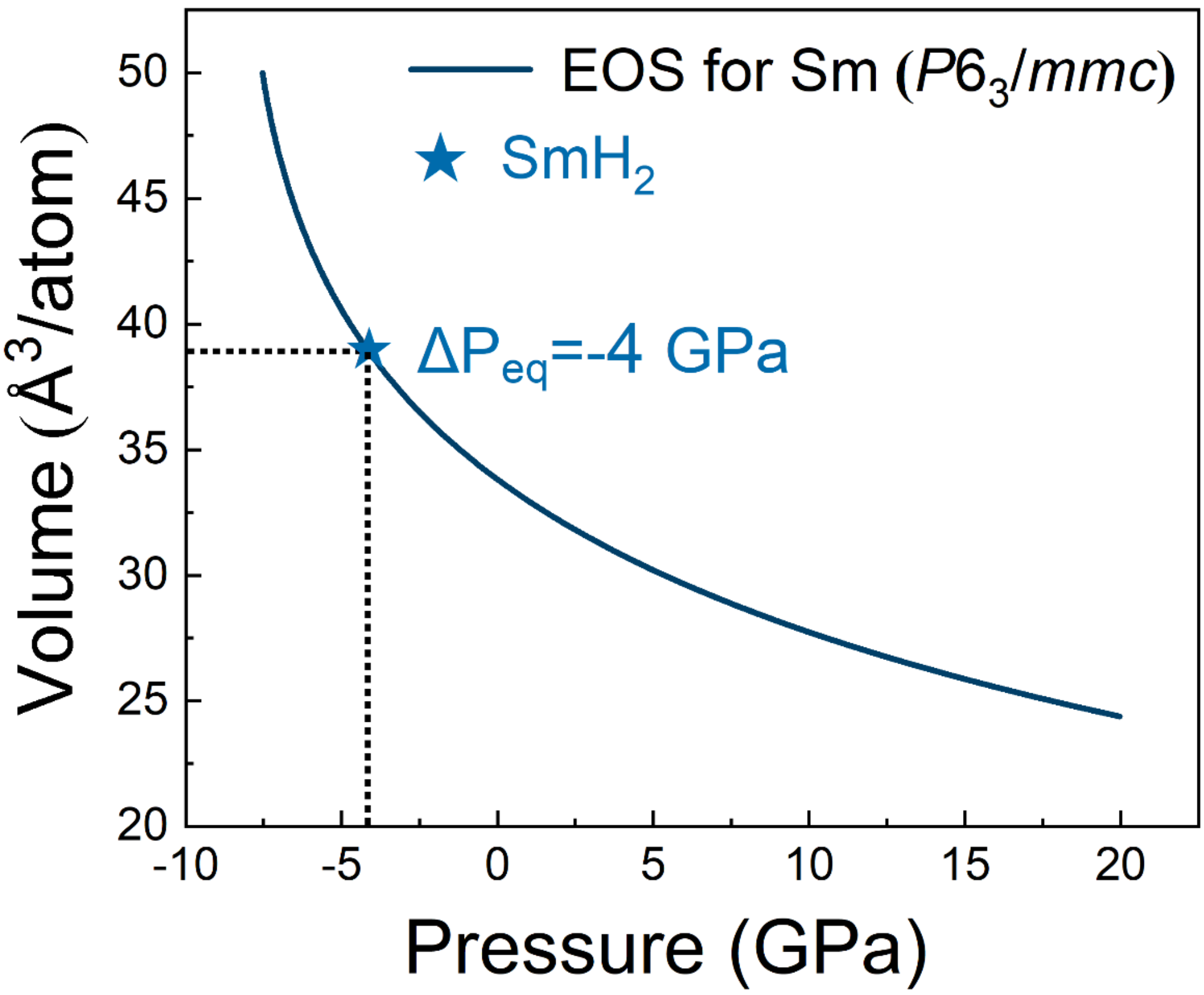

**Fig. S36. Eos for Sm (*P*6$_3$*/mmc*) and $\Delta P_{eq}$ of $SmH_2$.**

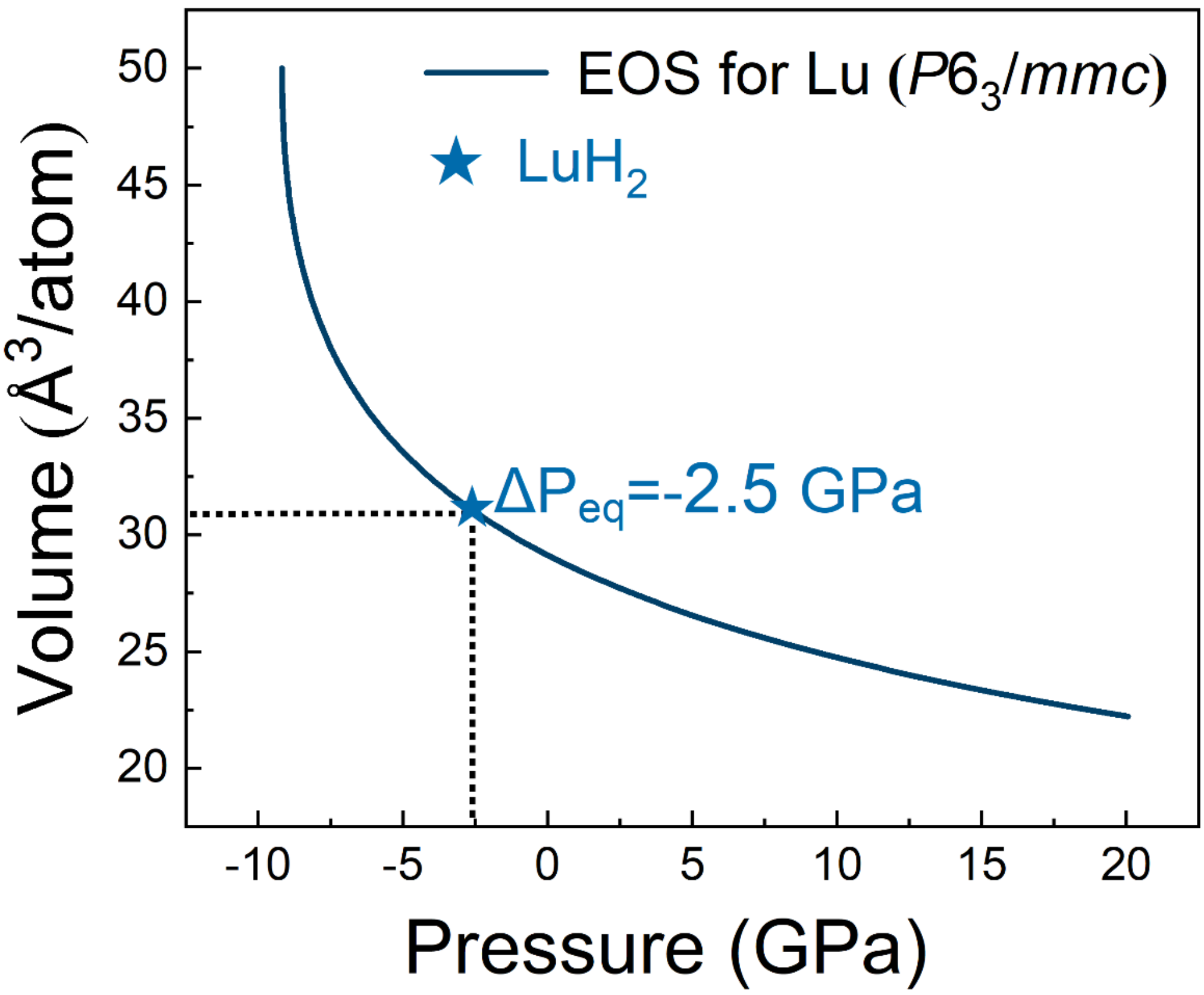

**Fig. S37. Eos for Lu (*P6₃/mmc*) and $\Delta P_{eq}$ of $LuH_2$.**

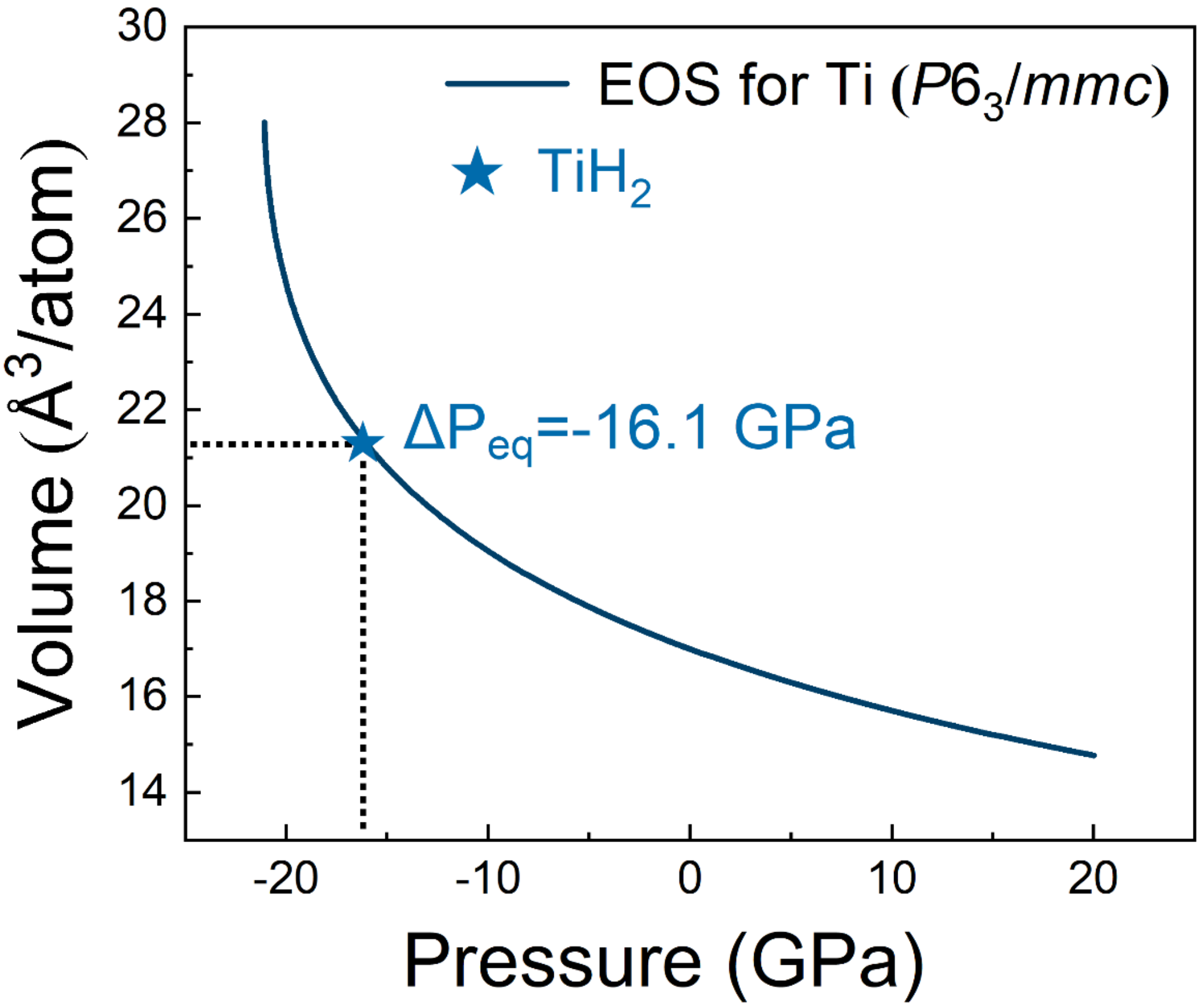

**Fig. S38. Eos for Ti (*P6₃/mmc*) and $\Delta P_{eq}$ of $TiH_2$.**

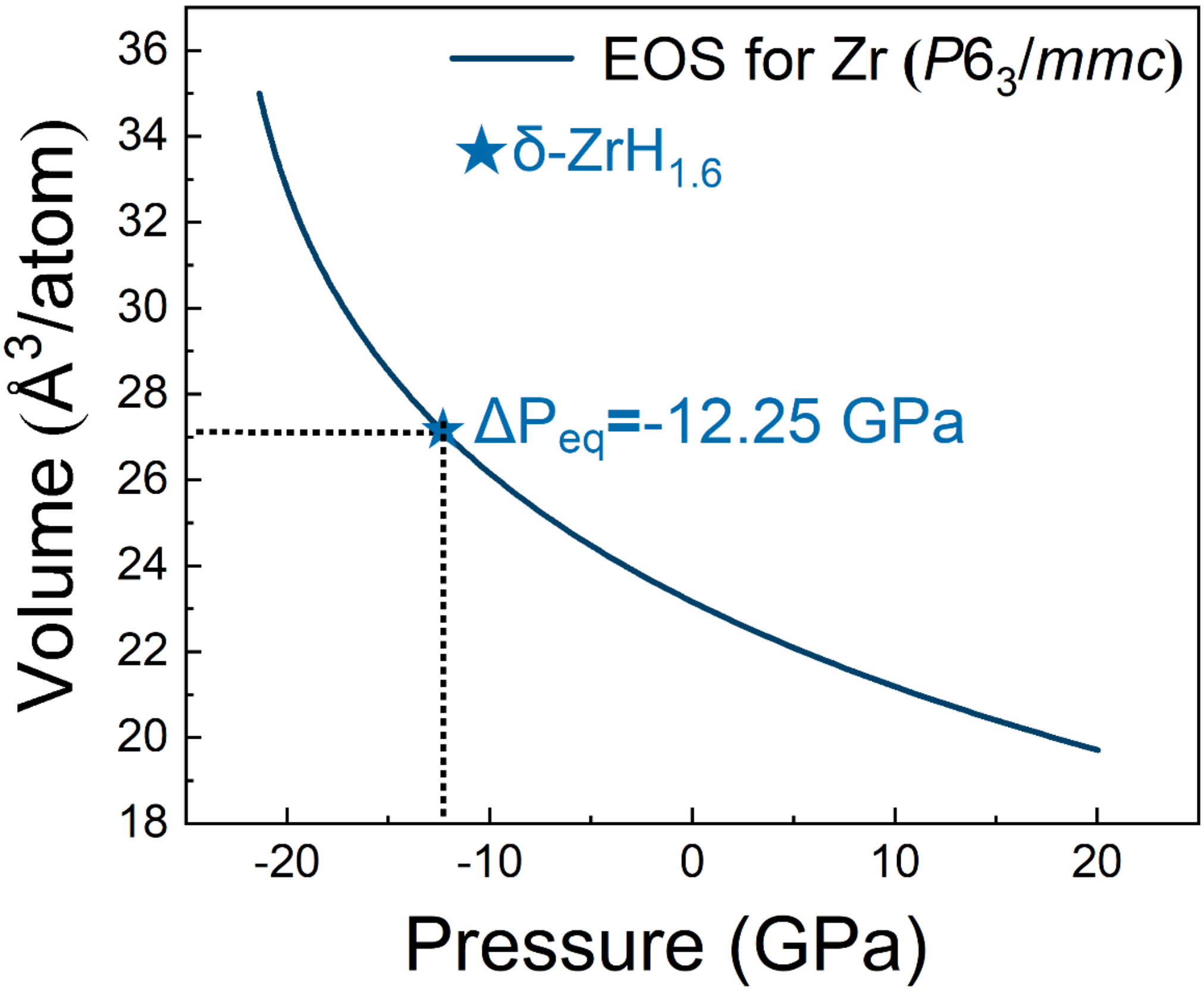

**Fig. S39. Eos for Zr (*$P6_3/mmc$*) and $\Delta P_{eq}$ of δ-$ZrH_{1.6}$.**

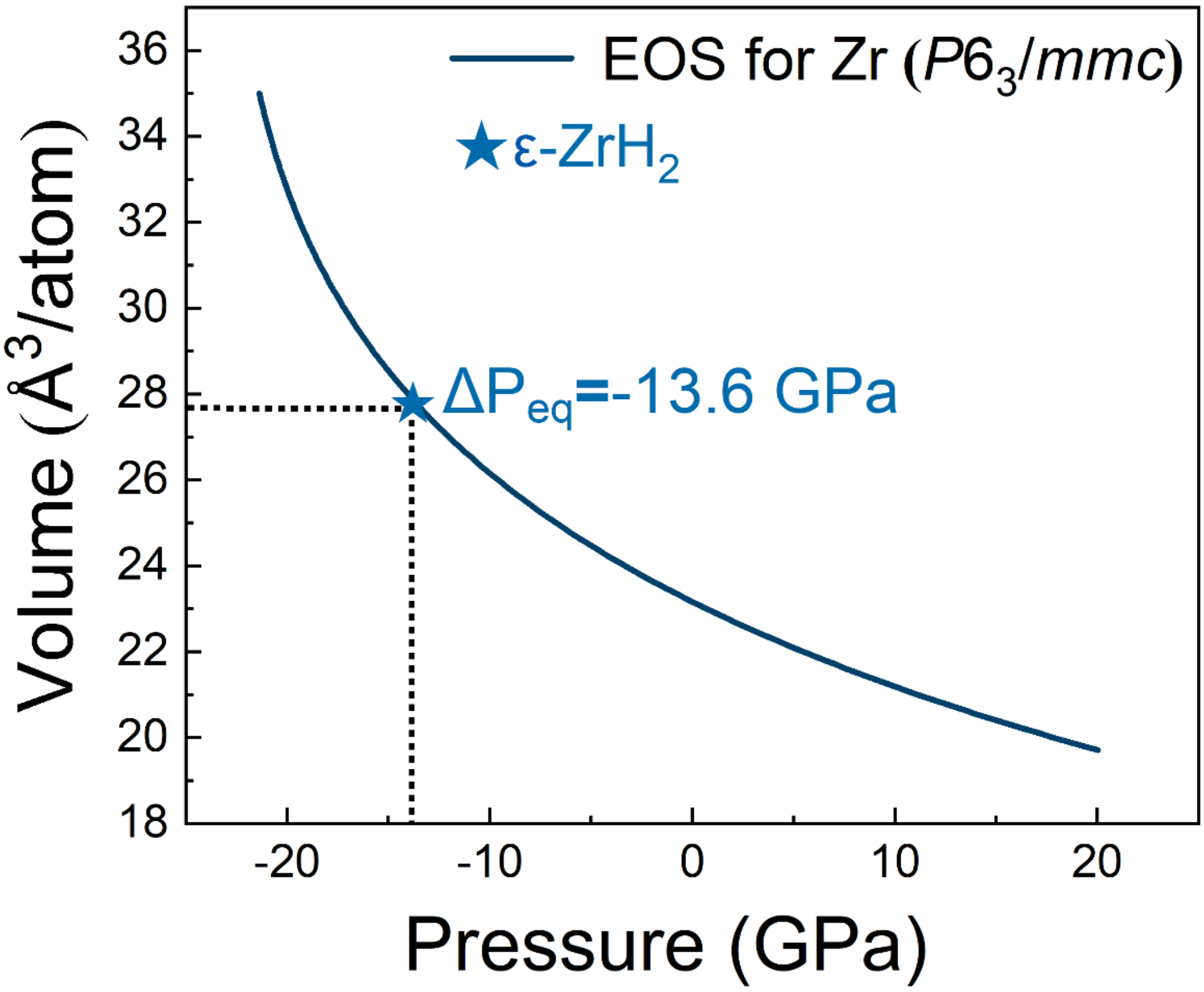

**Fig. S40. Eos for Zr (*P*6₃*/mmc*) and $\Delta P_{eq}$ of ε-$ZrH_2$.**

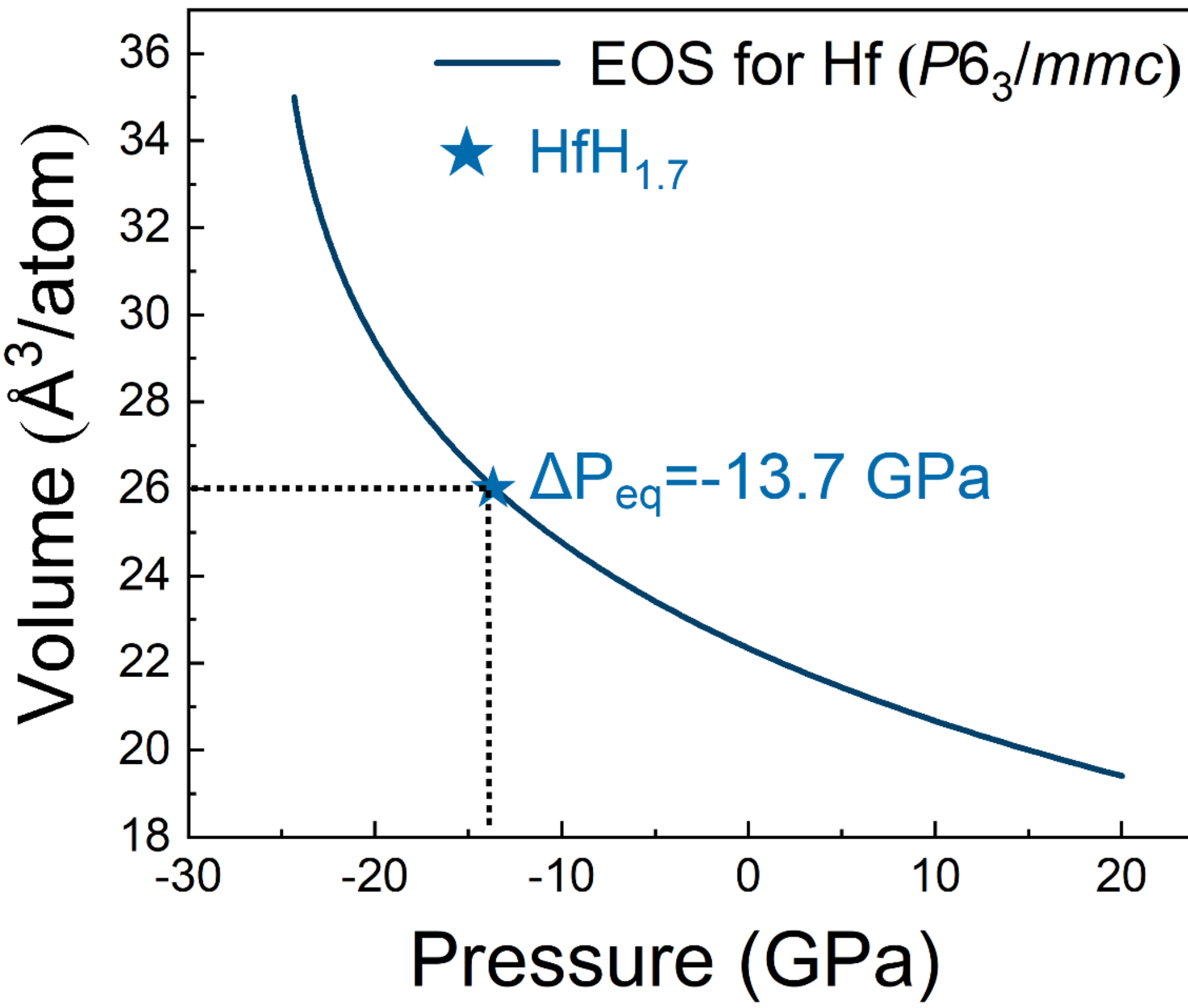

**Fig. S41. Eos for Hf (*$P6_3/mmc$*) and $\Delta P_{eq}$ of $HfH_{1.7}$.**

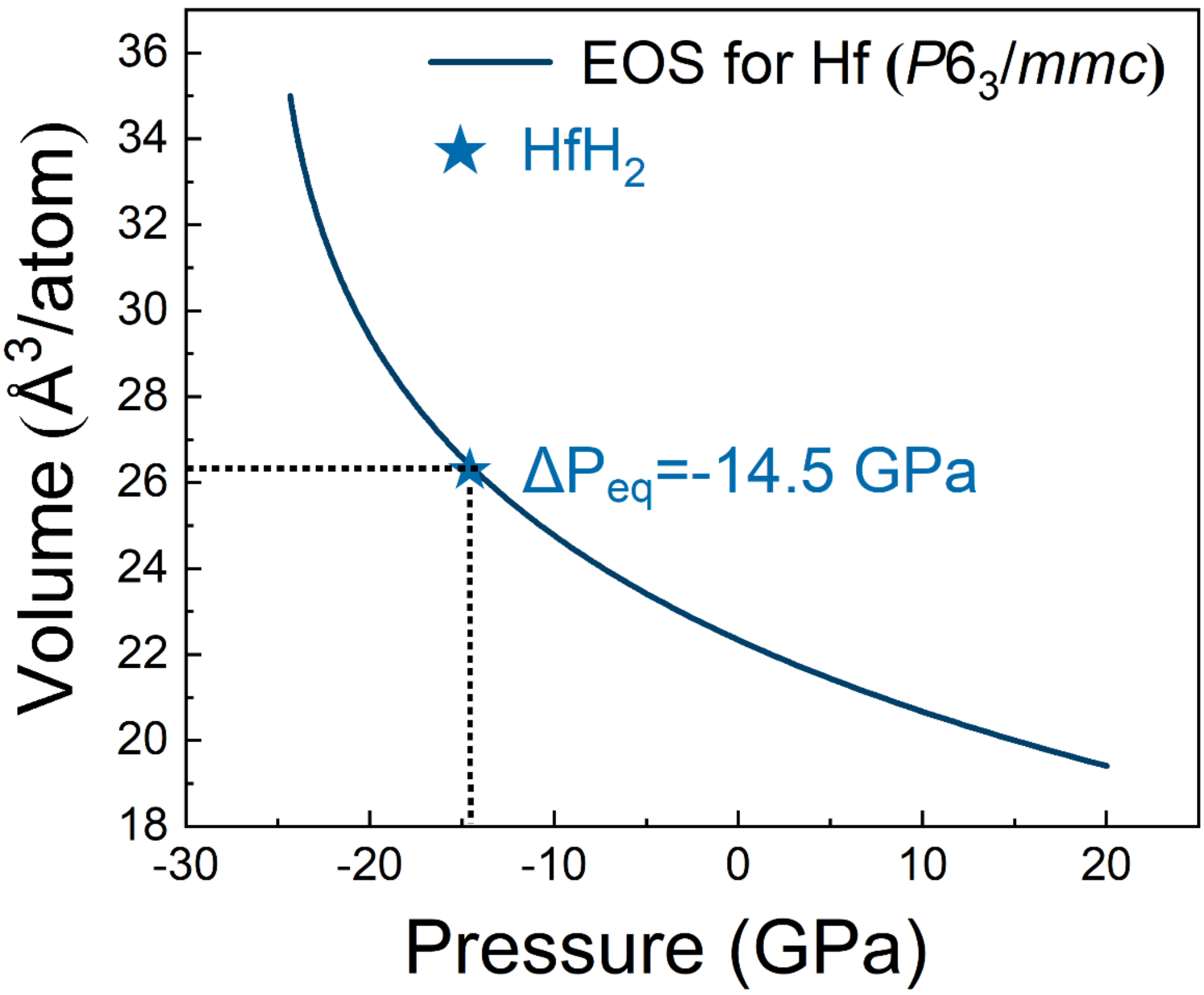

**Fig. S42. Eos for Hf (*P*6$_3$*/mmc*) and ΔP$_{eq}$ of $HfH_{1.7}$.**

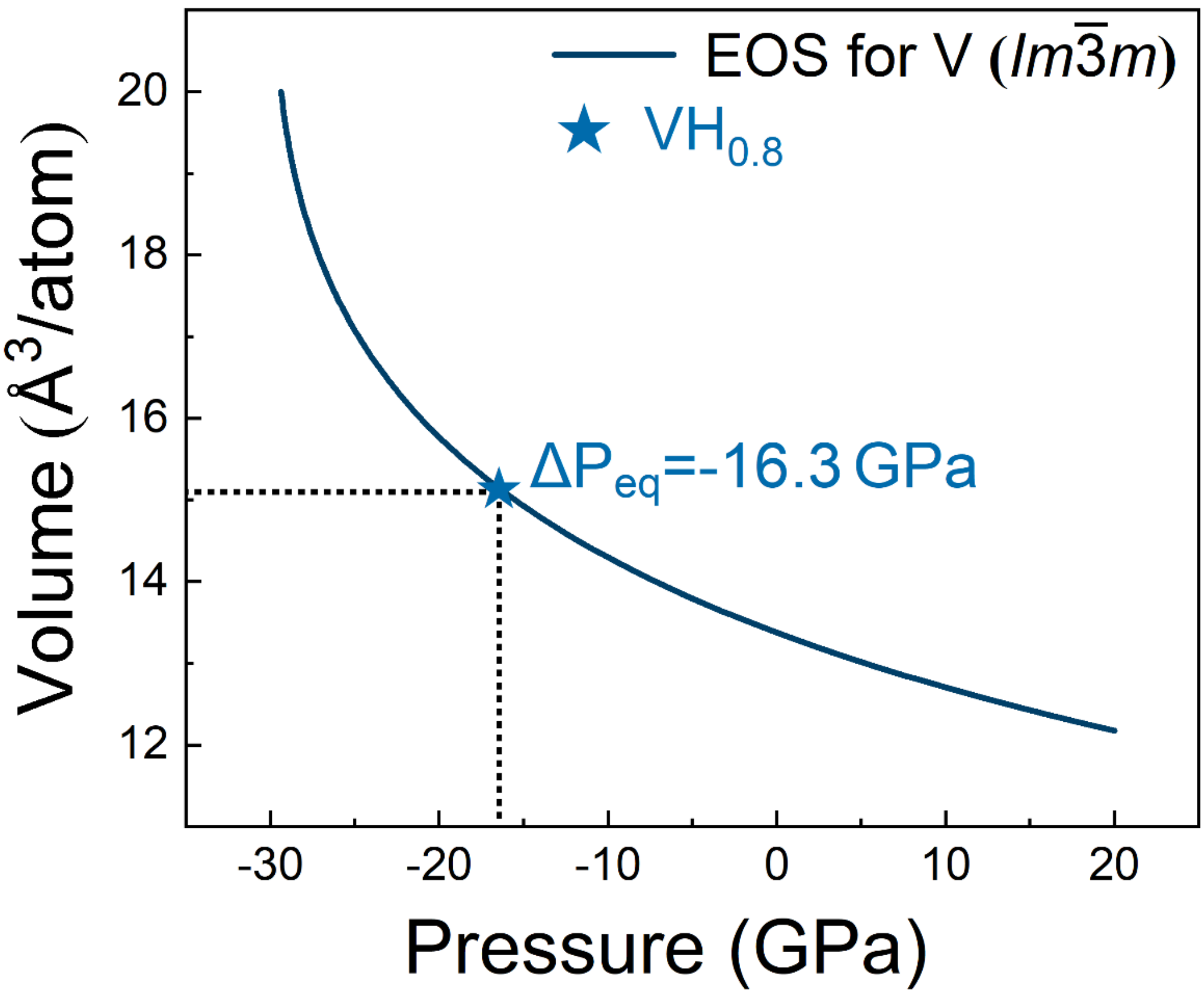

**Fig. S43. Eos for V ($Im\bar{3}m$) and $\Delta P_{eq}$ of $VH_{0.8}$.**

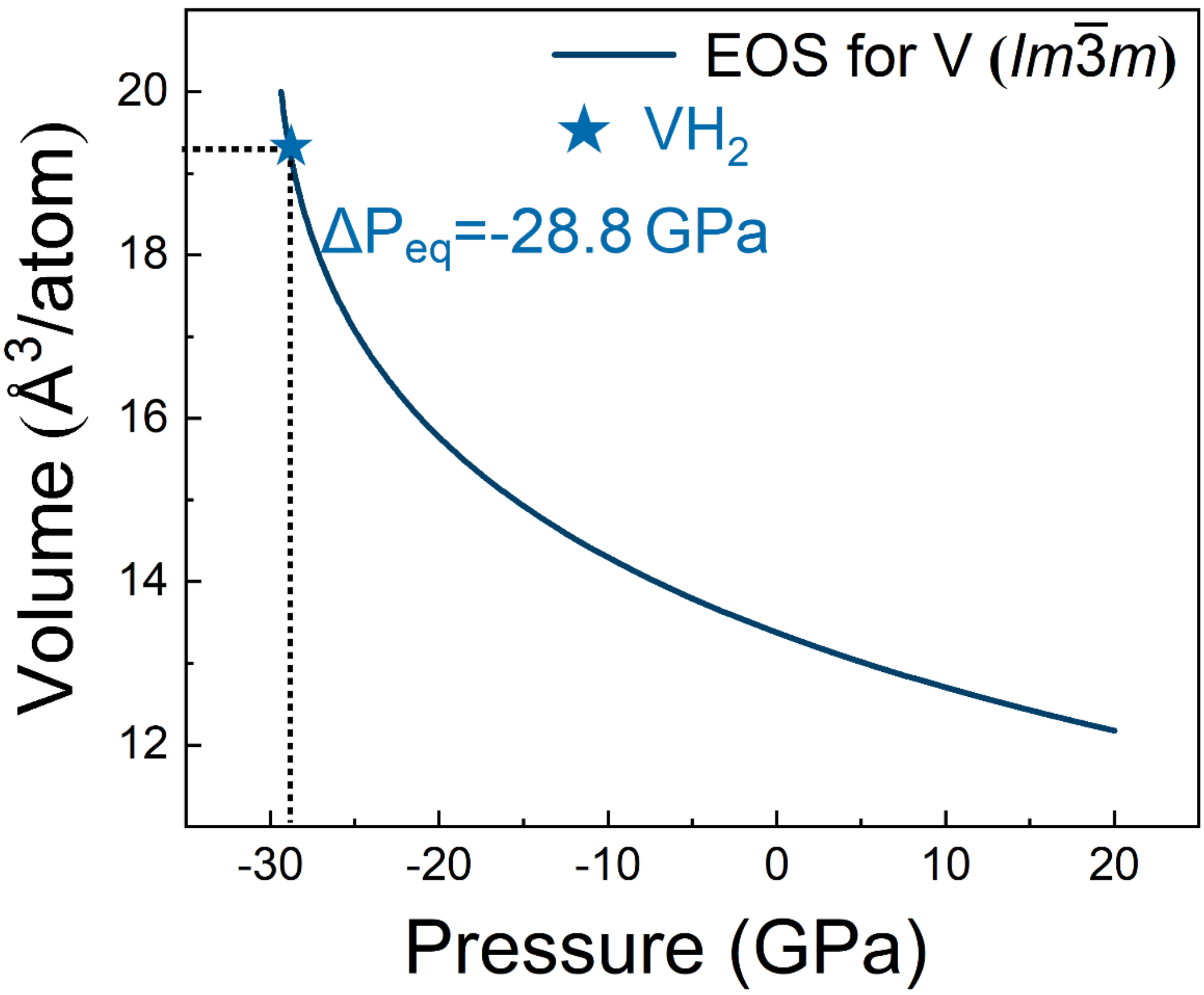

**Fig. S44. Eos for V ($Im\overline{3}m$) and $\Delta P_{eq}$ of $VH_2$.**

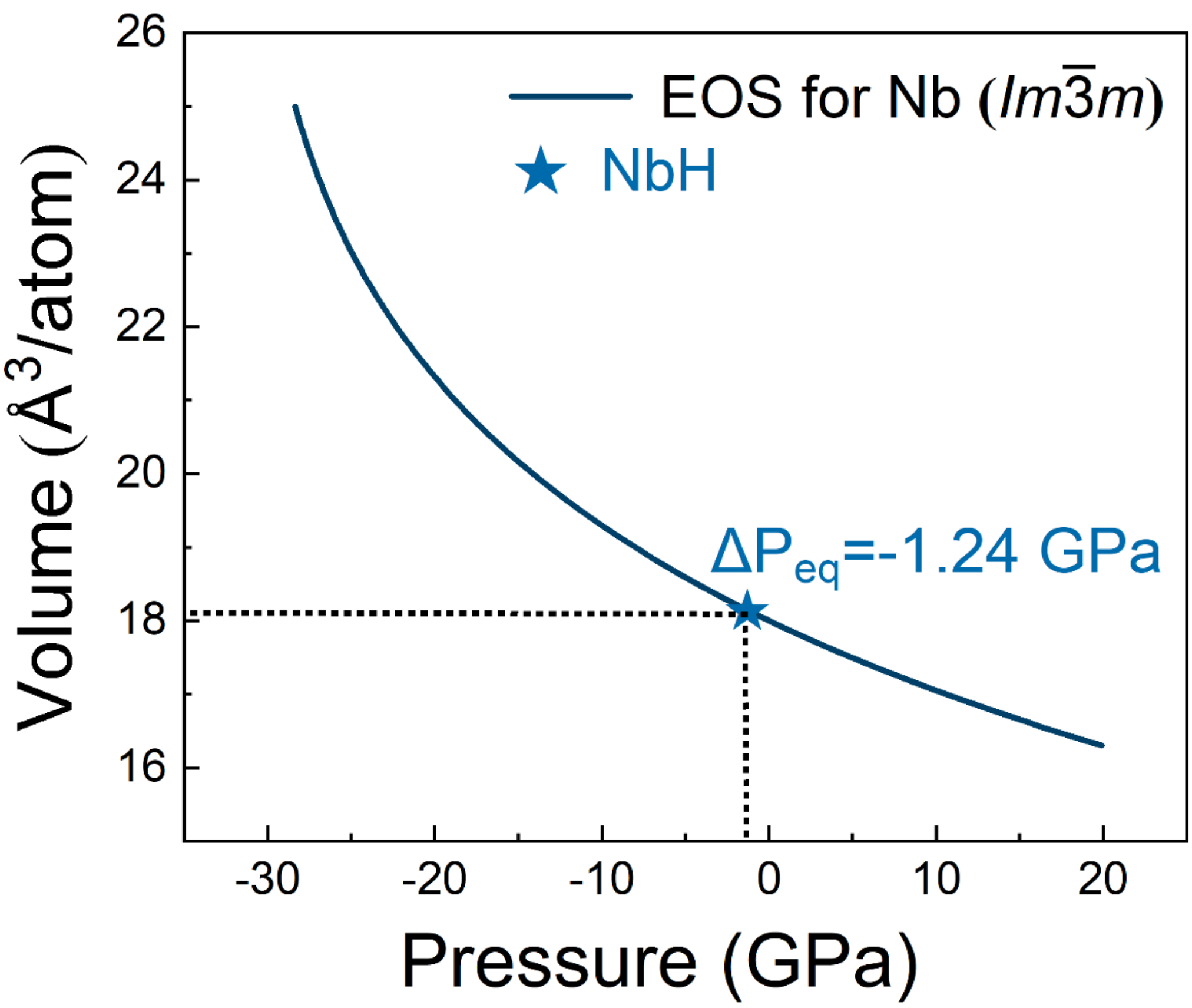

**Fig. S45. Eos for Nb (*Im*$\bar{3}$*m*) and $\Delta P_{eq}$ of NbH.**

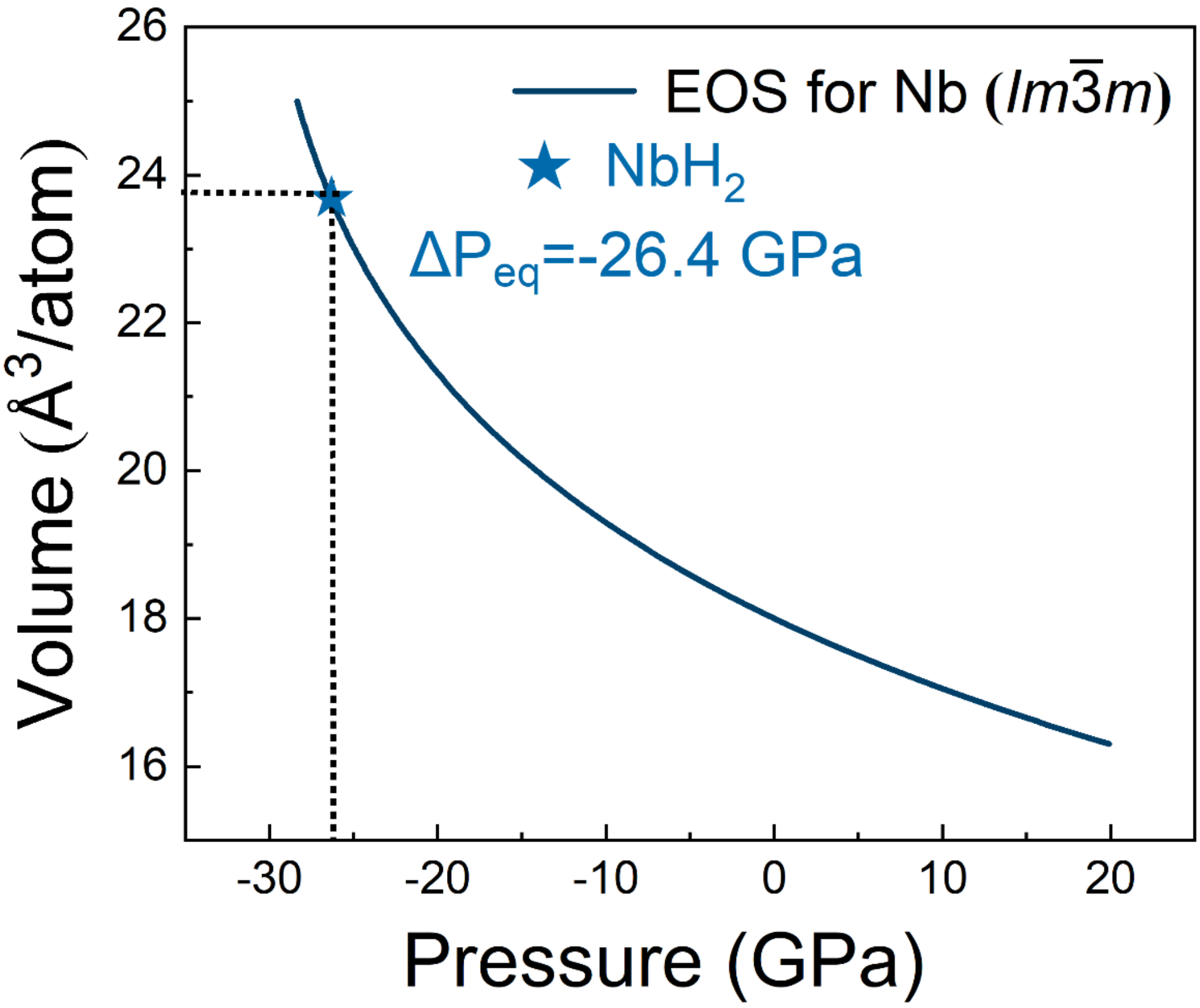

**Fig. S46. Eos for Nb ($Im\bar{3}m$) and $\Delta P_{eq}$ of $NbH_2$.**

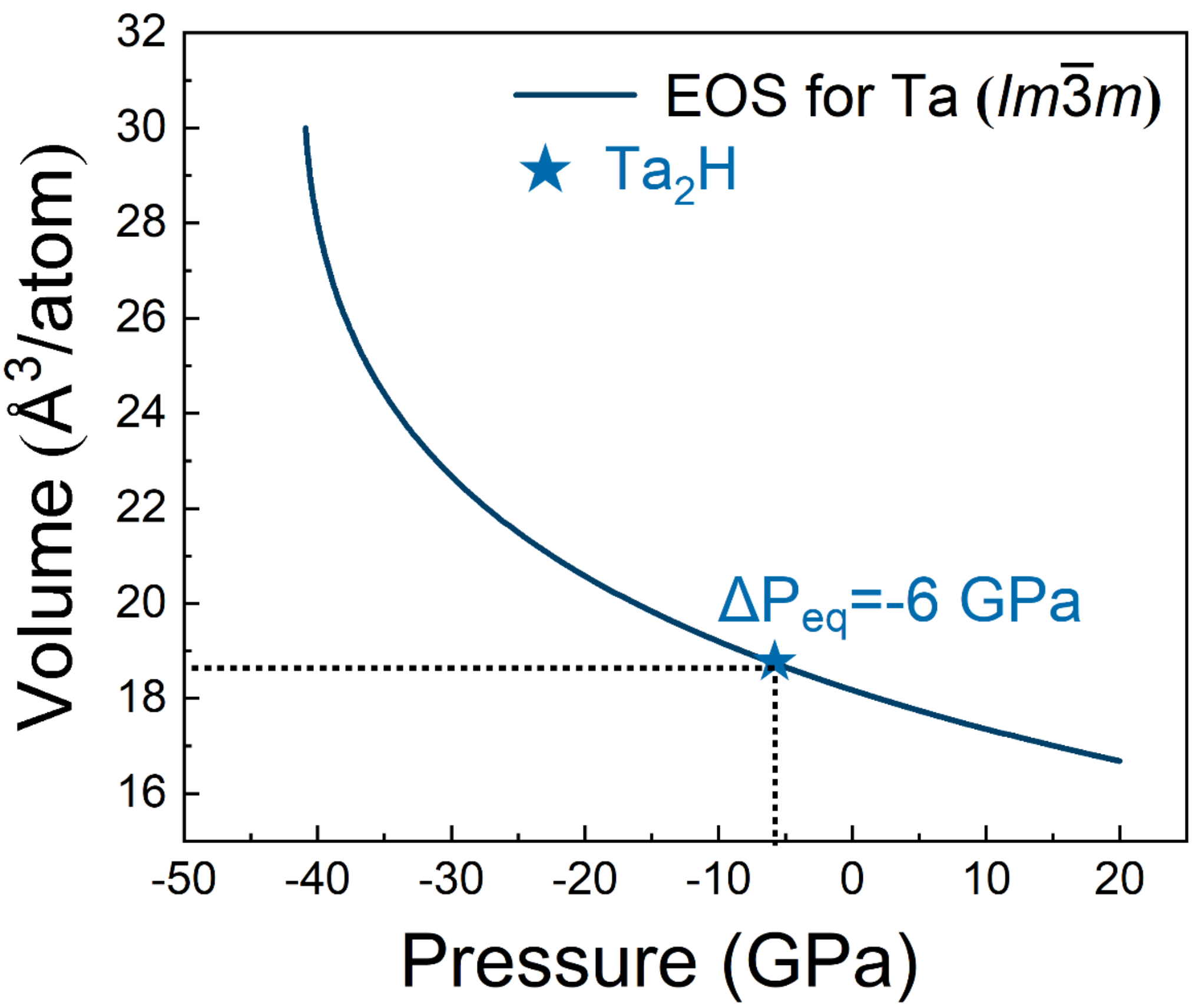

**Fig. S47. Eos for Ta (*Im*$\overline{3}$*m*) and $\Delta P_{eq}$ of $Ta_2H$.**

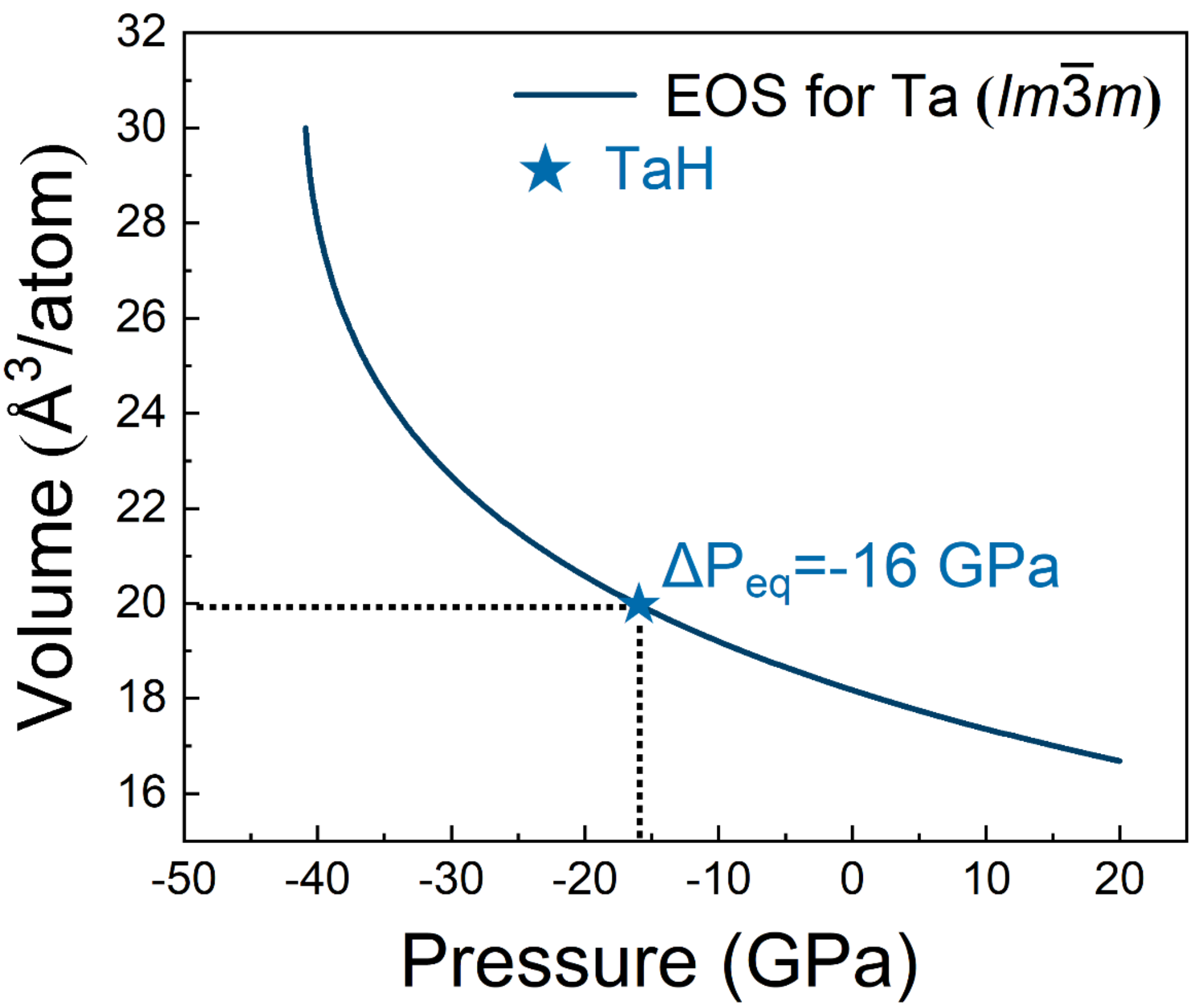

**Fig. S48. Eos for Ta ($Im\bar{3}m$) and $\Delta P_{eq}$ of TaH.**

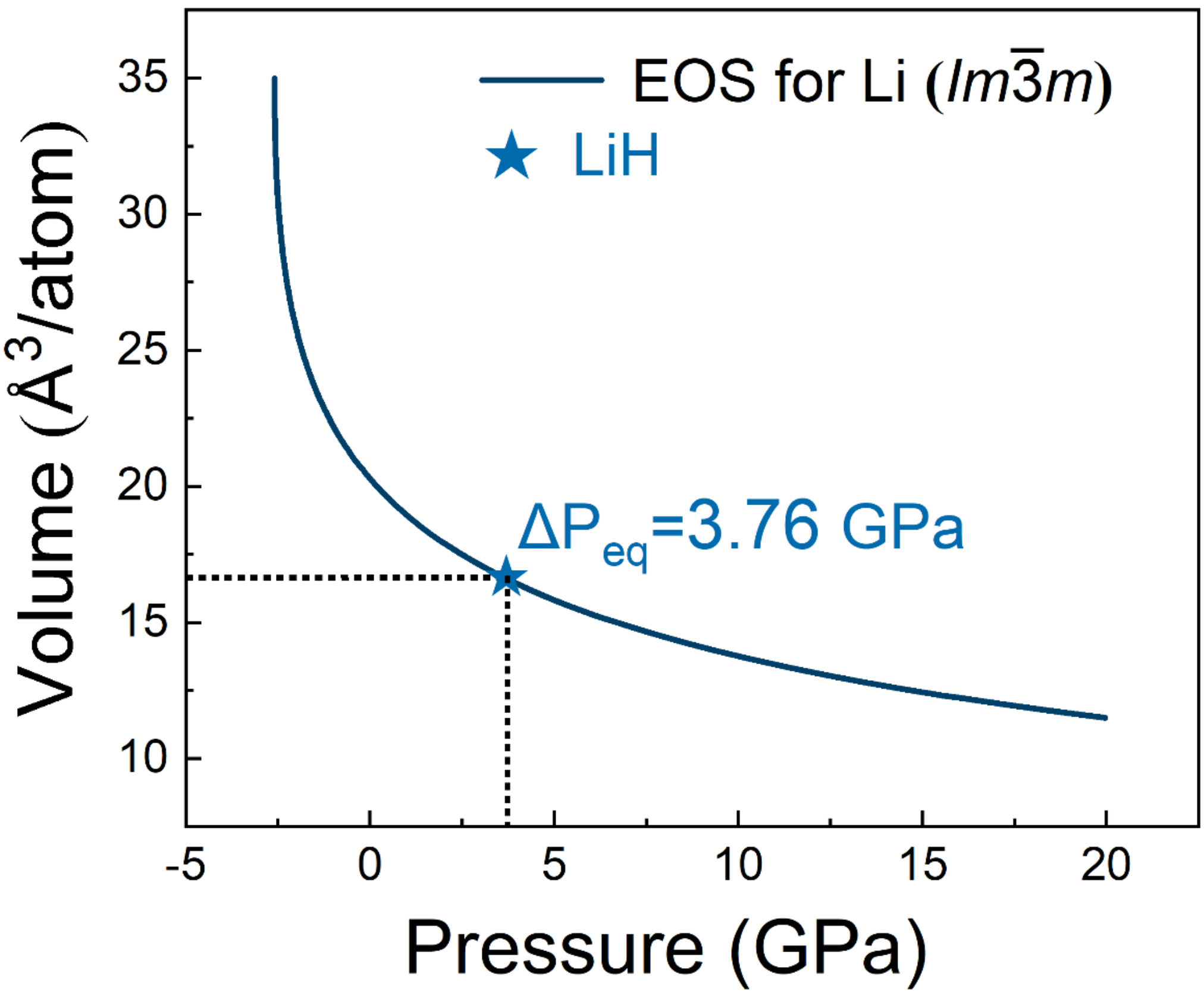

**Fig. S49. Eos for Li ($Im\overline{3}m$) and $\Delta P_{eq}$ of LiH.**

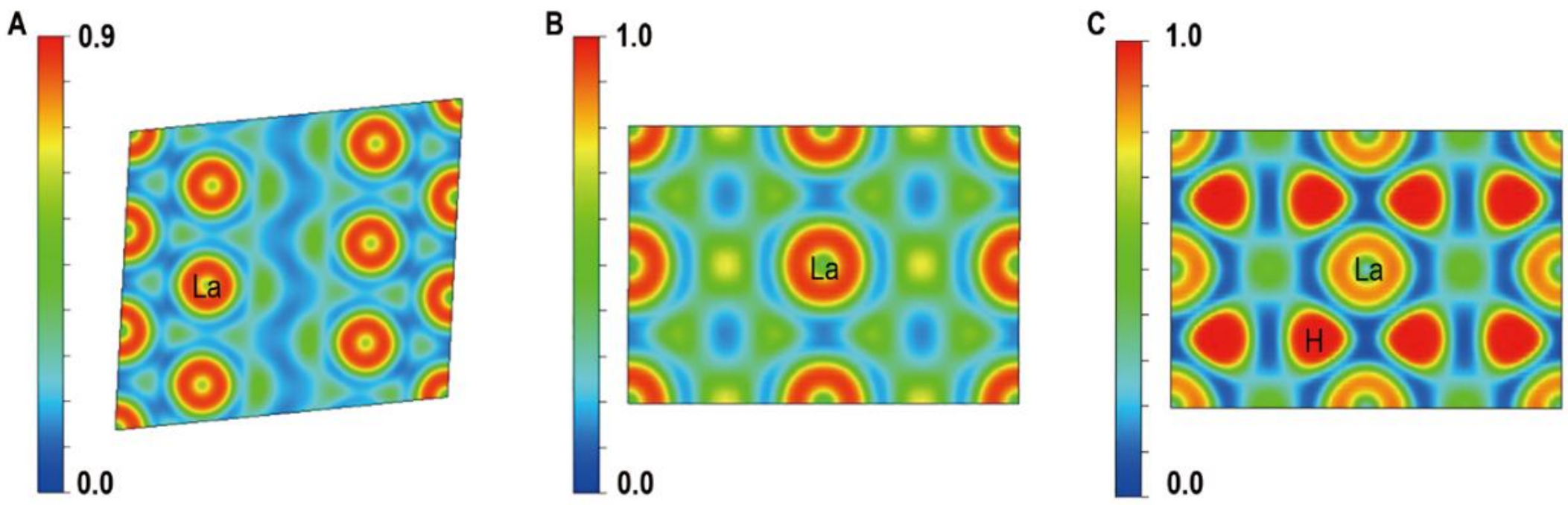

**Fig. S50. ELFs of La ($P6_3/mmc$), La ($Fm\bar{3}m$) and $LaH_2$ ($Fm\bar{3}m$).**

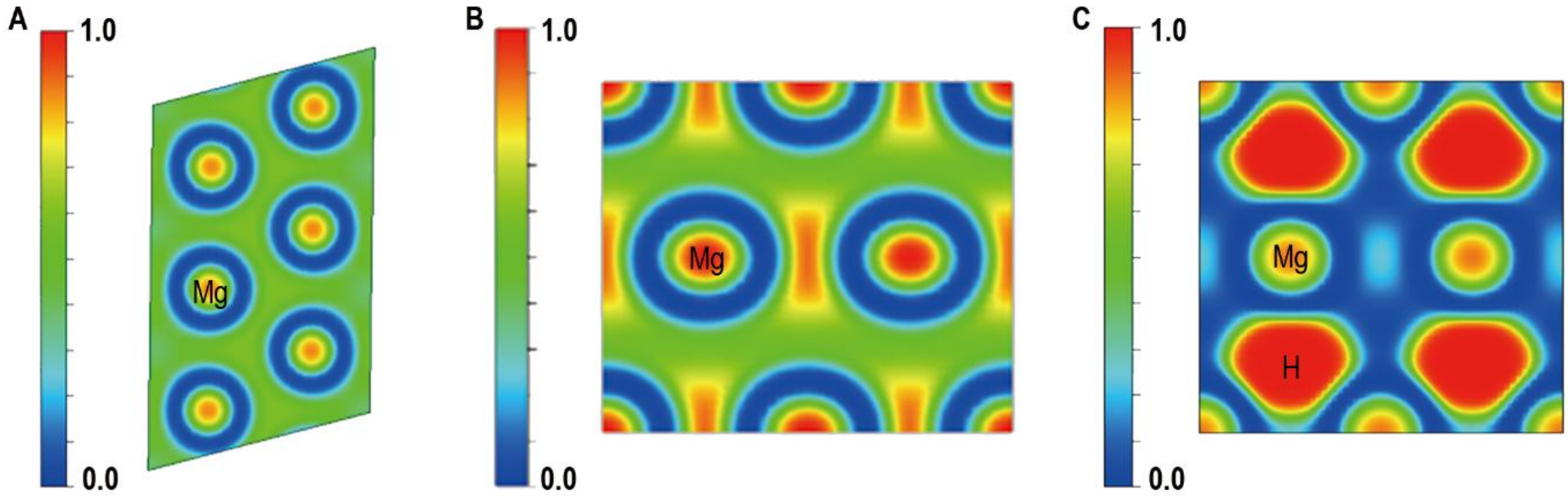

**Fig. S51. ELFs of Mg ($P6_3/mmc$), Mg ($P4_2/mnm$) and $MgH_2$ ($P4_2/mnm$).**

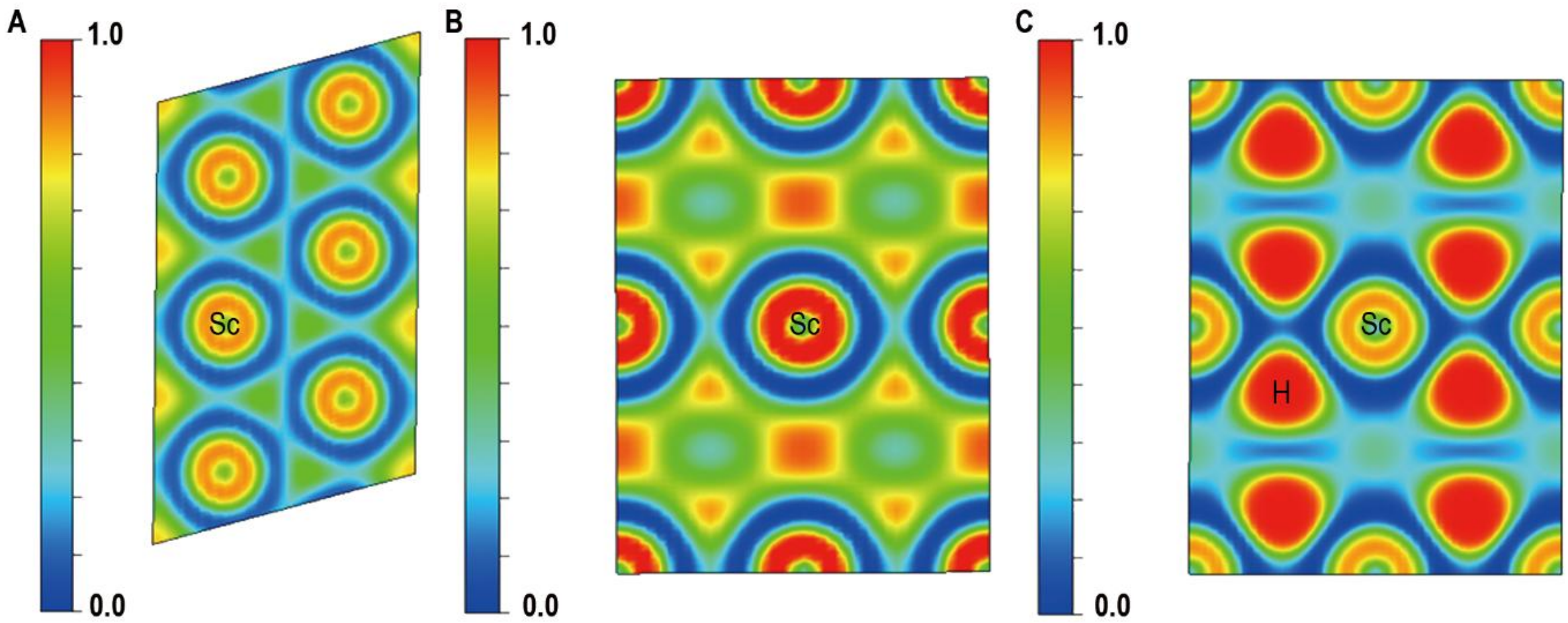

**Fig. S52. ELFs of Sc (*P*$6_3$*/mmc*), Sc (*Fm*$\bar{3}$*m*) and $ScH_2$ (*Fm*$\bar{3}$*m*).**

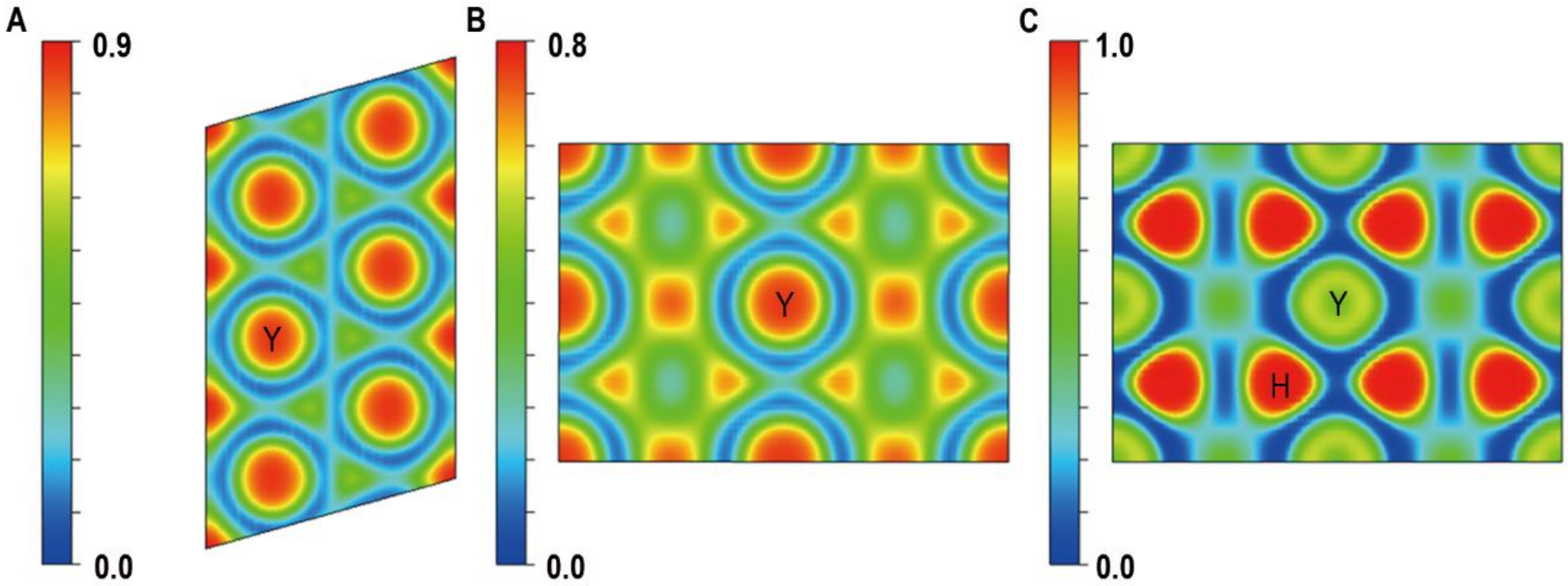

**Fig. S53. ELFs of Y (*P*6₃/*mmc*), Y (*Fm*$\bar{3}$*m*) and YH₂ (*Fm*$\bar{3}$*m*).**

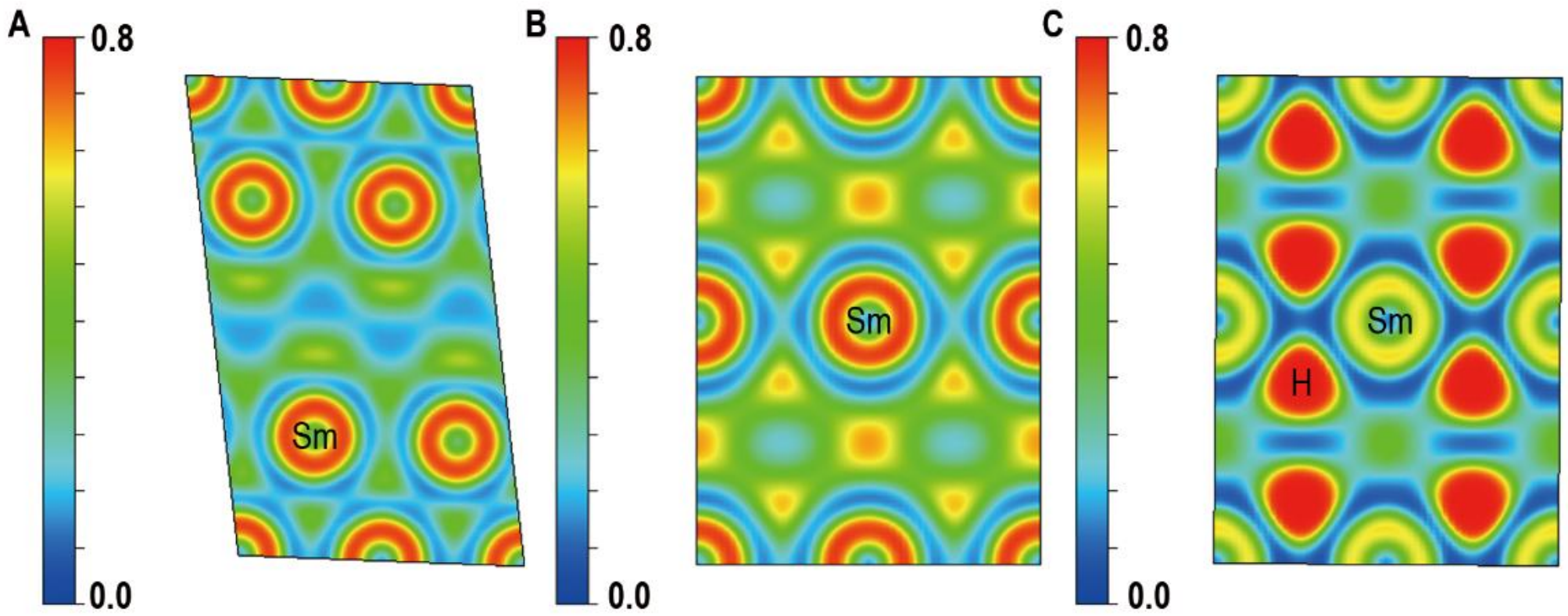

**Fig. S54. ELFs of Sm (*P*$6_3$/*mmc*), Sm (*Fm*$\bar{3}$*m*) and Sm$H_2$ (*Fm*$\bar{3}$*m*).**

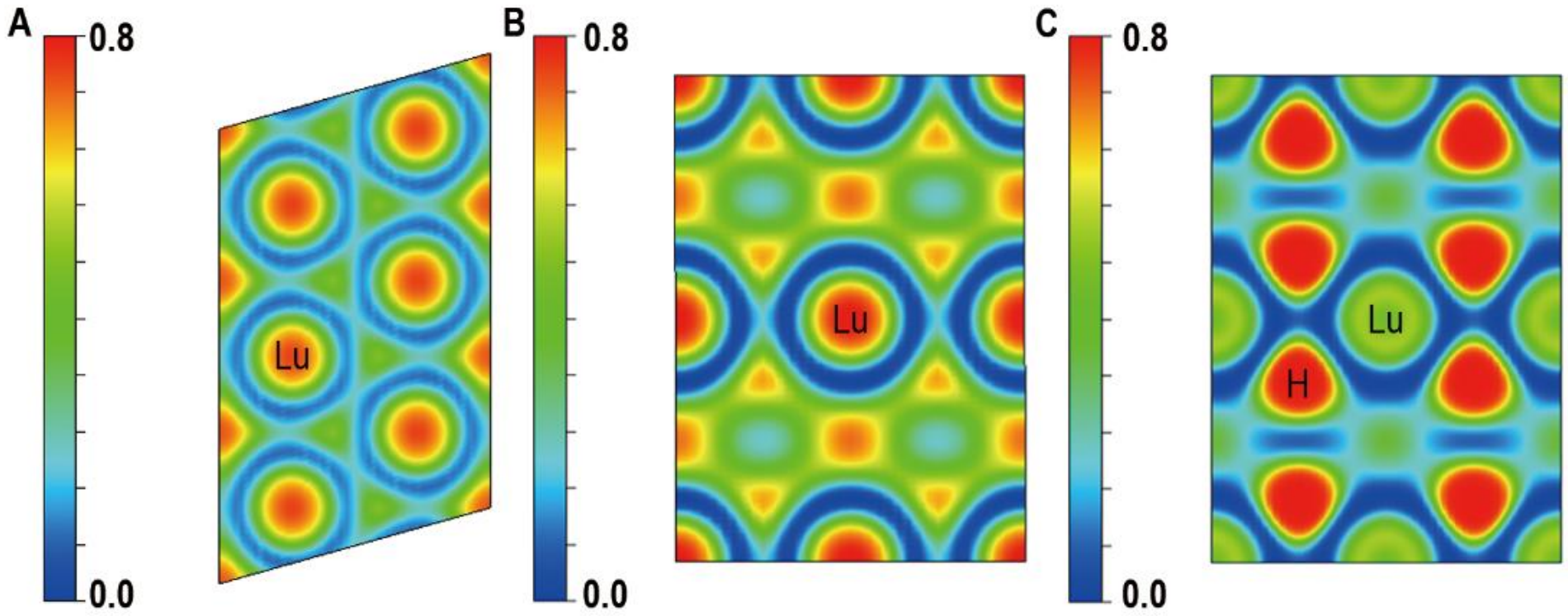

**Fig. S55. ELFs of Lu ($P6_3/mmc$), Lu ($Fm\bar{3}m$) and $LuH_2$ ($Fm\bar{3}m$).**

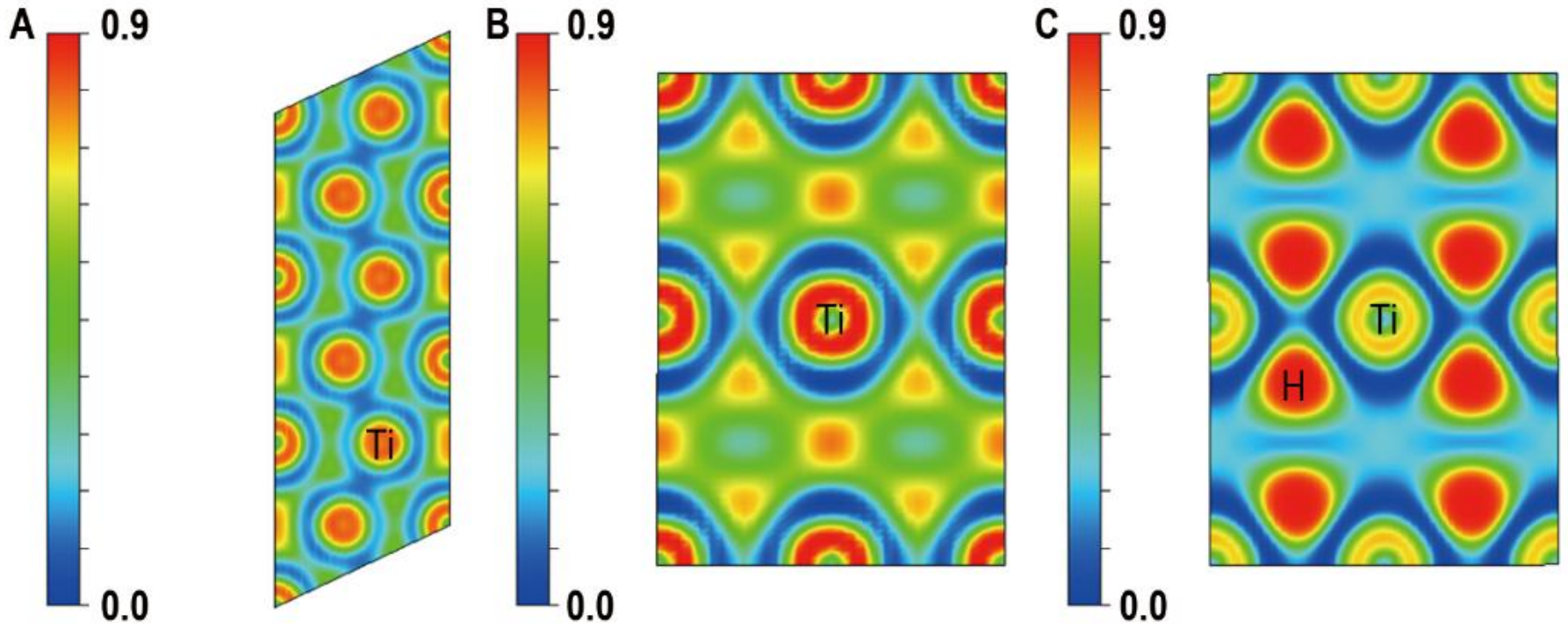

**Fig. S56. ELFs of Ti ($P6_3/mmc$), Ti ($Fm\bar{3}m$) and $TiH_2$ ($Fm\bar{3}m$).**

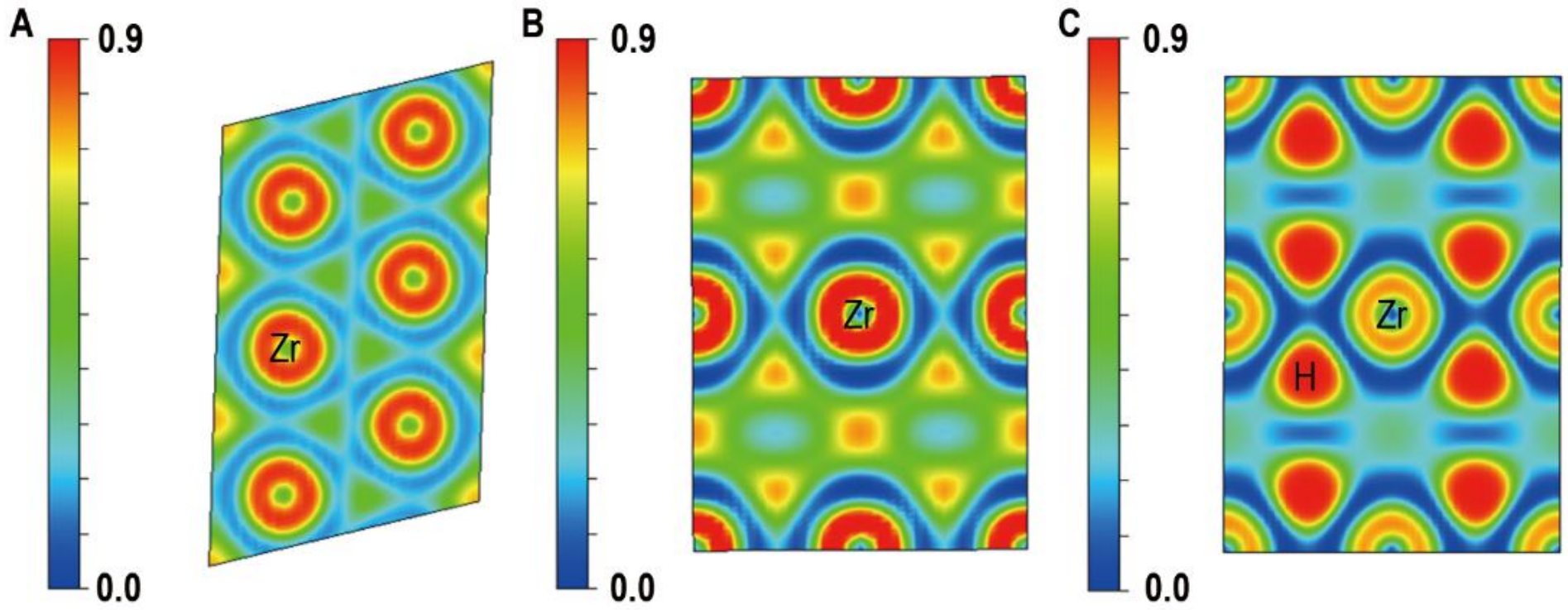

**Fig. S57. ELFs of Zr (*P*$6_3$/*mmc*), Zr (*Fm*$\bar{3}$*m*) and $ZrH_2$ (*Fm*$\bar{3}$*m*).**

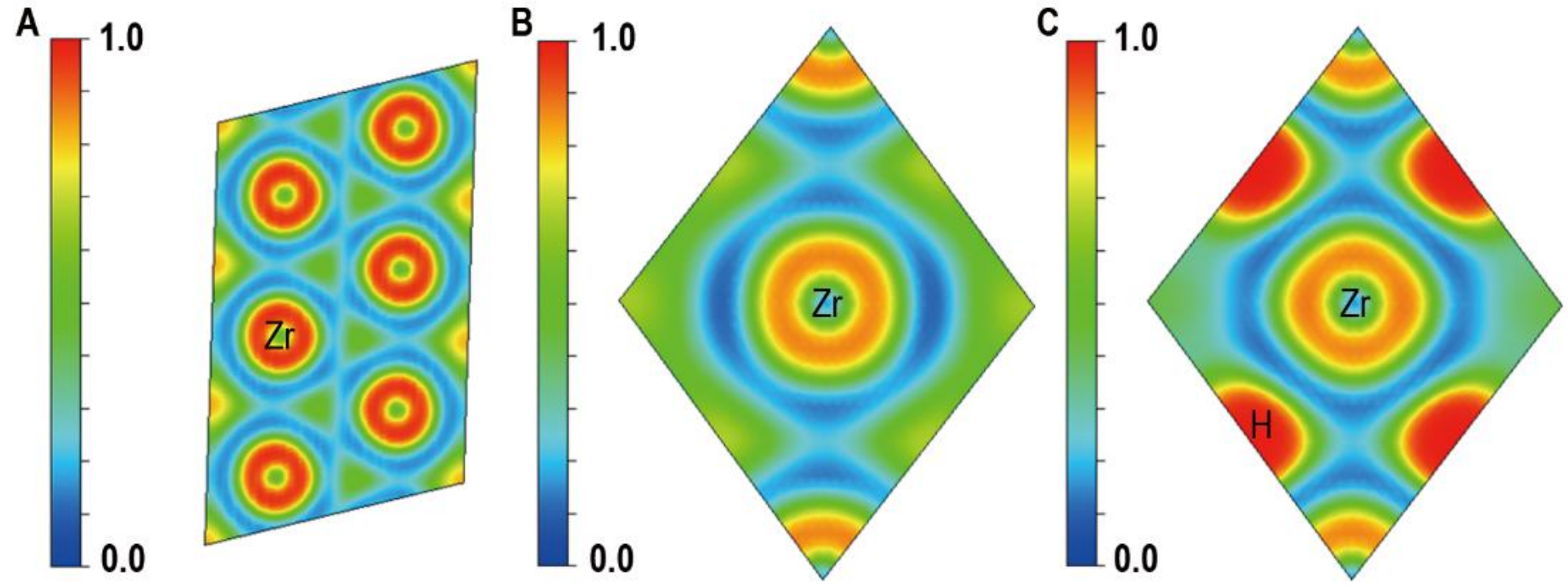

**Fig. S58. ELFs of Zr (*P*$6_3$/*mmc*), Zr (*I*4/*mmm*) and $ZrH_2$ (*I*4/*mmm*).**

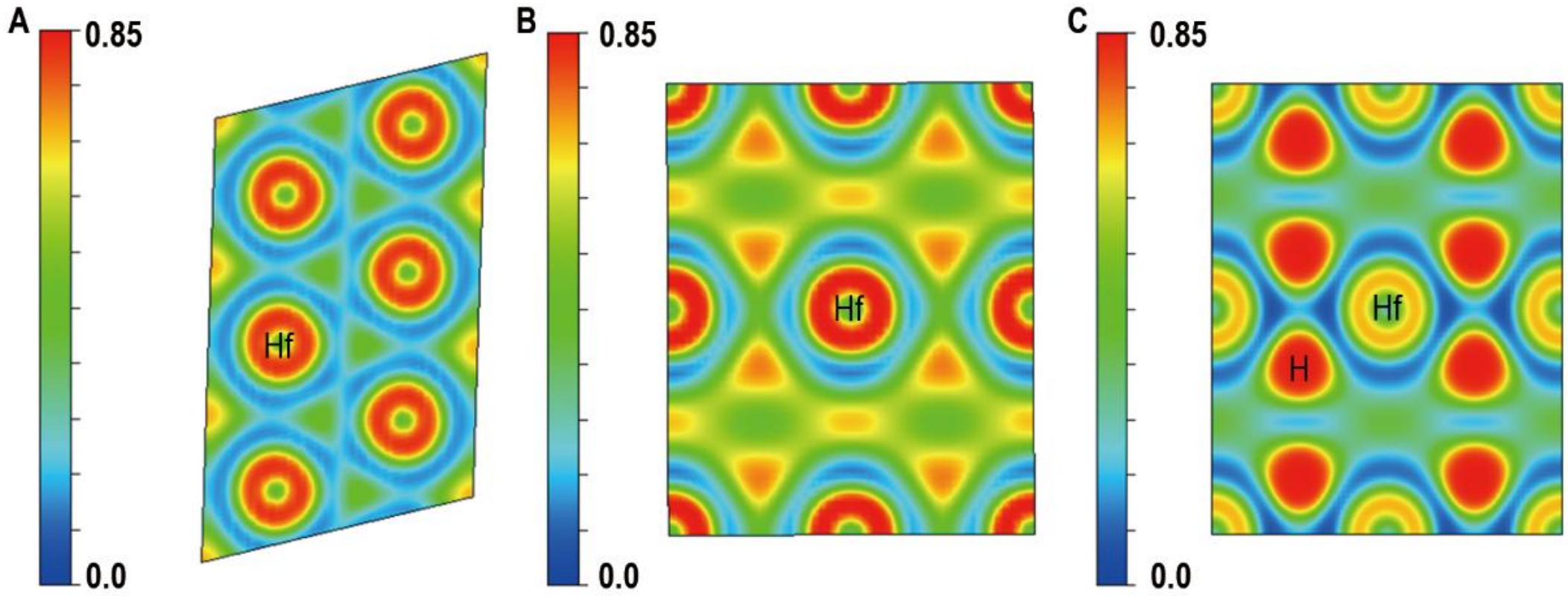

**Fig. S59. ELFs of Hf ($P6_3/mmc$), Hf ($I4/mmm$) and $HfH_2$ ($I4/mmm$).**

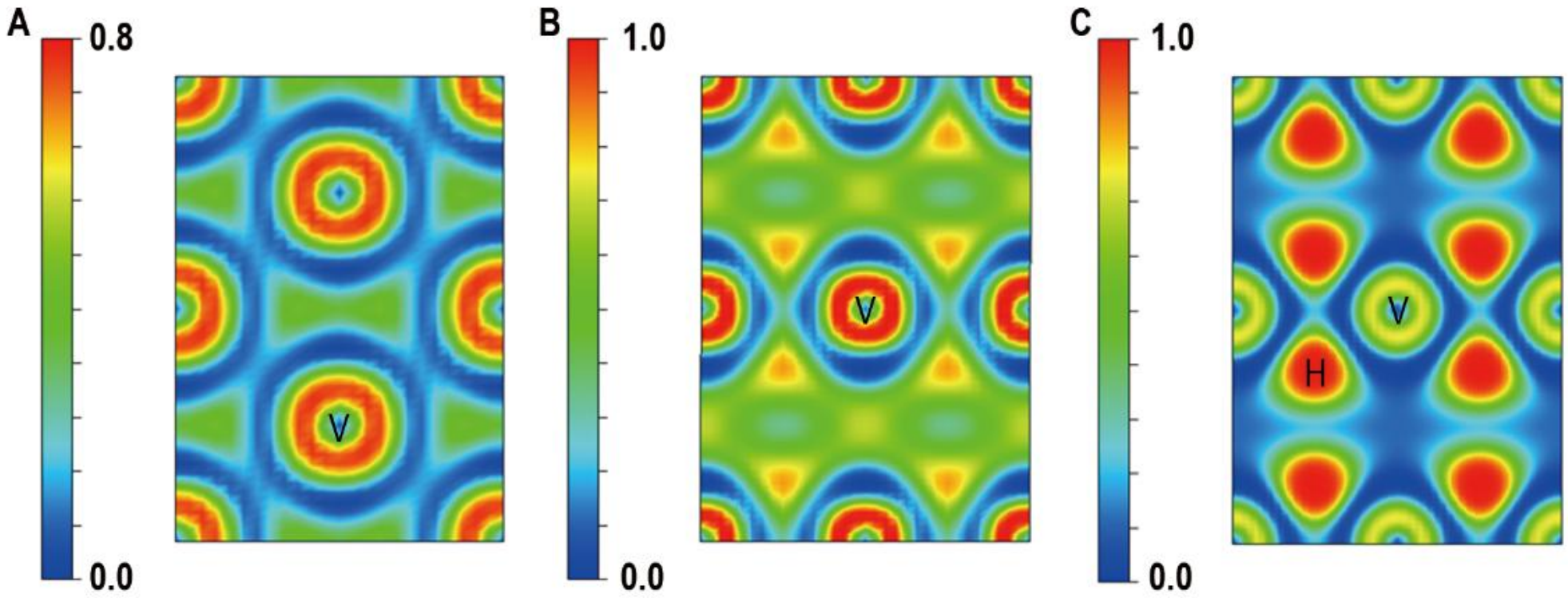

**Fig. S60. ELFs of V (*Im*$\overline{3}$*m*), V (*Fm*$\overline{3}$*m*) and $VH_2$ (*Fm*$\overline{3}$*m*).**

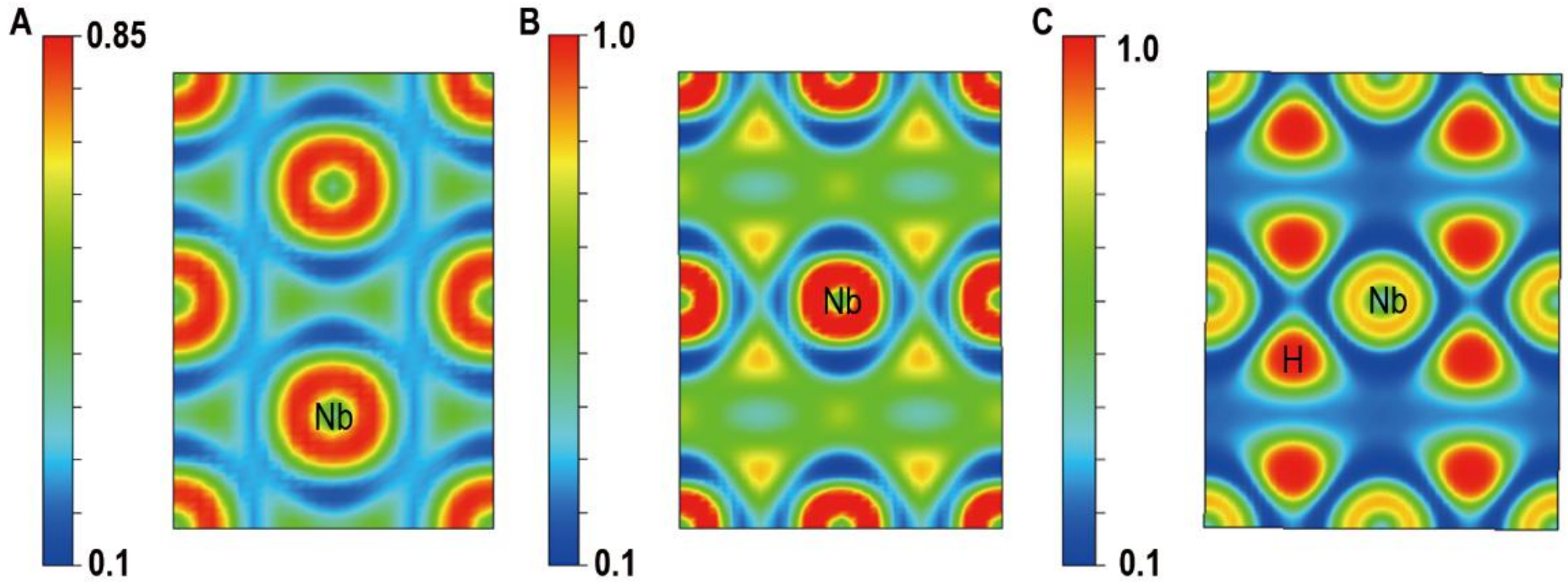

**Fig. S61. ELFs of Nb ($Im\overline{3}m$), Nb ($Fm\overline{3}m$) and $NbH_2$ ($Fm\overline{3}m$).**

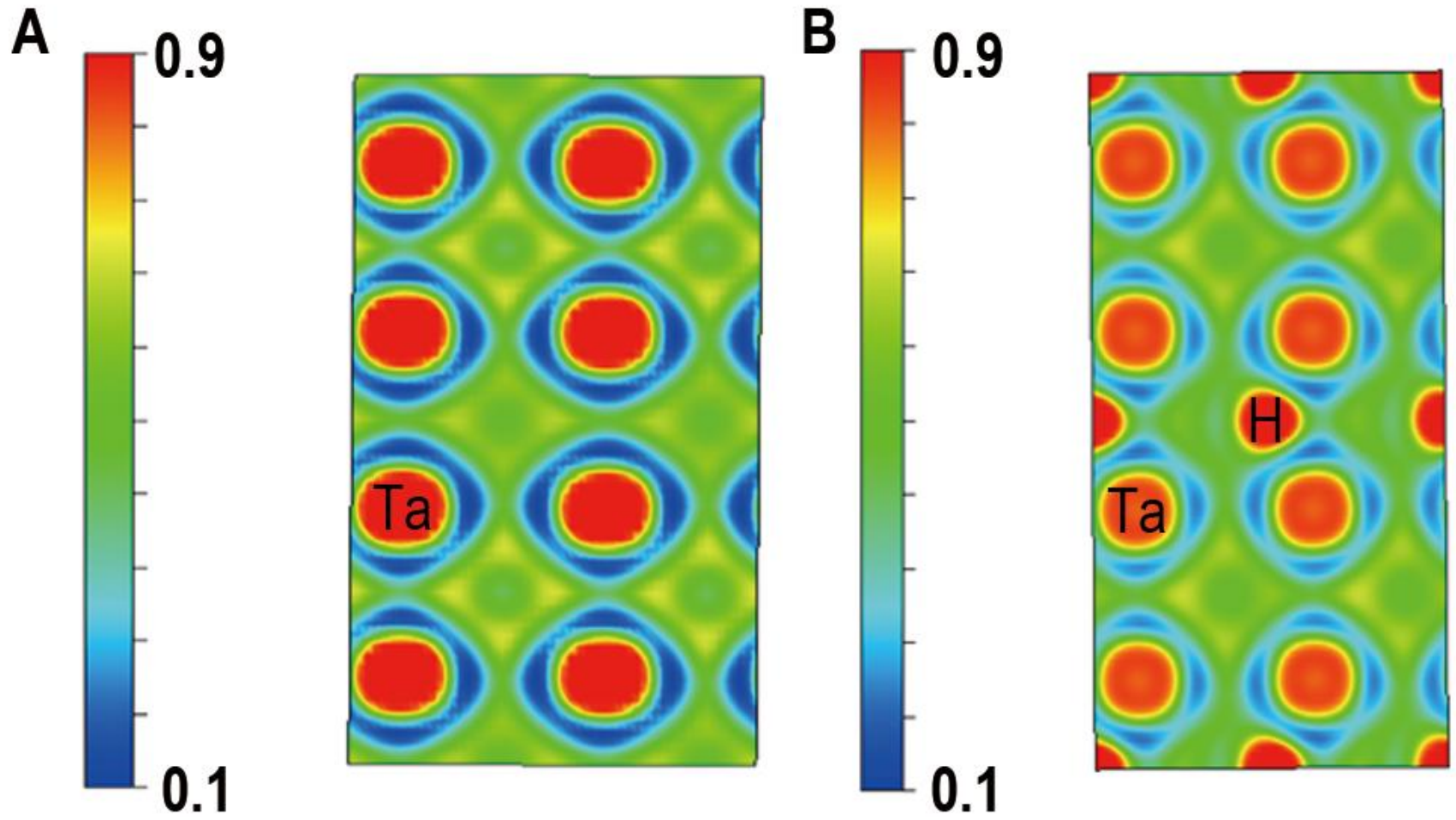

**Fig. S62. ELFs of Ta ($Im\overline{3}m$) and TaH ($C222$).**

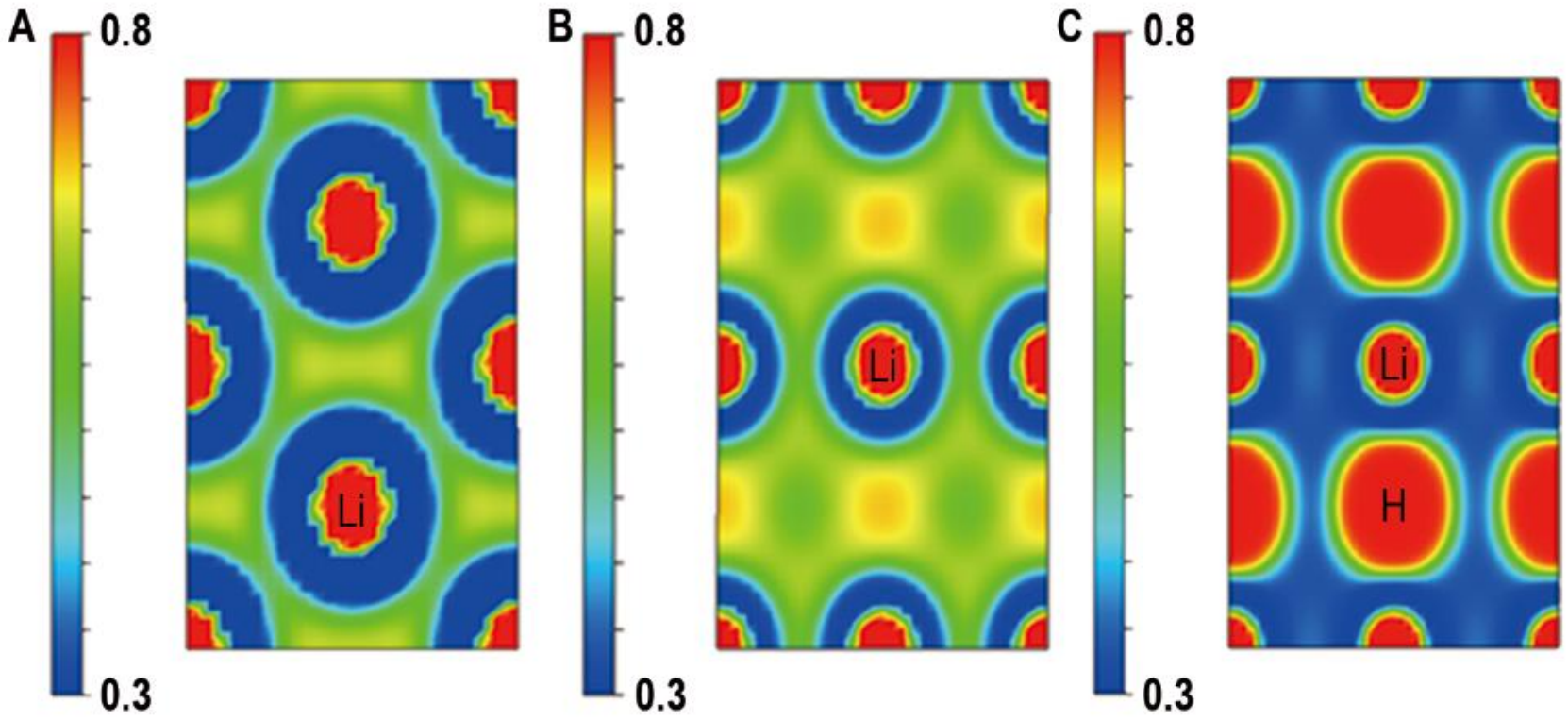

**Fig. S63. ELFs of Li (*Im*$\overline{3}$*m*), Li (*Fm*$\overline{3}$*m*) and LiH (*Fm*$\overline{3}$*m*).**

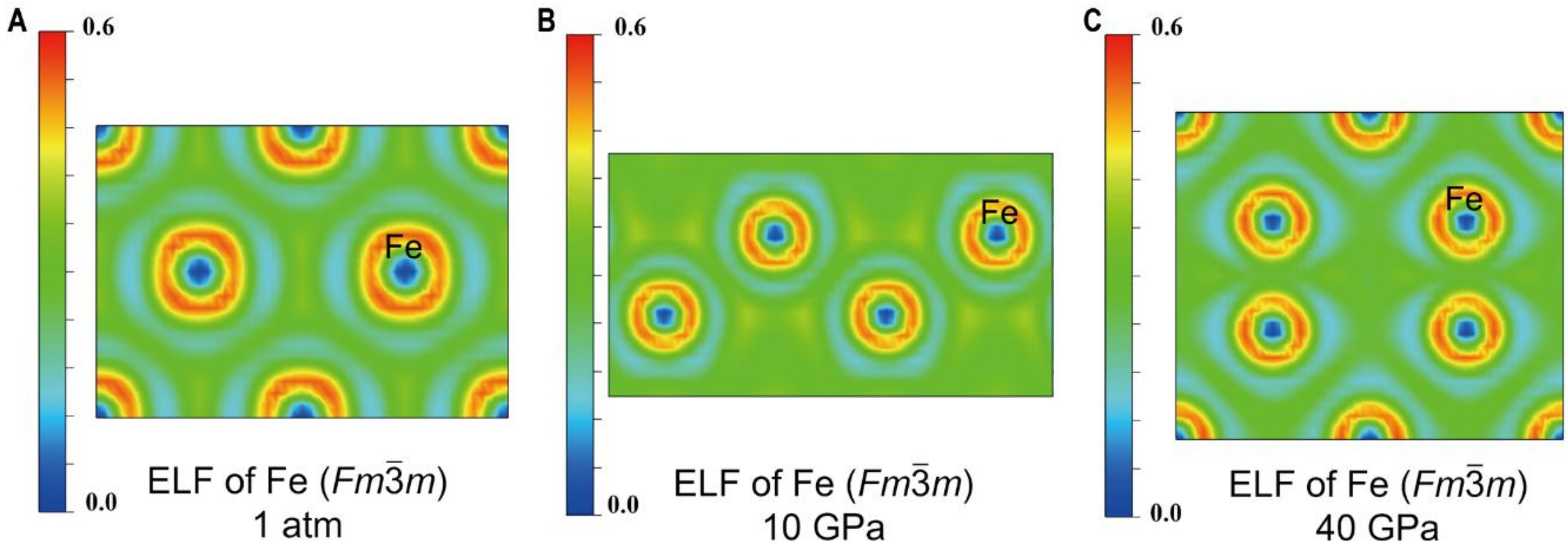

**Fig. S64. ELFs of Fe in 1 atm, 10 GPa and 40 GPa.**

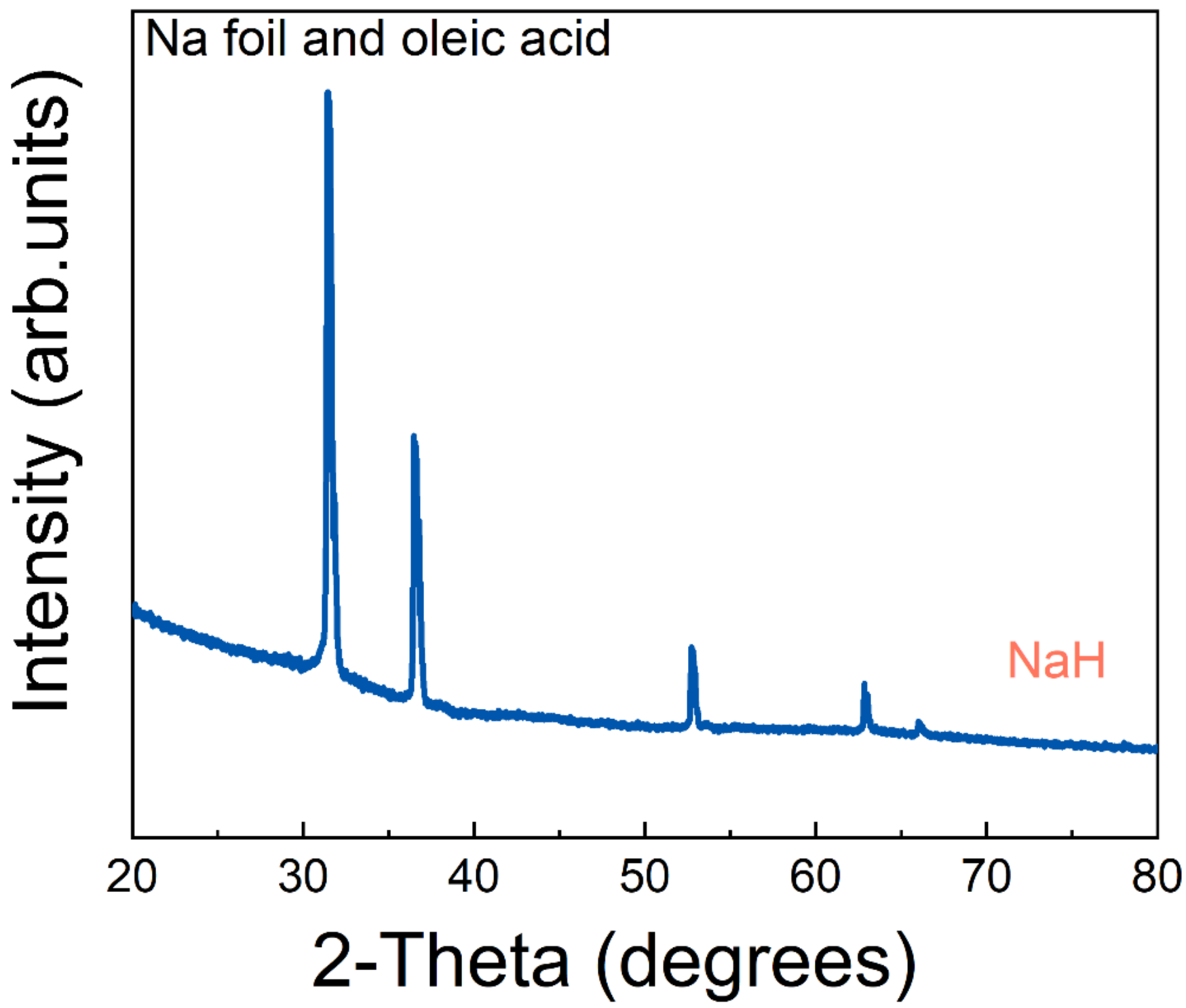

**Fig. S65. XRD of HTC-NaH.**

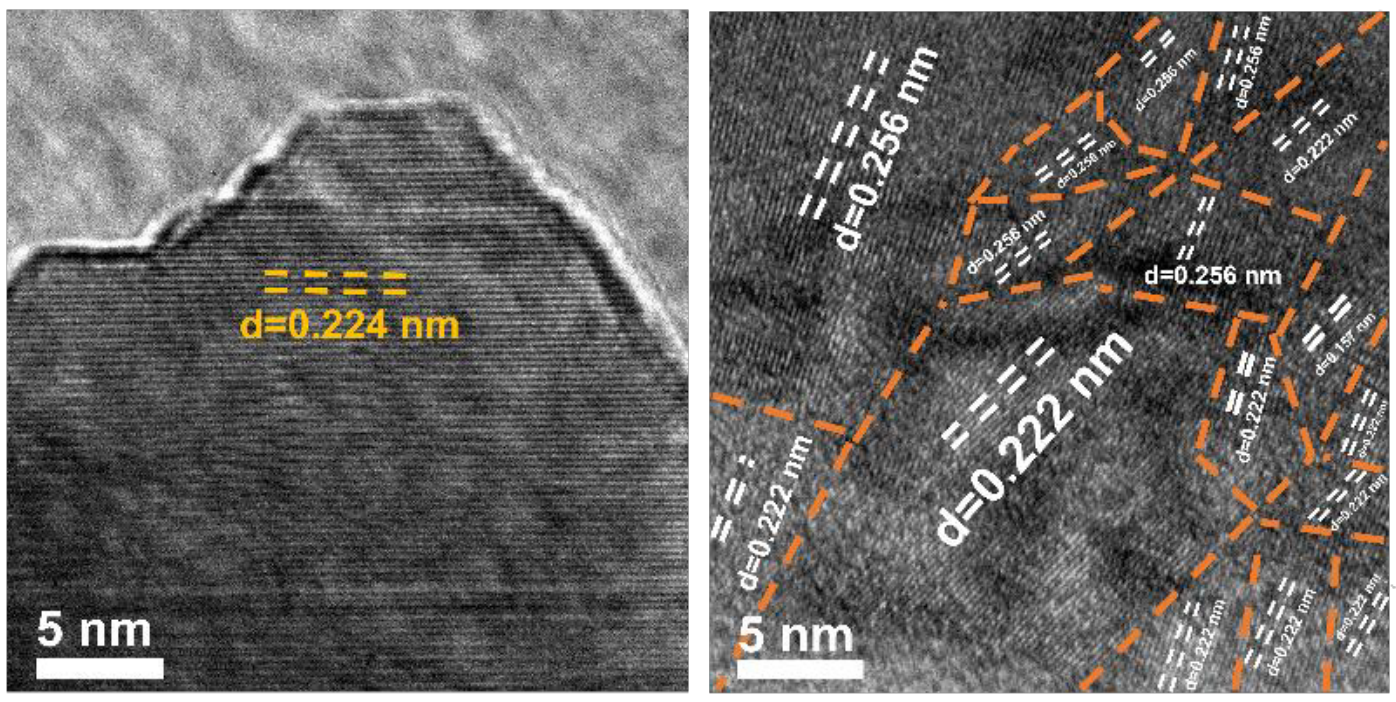

**Fig. S66. Representative TEM images of Ti FP and HTC-$TiH_2$.**

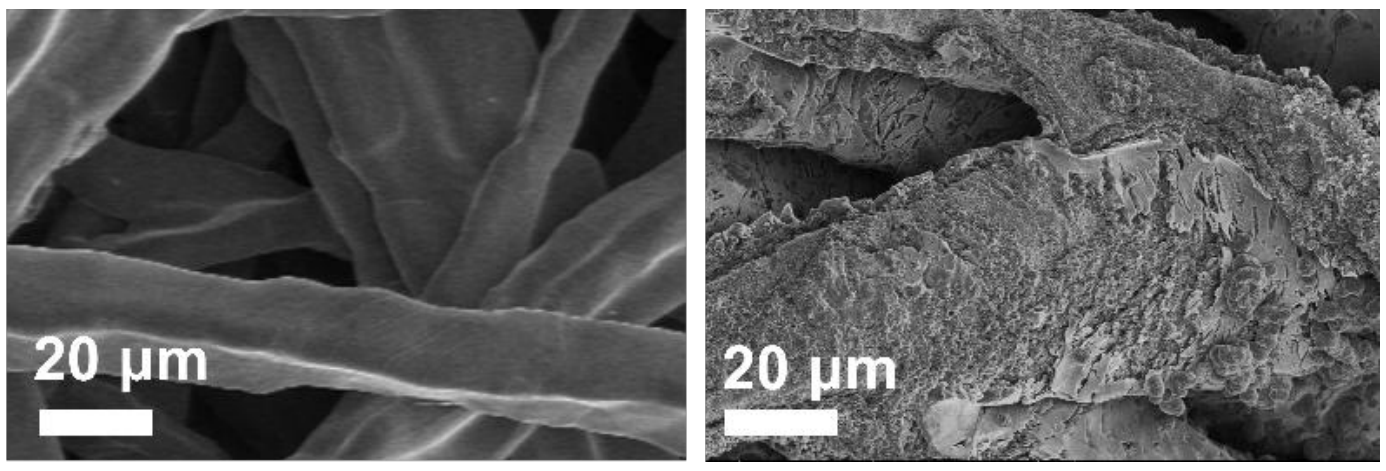

**Fig. S67. Representative SEM images of Ti FP and HTC-TiH2.**

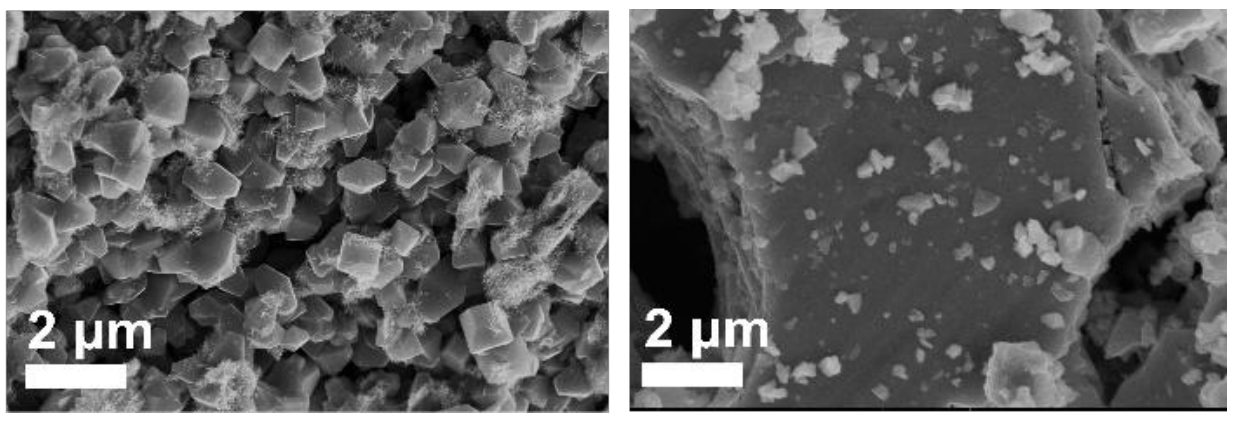

**Fig. S68. Representative SEM images of HTC-$TiH_2$ and $TiH_2$ (con.).**

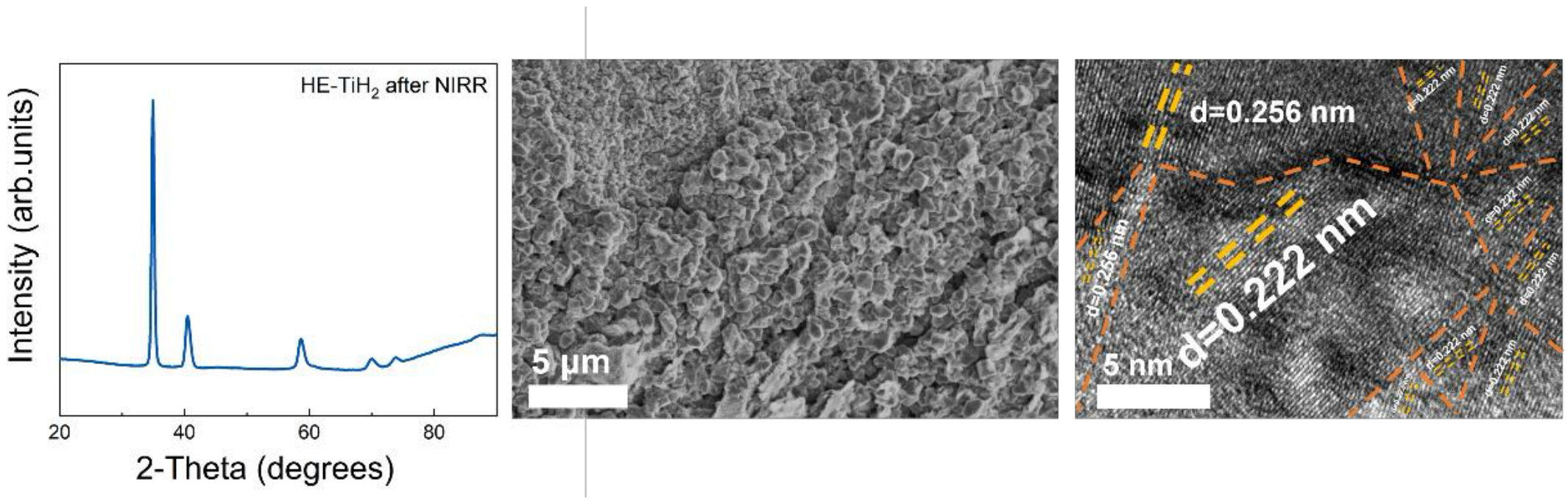

**Fig. S69. XRD phase analysis and Representative SEM and TEM images of HTC-TiH$_2$ after NIRR.**

| $MH_x$ | $\Delta P_{eq}$ | $\Delta P_{ph}$ | Stable under 1 atm |
|---|---|---|---|
| $MgH_2$ | -4.7 GPa | ≤0.1 GPa [10] | √ *exp.* |
| LaH | 0.65 GPa | 11 GPa [11] | ×, 8 GPa [11] |
| $LaH_2$ | -2.9 GPa | ≤1.7 GPa [12]*exp.* | √ *exp.* |
| $LaH_{2.3}$ | -3.0 GPa | ≤1.7 GPa [12]*exp.* | √ *exp.* |
| $ScH_2$ | -2.6 GPa | ≤0.01 GPa [13] | √ *exp.* |
| $YH_2$ | -1.33 GPa | ≤0.01 GPa [14] | √ *exp.* |
| $YH_3$ | -3.46 GPa | ≤4 GPa [14] | √ *exp.* |
| $SmH_2$ | -4 GPa | ≤0.1 GPa [15] | √ *exp.* |
| $LuH_2$ | -2.5 GPa | ≤2 GPa [8] | √ *exp.* |
| $TiH_2$ | -16.1 GPa | ≤0.01 GPa [16,17] | √ *exp.* |
| $ZrH_{1.6}$ | -12.25 GPa | ≤1 GPa [18] *exp.* | √ *exp.* |
| $ZrH_2$ | -13.6 GPa | ≤1 GPa [18] *exp.* | √ *exp.* |
| $HfH_{1.7}$ | -4.6 GPa | ≤4 GPa [19] | √ *exp.* |
| $HfH_2$ | -5.6 GPa | ≤4 GPa [19] | √ *exp.* |
| $VH_{0.8}$ | -16.3 GPa | ≤0.01 GPa [20] | √ *exp.* |
| $VH_2$ | -28.8 GPa | ≤0.01 GPa [20] | √ *exp.* |
| NbH | -1.24 GPa | ≤1 GPa [21] | √ *exp.* |
| $NbH_2$ | -26.4 GPa | ≤22 GPa [21] | √ *exp.* |
| $Ta_2H$ | -6 GPa | ≤5 GPa [22] | √ *exp.* |
| TaH | -16 GPa | ≤5 GPa [22] | √ *exp.* |
| FeH | -13.6 | 3.5 GPa [23] | ×, 3.2 GPa [23] |
| $FeH_2$ | -40.7 GPa | 67 GPa [24] | ×, 23 GPa [24] |
| LiH | -3.76 GPa | 0.05 GPa [25] *exp.* | √ *exp.* |

**Table. S1. $\Delta P_{eq}$ and $\Delta P_{ph}$ of $MH_x$.**

## References and Notes

1 Pickard, C. J. & Needs, R. Ab initio random structure searching. *J. Phys.: Condens. Matter* **23**, 053201 (2011).

2 Segall, M. *et al.* First-principles simulation: ideas, illustrations and the CASTEPcode. *J. Phys.: Condens. Matter* **14**, 2717 (2002).

3 Blöchl, P. E. Projector augmented-wave method. *Phys. Rev. B* **50**, 17953 (1994).

4 Kresse, G. G. kresse and d. joubert, phys. rev. b 59, 1758 (1999). *Phys. Rev. B* **59**, 1758 (1999).

5 Kresse, G. & Furthmüller, J. Efficiency of ab-initio total energy calculations for metals and semiconductors using a plane-wave basis set. *Comput. Mater. Sci.* **6**, 15 (1996).

6 Perdew, J. P., Burke, K. & Ernzerhof, M. Generalized Gradient Approximation Made Simple. *Phys. Rev. Lett.* **77**, 3865 (1996).

7 Giannozzi, P. *et al.* QUANTUM ESPRESSO: a modular and open-source software project for quantumsimulations of materials. *J. phys.: Condens. matter* **21**, 395502 (2009).

8 Cai, S. *et al.* No evidence of superconductivity in a compressed sample prepared from lutetium foil and $H_2/N_2$ gas mixture. *Matter Radiat. Extremes* **8** (2023).

9 Silvi, B. & Savin, A. Classification of chemical bonds based on topological analysis of electron localization functions. *Nature* **371**, 683 (1994).

10 Edalati, K., Yamamoto, A., Horita, Z. & Ishihara, T. High-pressure torsion of pure magnesium: Evolution of mechanical properties, microstructures and hydrogen storage capacity with equivalent strain. *Scr. Mater.* **64**, 880 (2011).

11 Machida, A. *et al.* Formation of NaCl-type monodeuteride LaD by the disproportionation reaction of LaD2. *Phys. Rev. Lett.* **108**, 205501 (2012).

12 Mulford, R. & Holley Jr, C. E. Pressure−temperature−composition of rare earth−hydrogen, systems. *The J. Phys. Chem.* **59**, 1222 (1955).

13 Manchester, F. & Pitre, J. The H-Sc (hydrogen-scandium) system. *J. phase equilib.* **18**, 194 (1997).

14 Wijngaarden, R. J. *et al.* Towards a metallic YH3 phase at high pressure. *J. Alloys Compd.* **308**, 44 (2000).

15 Manchester, F. & Pitre, J. The H-Sm system (hydrogen-samarium). *J. phase equilib.* **17**, 432 (1996).

16 Takasaki, A., Furuya, Y. & Taneda, Y. Hydrogen uptake in titanium aluminides in high pressure hydrogen. *Mater. Sci. Eng.: A* **239-240**, 265 (1997).

17 Gibb Jr, T. R., McSharry, J. J. & Bragdon, R. W. The titanium-hydrogen system and titanium hydride. II. Studies at high pressure. *J. Am. Chem. Soc.* **73**, 1751 (1951).

18 Kuzovnikov, M. A. *et al.* Synthesis of superconducting hcp-$ZrH_3$ under high hydrogen pressure. *Phys. Rev. Mater.* **7**, 024803 (2023).

19 Edalati, K., Horita, Z. & Mine, Y. High-pressure torsion of hafnium. *Mater. Sci. Eng.: A* **527**, 2136 (2010).

20 Kumar, S., Jain, A., Ichikawa, T., Kojima, Y. & Dey, G. K. Development of vanadium based hydrogen storage material: A review. *Renew. Sust. Energ. Rev.* **72**, 791 (2017).

21 Liu, G. *et al.* Nb-H system at high pressures and temperatures. *Phys. Rev. B* **95**, 104110 (2017).

22 Kuzovnikov, M. A. *et al.* Neutron scattering study of tantalum dihydride. *Phys. Rev. B* **102**, 024113 (2020).

23 Ishimatsu, N. *et al.* Hydrogen-induced modification of the electronic structure and magnetic states in Fe, Co, and Ni monohydrides. *Phys. Rev. B* **86**, 104430 (2012).

24 Pépin, C. M., Dewaele, A., Geneste, G., Loubeyre, P. & Mezouar, M. New iron hydrides under high pressure. *Phys. rev. lett.* **113**, 265504 (2014).

25 Howie, R. T., Narygina, O., Guillaume, C. L., Evans, S. & Gregoryanz, E. High-pressure synthesis of lithium hydride. *Phys. Rev. B-Cond. Matter Mater. Phys.* **86**, 064108 (2012).